\documentclass[10pt]{JHEP3}
\pdfoutput=1

\pdfmapfile{+fourier.map}

\usepackage{fourier}
\usepackage[scaled=0.875]{helvet}

\usepackage{slashed,cancel}
\usepackage{amsmath}
\usepackage{amsfonts}
\usepackage{amssymb}
\usepackage{graphicx}
\usepackage{bbm}

\numberwithin{equation}{section}

\DeclareMathOperator\Trace{tr}
\DeclareMathOperator\sign{sgn}
\DeclareMathOperator{\re}{Re}
\DeclareMathOperator{\im}{Im}
\DeclareMathOperator{\PV}{PV}
\DeclareMathOperator*{\Res}{Res}
\newcommand{\Csing}{\Delta}
\newcommand{\I}[1]{{\cal H}_{\nu,#1}(\ha,\hm)} 
\newcommand{\Ith}[1]{{\cal H}_{\nu,#1}^{\text{th}}(\ha,\hm)}
\newcommand\ev[1]{\left\langle{#1}\right\rangle}
\newcommand{\ha}{\hat \alpha}
\newcommand{\hm}{\hat m}
\newcommand{\hz}{\hat z}
\newcommand{\hp}{p_s}
\newcommand{\hr}{r_s}
\newcommand{\hw}{w_s}
\newcommand{\ph}{\bigl\langle e^{2i\theta} \bigr\rangle}
\newcommand{\phs}{\bigl\langle e^{2i\theta}_s \bigr\rangle}
\newcommand{\chrmt}{chRMT}
\newcommand{\Nf}{{N_f}}
\newcommand{\DI}{{\mathcal D}_\nu^{N_f}(\{\hm_f\})}
\renewcommand{\epsilon}{\varepsilon}

\title{Random matrix analysis of the QCD sign problem for general
  topology}

\author{Jacques Bloch and Tilo Wettig\\
  Institute for Theoretical Physics, University of Regensburg, 93040
  Regensburg, Germany\\ 
  \mbox{Email: \email{jacques.bloch@physik.uni-regensburg.de},
  \email{tilo.wettig@physik.uni-regensburg.de}}}

\preprint{December 1, 2008; revised January 15, 2009}

\abstract{ Motivated by the important role played by the phase of the
  fermion determinant in the investigation of the sign problem in
  lattice QCD at nonzero baryon density, we derive an analytical
  formula for the average phase factor of the fermion determinant for
  general topology in the microscopic limit of chiral random matrix
  theory at nonzero chemical potential, for both the quenched and the
  unquenched case. The formula is a nontrivial extension of the
  expression for zero topology derived earlier by Splittorff and
  Verbaarschot.  Our analytical predictions are verified by detailed
  numerical random matrix simulations of the quenched theory.  }


\begin{document}

\section{Introduction}

The theory of the strong interactions, also called Quantum
Chromodynamics (QCD), describes the interactions between quarks and
gluons and is responsible for the existence of hadrons.
Lattice-regularized QCD allows for the description of low-energy
properties and other nonperturbative phenomena in QCD and has the
salient property that it can be systematically improved towards the
continuum limit. In lattice QCD, space-time is discretized and the
functional integral of the quantum field theory is performed by a
Markov-chain Monte-Carlo method.

An important subject of study is the behavior of QCD in an environment
exhibiting an abundance of particles over anti-particles. Such
conditions arise, e.g., in astrophysical objects like neutron stars,
and can be reproduced in heavy-ion collision experiments. Part of the
interest comes from the existence of various phases in QCD, which are
usually exemplified by means of the QCD phase diagram
\cite{Stephanov:2007fk}. To study QCD at nonzero baryon density a
quark chemical potential is introduced in the QCD Lagrangian. (In the
following, we will omit the qualifier ``quark'' and only speak of a
chemical potential.) In the presence of a chemical potential the QCD
Dirac operator is no longer anti-Hermitian, i.e., its eigenvalues
spread into the complex plane and its determinant will generically be
complex.

In lattice QCD the effect of dynamical fermions can be integrated out,
leaving behind the determinant of the Dirac operator. Dynamical
lattice simulations for QCD at nonzero chemical potential are
problematic because the fermion determinant is complex and hence its
real part can be negative, which prohibits its incorporation in the
weight of the Monte-Carlo sampling.  This is the so-called sign
problem, which also occurs in other theories and has been the subject
of a large number of investigations in recent years (for an incomplete
list see, e.g.,
Refs.~\cite{Ambjorn:2002pz,Ambjorn:2004jk,Troyer:2004,Osborn:2005ss,Imachi:2006mw,Fukushima:2006uv,Ejiri:2007ga,Aarts:2008wh}).
While many of these works are concerned with a solution of the sign
problem by various clever ideas, here we concentrate on an analytical
study of the sign problem, in the hope that the results we derive will
contribute to its solution.  The severeness of the sign problem
depends on the magnitude of the chemical potential, and it is
therefore illuminating to investigate the relation between the phase
factor of the determinant and the chemical potential.

Chiral random matrix theory (\chrmt) is a useful auxiliary in the
study of the spectral properties of the Dirac operator in QCD
\cite{Shuryak:1992pi,Verbaarschot:2000dy,Verbaarschot:2005rj}. Indeed,
to leading order in the $\epsilon$-regime of QCD the spectral
properties of the Dirac operator are universal and can be described by
\chrmt\ \cite{Basile:2007ki}.  In the presence of a chemical potential
this correspondence is still valid even though the Dirac operator is
now non-Hermitian. Appropriate random matrix models have been
developed
\cite{Stephanov:1996ki,Akemann:2002ym,Akemann:2002js,Osborn:2004rf},
and their correspondence with QCD at nonzero chemical potential was
verified successfully, see Ref.~\cite{Akemann:2007rf} for a review.
The agreement of the microscopic spectral properties of the Dirac
operator with the predictions of chiral random matrix theory has been
confirmed for quenched lattice QCD simulations with chemical potential
using the staggered operator \cite{Akemann:2003wg}, and more recently
using the overlap operator \cite{Bloch:2006cd,Akemann:2007yj}. The
latter operator has the interesting property that it satisfies the
Ginsparg-Wilson relation and the trace anomaly at finite lattice
spacing and can therefore have exact zero modes
\cite{Ginsparg:1981bj,Narayanan:1993sk,Narayanan:1994gw,Neuberger:1997fp,Luscher:1998pqa,Bloch:2007xi}.
This allowed us to verify the predictions of \chrmt\ at nonzero
chemical potential for both zero and nonzero topology.  The comparison
between lattice QCD and {\chrmt} also allows for a determination of
the low-energy constants $\Sigma$ and $F$ of chiral perturbation
theory.

Motivated by this agreement, we expect that a study of the behavior of
the fermion determinant in \chrmt\ will give us, in certain
well-defined limits, important information about the sign problem that
is encountered in dynamical QCD simulations at nonzero chemical
potential.  In Ref.~\cite{Splittorff:2007ck} Splittorff and
Verbaarschot derived a solution for the average phase factor of the
determinant in the microscopic limit of QCD (see Sec.~\ref{sec:micro}
for a description of this limit) for the case of trivial topology
using \chrmt\ at nonzero chemical potential.  However, to compare the
overlap data of Ref.~\cite{Bloch:2006cd} with {\chrmt} one also needs
the RMT predictions for the average phase factor for general topology.
The derivation of a formula for general topology is the main goal of
this paper.  As will be seen, the final expression contains two
distinct parts.  The first part is the generalization of the integrals
representing the solution in Ref.~\cite{Splittorff:2007ck} from zero
to arbitrary topology. The second part is a low-degree bivariate
polynomial in mass and chemical potential which is absent for
topological charge $\nu=0$.  For $\nu \neq 0$ it gives an important
contribution to the average phase factor of the fermion determinant,
especially for small mass.  As the mass goes to zero only this term
remains and completely determines the value of the average phase.
  
An important ingredient of the derivation is the ability to write the
phase factor of the determinant as a ratio of characteristic
polynomials. This quantity is recurrent in random matrix studies, both
for theories with real \cite{Fyodorov:2002jw,Strahov:2002zu} and with
complex eigenvalues \cite{Akemann:2004zu}, and its average can be
computed in terms of Cauchy transforms of the orthogonal polynomials
of the theory. To determine the phase factor of the determinant, the
relevant Cauchy transform was computed in
Ref.~\cite{Splittorff:2007ck} and expressed in terms of
one-dimensional integrals for zero topology, i.e., for square random
matrices. In the present paper we extend the solution of the Cauchy
transform to the case of rectangular matrices.  This solution could
also be relevant for other applications, like those involving time
series, where one matrix dimension is typically much larger than the
other.

The structure of this paper is as follows.  In Sec.~\ref{Sec:RMT} we
describe the chiral random matrix model at nonzero chemical potential.
In Sec.~\ref{Sec:Phase} we show how the microscopic limit of the phase
of the fermion determinant, in both the quenched and the unquenched
case, can be formally computed for such a matrix model in terms of a
complex Cauchy transform integral.  This two-dimensional integral is
strongly oscillating, and in Sec.~\ref{Sec:Analytic} we apply and
extend the method of Ref.~\cite{Splittorff:2007ck} to transform this
integral into a much simpler and better behaved expression, involving
only one-dimensional integrals and a short double sum (or bivariate
polynomial). Explicit results for the quenched and unquenched cases as
well as for the chiral and thermodynamic limits are given in
Sec.~\ref{sec:results}.  In that section we also verify the analytical
predictions for the quenched case by random matrix simulations for
various values of the topological charge.  We conclude in
Sec.~\ref{Sec:concl}.  A number of technical details are worked out in
several appendices.

\section{Non-Hermitian chiral random matrix model}
\label{Sec:RMT}

To leading order in the $\epsilon$-regime of QCD the spectral
properties of the Dirac operator can be described by \chrmt. In the
presence of a chemical potential $\mu$ the Dirac operator $D$ is no
longer anti-Hermitian, and in the non-Hermitian chiral random matrix
model introduced by Osborn \cite{Osborn:2004rf} it takes the form
\begin{equation} D(\mu) =
  \begin{pmatrix}
    0 & i\Phi + \mu \Psi \\
    i \Phi^\dagger + \mu \Psi^\dagger & 0
  \end{pmatrix}\:,
  \label{Dran}
\end{equation}
where the matrices $\Phi$ and $\Psi$ are complex random matrices of
dimension $(N+\nu) \times N$, distributed according to the Gaussian
weight function
\begin{equation} 
  w(X) = (N/\pi)^{N(N+\nu)} \exp(-N \Trace X^\dagger X)\: .
\label{rmtdis}
\end{equation} 
For a detailed analysis of this model, see also
Ref.~\cite{Akemann:2004dr}.  For the conversion of random matrix units
to physical units, see the beginning of Sec.~\ref{sec:micro}.

The parameter $N$ will be taken to infinity when computing the
microscopic limit (see Sec.~\ref{sec:micro}).  The matrix in
Eq.~\eqref{Dran} has $|\nu|$ exact zero modes, which allows us to
identify $\nu$ with the topological charge.  In the following, we keep
$\nu$ fixed as $N\to\infty$ and assume without loss of generality that
$\nu\ge0$.  (For $\nu<0$ we can simply replace $\nu$ by $|\nu|$ in the
analytical results that will be computed below in the large-$N$
limit.)  The nonzero eigenvalues of $D(\mu)$ come in $N$ pairs
$(z_k,-z_k)$.  For $\mu=0$, the $z_k$ are purely imaginary.

For fixed $\nu$, the partition function of the random matrix model is
given by
\begin{equation}
  Z_\nu^{N_f}(\mu;\{m_f\}) = \int d\Phi d\Psi \, w(\Phi) w(\Psi)
  \prod_{f=1}^{N_f}\det(D(\mu)+m_f)\:, 
\end{equation}
where the integration measure is defined by
\begin{equation}
  dX=\prod_{k=1}^{N+\nu}\prod_{\ell=1}^Nd\re X_{k\ell}d\im X_{k\ell}\:,
\end{equation}
$N_f$ is the number of dynamical quarks, and the $m_f$ are the quark
masses.  The quenched case corresponds to $N_f=0$, i.e., the fermion
determinants are absent.

To perform the integration over $\Phi$ and $\Psi$, it is convenient to
go to an eigenvalue representation of the random matrix $D(\mu)$.  As
shown in Ref.~\cite{Osborn:2004rf}, the partition function can be
rewritten, up to a normalization constant that depends on $\mu$ and
$\nu$, as an integral over the $z_k$,
\begin{align}
  Z_\nu^{N_f}(\alpha;\{m_f\}) = \int_\mathbb{C} \prod_{k=1}^{N} d^2 z_k \: 
  w^\nu(z_k, z_k^*;\alpha) \: |\Delta_N(\{z^2\})|^2  \prod_{f=1}^{N_f}
  (m_f^2-z_k^2)\:,
  \label{eq:ZNf}
\end{align}
where we introduced $\alpha = \mu^2$, the integrals over the $z_k$ are
over the entire complex plane,
\begin{align}
  \Delta_N(\{z^2\}) \equiv \prod_{k>\ell} (z_k^2-z_\ell^2) 
\end{align}
is a Vandermonde determinant, the weight function is given by
\begin{equation}
  w^\nu(z,z^*;\alpha)  = |z|^{2\nu+2}
  \exp\left(-\frac{N(1-\alpha)}{4\alpha}(z^2+{z^*}^2)\right) 
  K_\nu\left(\frac{N(1+\alpha)}{2\alpha} |z|^2\right)\:,
\label{wnu}
\end{equation} 
and $K_\nu$ is a modified Bessel function.  The quenched partition
function will be denoted by $Z_\nu(\alpha)$.

The ensemble average of an observable $\mathcal O$ is given by
\begin{equation}
  \label{eq:qav}
  \ev{\mathcal O}_{\nu,N_f}=\frac1{Z_\nu^{N_f}}
  \int_\mathbb{C} \prod_{k=1}^{N} d^2 z_k \:      
  w^\nu(z_k, z_k^*;\alpha) \: |\Delta_N(\{z^2\})|^2  \prod_{f=1}^{N_f}
  (m_f^2-z_k^2)\: \mathcal O(z_1,\ldots,z_N) \:.
\end{equation}
When there is no danger of confusion we will omit one or both of the
subscripts on $\ev{\mathcal O}$.  

Our derivation will follow the general line of arguments given in
Ref.~\cite{Splittorff:2007ck} for $\nu=0$, with the necessary
generalizations to arbitrary topology $\nu$. To analyze the spectral
properties of the random matrix model it is useful to introduce the
orthogonal polynomials corresponding to the weight function
\eqref{wnu} \cite{Osborn:2004rf},
\begin{align}
  p_k^\nu(z;\alpha) = \left(\frac{1-\alpha }{N}\right)^k k! \:
  L_k^\nu\left(-\frac{N z^2}{1-\alpha }\right) \:, 
\label{pknu}
\end{align}
where $L_k^\nu(z)$ is the generalized or associated Laguerre
polynomial of order $\nu$ and degree $k$. These orthogonal
polynomials satisfy the orthogonality relation
\begin{align}
 \int_\mathbb{C} d^2z \: w^\nu(z,z^*;\alpha) p_k^\nu(z;\alpha)
 p_\ell^\nu(z;\alpha)^* = r_k^\nu(\alpha) \delta_{k\ell}  
\end{align}
with norm
\begin{align}
  r_k^\nu(\alpha) = \frac{\pi\alpha(1+\alpha)^{2k+\nu}k!(k+\nu)!} 
  {N^{2k+\nu+2}} \:. 
\label{rknu}
\end{align}
For later use we also introduce the Cauchy transform of the orthogonal
polynomials defined by
\begin{align}
  h_k^\nu(m;\alpha) = \int_\mathbb{C} \frac{d^2z}{z^2-m^2}
  w^\nu(z,z^*;\alpha)  p_k^\nu(z;\alpha)^* \:. 
\label{hknu}
\end{align}

\section{Phase factor of the fermion determinant} 
\label{Sec:Phase}

\subsection{The phase factor as a complex Cauchy transform}

The Dirac operator describing a massive fermion is defined as
$D(m;\mu) = D(\mu) + m \mathbbm{1}$, where we assume that $m$ is real.
If we write its determinant as $\det D(m;\mu) = r e^{i\theta}$, the
phase factor can be extracted by forming the ratio
\begin{align}
  \label{eq:ph}
  e^{2i\theta} = \frac{\det(D(\mu)+m)}{\det(D^\dagger(\mu)+m)} 
  =  \prod_{k=1}^{N} \frac{m^2-z_k^2}{m^2-{z_k^*}^2} \:.
\end{align}
In this expression, $m$ is the mass of a valence quark. From the
physics point of view, an interesting quantity is the ensemble average
of $e^{2i\theta}$ with two light dynamical quarks that have the same
mass as the valence quark.  This quantity tells us how the two-flavor
determinant in the weight function oscillates.  For simplicity, we
shall refer to $e^{2i\theta}$ as the phase factor of the determinant,
even though it is really the phase of the square of the determinant.

Because of the symmetries of \eqref{rmtdis}, each matrix appears in
the ensemble average with the same probability as its Hermitian
conjugate. As the corresponding determinants are complex conjugate,
the ensemble average of the phase factor is real.  For strongly
oscillating determinants the average phase factor will be close to
zero, and the sign problem will be severe. On the other hand, for
values of the chemical potential for which the average phase factor is
close to unity one should still be able to perform dynamical
simulations.

For each topic that is treated here and in the following sections, we
will first address the quenched case and then generalize to the
unquenched case.  The virtue of this approach is that the quenched
case already contains the essential ingredients, but the arguments and
the notation can be kept simple.  The generalization to the unquenched
case is straightforward but leads to somewhat more complicated
expressions.

\subsubsection{Quenched case}

The quenched ensemble average for the phase factor is given by
\begin{align}
  \ph_{N_f=0} &= \ev{\frac{\det(D(\mu)+m)}{\det(D^\dagger(\mu)+m)}}_{\!\!N_f=0}
  =  \frac{Z^{1|1^*}_\nu\!\!(\alpha,m)}{Z_\nu(\alpha)} \notag\\
  &= \frac{1}{Z_\nu(\alpha) }\int_\mathbb{C} 
  \prod_{k=1}^{N} d^2 z_k \: w^\nu(z_k, z_k^*;\alpha) \: 
  |\Delta_N(\{z^2\})|^2 \frac{m^2-z_k^2}{m^2-{z_k^*}^2} \:.
 \label{average}
\end{align}
The quantity $Z^{1|1^*}_\nu\!\!(\alpha,m)$ is the partition function
of a random matrix model with one fermionic quark and one conjugate
bosonic quark, see Ref.~\cite{Splittorff:2007ck} for a detailed
discussion.

Using the formalism developed in
Refs.~\cite{Bergere:2004cp,Akemann:2004zu}, the quenched average of
ratios of characteristic polynomials can be written in terms of the
orthogonal polynomials \eqref{pknu} and their Cauchy transforms
\eqref{hknu}.  Applying this formalism to the quenched average phase
factor \eqref{average} gives the compact expression
\begin{align}
  \label{averageAP}
  \ph_{N_f=0}
  = -\frac{1}{r_{N-1}^\nu(\alpha)}
  \begin{vmatrix}
    h_{N-1}^\nu(m;\alpha) & h_{N}^\nu(m;\alpha) \\[1mm]
    p_{N-1}^\nu(m;\alpha) & p_N^\nu(m;\alpha)
  \end{vmatrix} \:,
\end{align}
which is a complex integral over the orthogonal polynomials due to the
Cauchy transforms.  This expression (and its analog for the unquenched
case, see Sec.~\ref{Sec:Unquenched}) will prove to be very useful to
compute the phase factor.  Inserting the Cauchy transform \eqref{hknu}
in Eq.~\eqref{averageAP} yields an integral over orthogonal
polynomials,
\begin{align}
  \ph_{N_f=0}
  = -\frac{1}{r_{N-1}^\nu(\alpha)}
  \int_\mathbb{C} \frac{d^2z}{z^2-m^2} w^\nu(z,z^*;\alpha) 
  \begin{vmatrix}
    p_{N-1}^\nu(z^*;\alpha) & p_N^\nu(z^*;\alpha) \\[1mm]
    p_{N-1}^\nu(m;\alpha) & p_N^\nu(m;\alpha)
  \end{vmatrix} \:,
  \label{Z1Laguerre}
\end{align}
where we also used $p_k^\nu(z)^* = p_k^\nu(z^*)$.  The well-known
recurrence relation for the generalized Laguerre polynomials,
\begin{equation}
  (k+1) L^\nu_{k+1}(x) = (k+1+\nu) L^\nu_k(x) - x L^{\nu+1}_{k}(x) \:,
  \label{Lagrecur}
\end{equation}
translates into a recurrence relation for the $p_k^\nu$ defined in
Eq.~\eqref{pknu},
\begin{align}
  p_{k+1}^\nu(z;\alpha) &= (k+1+\nu) \left(\frac{1-\alpha }{N}\right)
  p_k^\nu(z;\alpha) + z^2 p_k^{\nu+1}(z;\alpha) \:.
  \label{pkrecur}
\end{align}
Since the determinant remains unchanged when forming linear
combinations of its columns, we can rewrite the phase factor
\eqref{Z1Laguerre} using the recurrence \eqref{pkrecur} as
\begin{align}
  \ph_{N_f=0}
  = -\frac{1}{r_{N-1}^\nu(\alpha)}
  \int_\mathbb{C} \frac{d^2z}{z^2-m^2} w^\nu(z,z^*;\alpha) 
  \begin{vmatrix}
    p_{N-1}^\nu(z^*;\alpha) & {z^*}^2 p_{N-1}^{\nu+1}(z^*;\alpha) \\[1mm]
    p_{N-1}^\nu(m;\alpha) & m^2 p_{N-1}^{\nu+1}(m;\alpha)
  \end{vmatrix} \:,
  \label{Z1by1}
\end{align}
where all the orthogonal polynomials are now of equal degree.

\subsubsection{Unquenched case}
\label{Sec:Unquenched}

In the presence of $N_f$ dynamical fermion flavors with masses $m_1,
\ldots, m_{N_f}$, the phase factor for a valence quark of mass $m$ is
given by
\begin{align}
  \ph_{N_f}
  &=\left\langle
    \frac{\det(D(\mu)+m)}{\det(D^\dagger(\mu)+m)}\right\rangle_{\!\!N_f}
  \notag\\
  &= \frac{1}{Z^\Nf_\nu(\alpha;\{m_f\})}\int_\mathbb{C}
  \prod_{k=1}^{N} d^2 z_k \,  
  w^\nu(z_k, z_k^*;\alpha) \, |\Delta_N(\{z^2\})|^2
  \frac{m^2-z_k^2}{m^2-{z_k^*}^2} \prod_{f=1}^\Nf (m_f^2-z_k^2) \:,
\end{align}
where $Z^\Nf_\nu(\alpha;\{m_f\})$ is given by Eq.~\eqref{eq:ZNf}.
This can be written as a ratio of two partition functions,
\begin{align}
  \ph_{N_f}
  &= \frac{Z^{N_f+1|1^*}_\nu\!\!(\alpha,m;\{m_f\})}
  {Z^{N_f}_\nu(\alpha;\{m_f\})} \:,
  \label{unqphfac} 
\end{align}
where, in analogy to Eq.~\eqref{average}, $Z_\nu^{N_f+1|1^*}$ is the
chRMT partition function with $N_f+1$ fermionic quarks and one
conjugate bosonic quark.  Both partition functions can be computed
using the results of Ref.~\cite{Akemann:2004zu}, but to apply these
results we need to change the normalization and divide both
$Z_\nu^{N_f+1|1^*}$ and $Z_\nu^{N_f}$ by the quenched partition
function.  Then the partition functions can be interpreted as averages
of ratios of characteristic polynomials in the quenched ensemble.
Applying the formalism of Ref.~\cite{Akemann:2004zu} to the numerator
of Eq.~\eqref{unqphfac} (with modified normalization) we find
\begin{align}
  Z^{N_f+1|1^*}_\nu(\alpha,m;\{m_f\})
  &= -\frac{1}{r_{N-1}^\nu(\alpha)\Delta_{N_f+1}(m^2,\{m_f^2\})} 
  \begin{vmatrix}
    h_{N-1}^\nu(m;\alpha) & h_{N}^\nu(m;\alpha) & \cdots & h_{N+N_f}^\nu(m;\alpha)\\[1mm]
    p_{N-1}^\nu(m;\alpha) & p_{N}^\nu(m;\alpha) & \cdots & p_{N+N_f}^\nu(m;\alpha)\\[1mm]
    p_{N-1}^\nu(m_1;\alpha) & p_{N}^\nu(m_1;\alpha) & \cdots & p_{N+N_f}^\nu(m_1;\alpha)\\
    \vdots & \vdots & \vdots & \vdots \\
    p_{N-1}^\nu(m_\Nf;\alpha) & p_{N}^\nu(m_\Nf;\alpha) & \cdots & p_{N+N_f}^\nu(m_\Nf;\alpha)
  \end{vmatrix} \:,
\end{align}
where the matrix in the determinant is of size $(\Nf+2) \times
(\Nf+2)$.  Substituting the definition of the Cauchy transform
\eqref{hknu} gives
\begin{align}
  Z^{N_f+1|1^*}_\nu(\alpha,m;\{m_f\})
  = -&\frac{1}{r_{N-1}^\nu(\alpha)\Delta_{N_f+1}(m^2,\{m_f^2\})}
  \int \frac{d^2 z}{z^2-m^2} w^\nu(z,z^*;\alpha) 
  \notag\\
  &\quad\times 
  \begin{vmatrix}
    p_{N-1}^\nu(z^*;\alpha) & p_{N}^\nu(z^*;\alpha) & \cdots & p_{N+N_f}^\nu(z^*;\alpha)\\[1mm]
    p_{N-1}^\nu(m;\alpha) & p_{N}^\nu(m;\alpha) & \cdots & p_{N+N_f}^\nu(m;\alpha)\\[1mm]
    p_{N-1}^\nu(m_1;\alpha) & p_{N}^\nu(m_1;\alpha) & \cdots & p_{N+N_f}^\nu(m_1;\alpha)\\
    \vdots & \vdots & \vdots & \vdots \\
    p_{N-1}^\nu(m_\Nf;\alpha) & p_{N}^\nu(m_\Nf;\alpha) & \cdots & p_{N+N_f}^\nu(m_\Nf;\alpha)
  \end{vmatrix} \:.
\end{align}
With the recurrence relation \eqref{pkrecur} for the orthogonal
polynomials this becomes
\begin{align}
  Z^{N_f+1|1^*}_\nu(\alpha,m;\{m_f\})
  = -&\frac{1}{r_{N-1}^\nu(\alpha)\Delta_{N_f+1}(m^2,\{m_f^2\})}
  \int \frac{d^2 z}{z^2-m^2} w^\nu(z,z^*;\alpha) \notag\\
  &\times \begin{vmatrix}
    p_{N-1}^\nu(z^*;\alpha) & z^{*2} p_{N-1}^{\nu+1}(z^*;\alpha) & \cdots & (z^{*2})^{\Nf+1} p_{N-1}^{\nu+\Nf+1}(z^*;\alpha)\\[1mm]
    p_{N-1}^\nu(m;\alpha) & m^2 p_{N-1}^{\nu+1}(m;\alpha) & \cdots & (m^2)^{\Nf+1} p_{N-1}^{\nu+\Nf+1}(m;\alpha)\\[1mm]
    p_{N-1}^\nu(m_1;\alpha) & m_1^2 p_{N-1}^{\nu+1}(m_1;\alpha) & \cdots & (m_1^2)^{\Nf+1} p_{N-1}^{\nu+\Nf+1}(m_1;\alpha)\\
    \vdots & \vdots & \vdots & \vdots \\
    p_{N-1}^\nu(m_\Nf;\alpha) & m_\Nf^2 p_{N-1}^{\nu+1}(m_\Nf;\alpha) & \cdots & (m_\Nf^2)^{\Nf+1} p_{N-1}^{\nu+\Nf+1}(m_\Nf;\alpha)
  \end{vmatrix} \:.
  \label{eq:ZNf11}
\end{align}
The denominator in Eq.~\eqref{unqphfac} (with modified normalization)
can be written as \cite{Akemann:2004zu}
\begin{align}
  Z^{N_f}_\nu(\alpha;\{m_f\})
  = \frac{1}{\Delta_{N_f}(\{m_f^2\})}
  \begin{vmatrix}
    p_{N}^\nu(m_1;\alpha) & p_{N+1}^\nu(m_1;\alpha) & \cdots & p_{N+N_f-1}^\nu(m_1;\alpha)\\[1mm]
    p_{N}^\nu(m_2;\alpha) & p_{N+1}^\nu(m_2;\alpha) & \cdots & p_{N+N_f-1}^\nu(m_2;\alpha)\\
    \vdots & \vdots & \vdots & \vdots \\
    p_{N}^\nu(m_\Nf;\alpha) & p_{N+1}^\nu(m_\Nf;\alpha) & \cdots & p_{N+N_f-1}^\nu(m_\Nf;\alpha)
  \end{vmatrix} \:,
\end{align}
and using the recurrence relation \eqref{pkrecur} for the orthogonal
polynomials this can be rewritten as
\begin{align}
  Z^{N_f}_\nu(\alpha;\{m_f\})
  = \frac{1}{\Delta_{N_f}(\{m_f^2\})}
  \begin{vmatrix}
    p_{N}^\nu(m_1;\alpha) & m_1^2 p_{N}^{\nu+1}(m_1;\alpha) & \cdots & (m_1^2)^{\Nf-1} p_{N}^{\nu+\Nf-1}(m_1;\alpha)\\[1mm]
    p_{N}^\nu(m_2;\alpha) & m_2^2 p_{N}^{\nu+1}(m_2;\alpha) & \cdots & (m_2^2)^{\Nf-1} p_{N}^{\nu+N_f-1}(m_2;\alpha)\\
    \vdots & \vdots & \vdots & \vdots \\
    p_{N}^\nu(m_\Nf;\alpha) & m_\Nf^2 p_{N}^{\nu+1}(m_\Nf;\alpha) & \cdots & (m_\Nf^2)^{\Nf-1} p_{N}^{\nu+N_f-1}(m_\Nf;\alpha)
  \end{vmatrix} \:.
  \label{eq:ZNf00}
\end{align}

\subsection{Microscopic limit}
\label{sec:micro}

Universal results, i.e., results that also apply to QCD, can be
obtained from chRMT in the so-called microscopic regime.  This regime
is obtained by taking $N\to\infty$ while keeping the rescaled
parameters $\hm=2N m$, $\hm_f = 2N m_f$, and $\ha = 2N \alpha$ fixed,
and rescaling the spectrum using $\hz=2N z$.  The rescaled random
matrix parameters can be converted to the physical parameters $z$,
$m$, and $\mu$ using the relations $\hz=zV\Sigma$, $\hm=mV\Sigma$, and
$\ha={\hat\mu}^2=\mu^2F^2V$, where $V$ is the four-volume.\footnote{To
  be more precise, one should distinguish random matrix parameters and
  physical parameters in the relations
  $2Nz_\text{RMT}=\hz=z_\text{phys}V\Sigma$,
  $2Nm_\text{RMT}=\hm=m_\text{phys}V\Sigma$, and
  $2N\mu_\text{RMT}^2={\hat\mu}^2=\mu_\text{phys}^2F^2V$.  This
  distinction makes it explicit that the limits $N\to\infty$ and
  $V\to\infty$ can be decoupled.}  Furthermore, the pion mass $m_\pi$
can be introduced through the combination
$\mu^2/m_\pi^2={\hat\mu}^2/2\hm$, where we have used the
Gell-Mann--Oakes--Renner relation $m_\pi^2F^2=2m\Sigma$ (assuming
equal quark masses).

We now introduce the microscopic limits (denoted by a subscript $s$)
of the orthogonal polynomials, the norm, and the weight function,
respectively.  They are worked out in App.~\ref{App:micpnu}, and we
obtain
\begin{align}
  \label{limp2}
  \hp^\nu(\hz;\ha) &\equiv \lim_{N\to\infty} \frac{e^N}{(2N)^{\nu+1/2}}
  p_{N-1}^\nu(\hz/2N;\ha/2N)
  =\sqrt{\pi}  e^{-\ha/2}
  \hz^{-\nu} I_\nu\left(\hz\right)\:,\\
  \label{micrknu}
  \hr^\nu(\ha) &\equiv \lim_{N\to\infty}(2N)^2e^{2N}r_{N-1}^\nu(\ha/2N) 
  = 4\pi^2 \ha e^{\ha} \:,\\
  \label{micw}
  \hw^\nu(\hz, \hz^*;\ha) 
  &\equiv\lim_{N\to\infty} (2N)^{2\nu+2}w^\nu(\hz/2N, \hz^*/2N;\ha/2N) 
  = |\hz|^{2(\nu+1)} 
  e^{-\textstyle\frac{{\hz}^2+{\hz}^{*2}}{8\ha}}
  K_\nu\left(\frac{|\hz|^2}{4\ha}\right)\:.
\end{align}

\subsubsection{Quenched case}

In terms of the above definitions, the microscopic limit of the
quenched average phase factor \eqref{Z1by1} is given by
\begin{align}
  \phs_{N_f=0} &\equiv \lim_{N\to\infty} \bigl\langle e^{2i\theta}
  \bigr\rangle_{N_f=0,\ha/2N,\hm/2N} \notag\\
  &= - \frac{1}{\hr^\nu(\ha)}
  \int_\mathbb{C} \frac{d^2 \hz}{\hz^2-\hm^2} \hw^\nu(\hz,\hz^*;\ha) 
  \begin{vmatrix}
    \hp^\nu(\hz^*;\ha) & \hz^{*2} \hp^{\nu+1}(\hz^*;\ha) \\[1mm]
    \hp^\nu(\hm;\ha) & \hm^2 \hp^{\nu+1}(\hm;\ha)
  \end{vmatrix} \:.
  \label{micZ1by1_0}
\end{align}
As expected, the dependence on $N$ has dropped out, leaving a finite
microscopic limit for the average phase factor.  Substituting the
asymptotic results from Eqs.~\eqref{limp2}--\eqref{micw} yields
\begin{align}
  \phs_{N_f=0}
  = -\frac{e^{-2\ha}}{4\pi\ha\hm^\nu}
  \int_\mathbb{C} \frac{d^2z}{z^2-\hm^2}
  |z|^{2(\nu+1)} 
  &e^{-\textstyle\frac{z^2+{z^*}^2}{8\ha}} 
  K_\nu\left(\frac{|z|^2}{4\ha}\right) 
  (z^*)^{-\nu}
  \begin{vmatrix}
    I_\nu\left(z^*\right) &  {z^*} I_{\nu+1}\left(z^*\right) \\
    I_\nu\left(\hm\right) &  \hm I_{\nu+1}\left(\hm\right) 
  \end{vmatrix} \:,
  \label{micZ1by1}
\end{align}
where we renamed the integration variable $\hz$ back to $z$.  This
equation can be rewritten compactly as
\begin{align}
  \phs_{N_f=0}
  = 
  \begin{vmatrix}
    \I{0} &  \I{1} \\
    I_\nu\left(\hm\right) &  \hm I_{\nu+1}\left(\hm\right) 
  \end{vmatrix} \:,
  \label{qu0}
\end{align}
where we defined the integral
\begin{align}
  \I{k}
  &\equiv -\frac{e^{-2\ha}}{4\pi\ha\hm^\nu}
  \int_\mathbb{C} \frac{d^2z}{z^2-\hm^2}
  |z|^{2(\nu+1)} 
  e^{-\textstyle\frac{z^2+{z^*}^2}{8\ha}}
  K_\nu\left(\frac{|z|^2}{4\ha}\right)
  (z^*)^{-\nu+k} I_{\nu+k}(z^*) \:,
  \label{calI}
\end{align}
which is closely related to the microscopic limit of the Cauchy
transform \eqref{hknu}. For the quenched case, this integral is only
needed for $k=0,1$, but as we shall see in the next subsection, in the
unquenched case it will be needed for $k=0,\ldots,N_f+1$.

\subsubsection{Unquenched case}

We now take the microscopic limit of Eqs.~\eqref{eq:ZNf11} and
\eqref{eq:ZNf00}.
In Eq.~\eqref{eq:ZNf11}, the Vandermonde determinant
$\Delta_{N_f+1}(m^2,\{m_f^2\})$ is a product of $\Nf(\Nf+1)/2$ factors
for which the microscopic limit yields a $(2N)^{\Nf(\Nf+1)}$
dependence on $N$, while the explicit mass and $z^*$ factors in the
determinant yield a factor $1/(2N)^{(\Nf+1)(\Nf+2)}$. After also
introducing the microscopic limits \eqref{limp2}, \eqref{micrknu}, and
\eqref{micw} for the orthogonal polynomials, their normalization
factor, and the weight function, respectively, one finds
\begin{align}
  &Z^{N_f+1|1^*}_{\nu,s}(\ha,\hm;\{\hm_f\})
  = -  (2N)^{(2\nu+N_f)N_f/2}e^{-NN_f} \frac{\pi^{\Nf/2}e^{-\ha(\Nf/2+2)}}
  { 4\pi\ha(\hm \hm_1 \hm_2 \ldots \hm_\Nf)^{\nu}
    \Delta_{N_f+1}(\hm^2,\{\hm_f^2\})}  \notag\\
  &\quad\times
  \int \frac{d^2 z}{z^2-\hm^2} 
  |z|^{2(\nu+1)} (z^*)^{-\nu}
  K_\nu\left(\frac{|z|^2}{4\ha}\right)
  e^{-\textstyle\frac{z^2+z^{*2}}{8\ha}}
  \begin{vmatrix}
    I_{\nu,0}(z^*) & I_{\nu,1}(z^*) & \cdots & I_{\nu,N_f+1}(z^*)\\[1mm]  
    I_{\nu,0}(\hm) & I_{\nu,1}(\hm) & \cdots & I_{\nu,\Nf+1}(\hm)\\[1mm]
    I_{\nu,0}(\hm_1) & I_{\nu,1}(\hm_1) & \cdots & I_{\nu,\Nf+1}(\hm_1)\\
    \vdots & \vdots & \vdots & \vdots \\
    I_{\nu,0}(\hm_\Nf) & I_{\nu,1}(\hm_\Nf) & \cdots & I_{\nu,\Nf+1}(\hm_\Nf)
  \end{vmatrix} \:,
  \label{ZNf+1|1*}
\end{align}
where we have again renamed the integration variable from $\hz$ back
to $z$ and introduced the notation 
\begin{align}
  \label{eq:Inuk}
  I_{\nu,k}(z) = z^k I_{\nu+k}(z)\:.  
\end{align}
In the microscopic limit of Eq.~\eqref{eq:ZNf00}, the Vandermonde
determinant yields a factor $(2N)^{(\Nf-1)\Nf}$ which exactly cancels
the factor $1/(2N)^{(\Nf-1)\Nf}$ coming from the explicit mass factors
in the determinant.  After introducing the microscopic limit
\eqref{limp2} of the orthogonal polynomials we find
\begin{align}
  Z^{N_f}_{\nu,s}(\ha;\{\hm_f\})
  &= (2N)^{(2\nu+N_f)N_f/2}e^{-NN_f}
  \frac{\pi^{\Nf/2}  e^{-\ha\Nf/2}}{(\hm_1 \hm_2 \ldots \hm_\Nf)^{\nu}
    \Delta_{N_f}(\{\hm_f^2\})}\:\DI
\label{micZNf}
\end{align}
with
\begin{align}    
  \DI&=
  \begin{vmatrix}
    I_{\nu,0}(\hm_1) & I_{\nu,1}(\hm_1) & \cdots & I_{\nu,\Nf-1}(\hm_1)\\[1mm]
    I_{\nu,0}(\hm_2) & I_{\nu,1}(\hm_2) & \cdots & I_{\nu,N_f-1}(\hm_2)\\
    \vdots & \vdots & \vdots & \vdots \\
    I_{\nu,0}(\hm_\Nf) & I_{\nu,1}(\hm_\Nf) & \cdots & I_{\nu,N_f-1}(\hm_\Nf)
  \end{vmatrix} 
  \:.
\end{align}
The microscopic limit of the average phase factor \eqref{unqphfac} is
given by the ratio of Eqs.~\eqref{ZNf+1|1*} and \eqref{micZNf}.  The
dependence on $N$ drops out to give
\begin{align}
  \phs_{N_f}
  &= -  \frac{e^{-2\ha}}{ 4\pi\ha\hm^\nu\prod_{f=1}^\Nf
    (\hm_f^2-\hm^2)\DI} \\
  &\quad\times\int\frac{d^2 z}{z^2-\hm^2} 
  |z|^{2(\nu+1)} (z^*)^{-\nu} 
  K_\nu\left(\frac{|z|^2}{4\ha}\right)
  e^{-\textstyle\frac{z^2+z^{*2}}{8\ha}}
  \begin{vmatrix}
    I_{\nu,0}(z^*) & I_{\nu,1}(z^*) & \cdots & I_{\nu,\Nf+1}(z^*)\\[1mm]
    I_{\nu,0}(\hm) & I_{\nu,1}(\hm) & \cdots & I_{\nu,\Nf+1}(\hm)\\[1mm]
    I_{\nu,0}(\hm_1) & I_{\nu,1}(\hm_1) & \cdots & I_{\nu,\Nf+1}(\hm_1)\\
    \vdots & \vdots & \vdots & \vdots \\
    I_{\nu,0}(\hm_\Nf) & I_{\nu,1}(\hm_\Nf) & \cdots & I_{\nu,\Nf+1}(\hm_\Nf)
  \end{vmatrix}
  \:,\notag
\end{align}
which can be rewritten using the integral definition \eqref{calI} as
\begin{align}
  \phs_{N_f}
  &=  \frac{1}{\prod_{f=1}^\Nf (\hm_f^2-\hm^2)\DI}
  \begin{vmatrix}
      \I{0} & \I{1} & \cdots & \I{\Nf+1}\\[1mm]
      I_{\nu,0}(\hm) & I_{\nu,1}(\hm) & \cdots & I_{\nu,\Nf+1}(\hm)\\[1mm]
      I_{\nu,0}(\hm_1) & I_{\nu,1}(\hm_1) & \cdots & I_{\nu,\Nf+1}(\hm_1)\\
      \vdots & \vdots & \vdots & \vdots \\
      I_{\nu,0}(\hm_\Nf) & I_{\nu,1}(\hm_\Nf) & \cdots & I_{\nu,\Nf+1}(\hm_\Nf)
    \end{vmatrix}
  \:.
  \label{phfacNfflavors_arbmass}
\end{align}
We have thus reduced the problem of calculating the phase factor to
the calculation of the two-dimen\-sional integral $\I{k}$ in
Eq.~\eqref{calI} for $k=0,\ldots,N_f+1$.  This integral will be
computed in Sec.~\ref{Sec:Analytic}.

\subsubsection{Equal mass fermions}

We now consider Eq.~\eqref{phfacNfflavors_arbmass} for the special
case in which all dynamical fermions have the same mass $\hm$ as the
valence quark. To simplify the general expression we perform a Taylor
expansion of the entries $I_{\nu,k}(\hm_f)$ of the determinant around
$\hm$,
\begin{align}
  I_{\nu,k}(\hm_f) = I_{\nu,k}(\hm) + \sum_{j=1}^{\infty}
  \frac{I_{\nu,k}^{(j)}(\hm)}{j!}(\hm_f-\hm)^j \:,
  \qquad f=1, \ldots, \Nf\:.
  \label{mTaylor}
\end{align}
Because a determinant remains unaltered when making linear
combinations of its rows, we see that for each additional fermion it
is sufficient to keep the next higher-order term in the expansion
\eqref{mTaylor}. The lower-order terms will not contribute as they are
identical to the contribution from one of the previous fermions in the
determinant, while the higher-order terms can be neglected as their
contribution will vanish when $\hm_f \to \hm$. After taking each
fermion mass in turn to $\hm$, this leads to the simplified expression
\begin{align}
  \phs_{N_f}
  &= \frac{1}{(2\hm)^\Nf \Nf!} 
  \frac{ \begin{vmatrix}
      \I{0} & \I{1} & \cdots & \I{\Nf+1}\\[1mm]
      I_{\nu,0}(\hm) & I_{\nu,1}(\hm) & \cdots & I_{\nu,\Nf+1}(\hm)\\[1mm]
      I'_{\nu,0}(\hm) & I'_{\nu,1}(\hm) & \cdots & I'_{\nu,\Nf+1}(\hm)\\
      \vdots & \vdots & \vdots & \vdots \\
      I^{(\Nf)}_{\nu,0}(\hm) & I^{(\Nf)}_{\nu,1}(\hm) & \cdots & I^{(\Nf)}_{\nu,\Nf+1}(\hm)
    \end{vmatrix} }
  {\begin{vmatrix}
      I_{\nu,0}(\hm) & I_{\nu,1}(\hm) & \cdots & I_{\nu,\Nf-1}(\hm)\\[1mm]
      I'_{\nu,0}(\hm) & I'_{\nu,1}(\hm) & \cdots & I'_{\nu,N_f-1}(\hm)\\
      \vdots & \vdots & \vdots & \vdots \\
      I^{(\Nf-1)}_{\nu,0}(\hm) & I^{(\Nf-1)}_{\nu,1}(\hm) & \cdots & I^{(\Nf-1)}_{\nu,N_f-1}(\hm)
    \end{vmatrix} }
  \:.
  \label{phfacNfflavors}
\end{align}
An alternative way to write this result is
\begin{align}
  \phs_{N_f} = \frac{1}{(2\hm)^\Nf \Nf!} \frac{{\cal
      W}_{\Nf}(\ha,\hm)}{W_\Nf(0, 1, \ldots, \Nf-1)} \:,
\label{phapp}
\end{align}
where we have defined
\begin{align}
{\cal W}_{\Nf}(\ha,\hm) = \sum_{k=0}^{\Nf+1} (-)^k \I{k} W_{\Nf+1}(0,
\ldots, k-
1, k+1, \ldots, \Nf+1)
\label{numerator}
\end{align}
as a sum of Wronskians of order $\Nf+1$ with indices ranging from $0$
to $\Nf+1$, where in each term a different index $k$ is absent.
The Wronskian
\begin{equation}
  W_{n}(I_{\nu,k_1}(\hm), \ldots, I_{\nu,k_n}(\hm))=
  {\begin{vmatrix}
      I_{\nu,k_1}(\hm) & I_{\nu,k_2}(\hm) & \cdots &
      I_{\nu,k_n}(\hm)\\[1mm]
      I'_{\nu,k_1}(\hm) & I'_{\nu,k_2}(\hm) & \cdots &
      I'_{\nu,k_n}(\hm)\\
      \vdots & \vdots & \vdots & \vdots \\
      I^{(n-1)}_{\nu,k_1}(\hm) & I^{(n-1)}_{\nu,k_2}(\hm) & \cdots &
      I^{(n-1)}_{\nu,k_n}(\hm) 
      \label{G.4}
    \end{vmatrix} }
\end{equation}
that appears in Eqs.~\eqref{phapp} and \eqref{numerator} has been
abbreviated by $W_{n}(k_1,\ldots,k_n)$.

\section{Evaluation of the complex Cauchy transform}
\label{Sec:Analytic}

\subsection{Asymptotic behavior}

To investigate the two-dimensional integral \eqref{calI} it is
instructive to first study the asymptotic behavior of the integrand.
For large values of its argument the $K$-Bessel function behaves like
\cite[Eq.~(9.7.2)]{Abram:1964}
\begin{equation}
K_\nu(z) \sim \sqrt{\frac{\pi}{2z}} e^{-z} \:.
\label{Kas}
\end{equation}
Here and in the rest of the paper we use the $\sim$ symbol for the
leading-order term in an expansion for small or large argument,
including all prefactors.  To find the asymptotic behavior of the
$I$-Bessel function we first note that the modified Bessel functions
satisfy the relation \cite[Eq.~(7.11.45)]{BatemanII}
\begin{align}
  K_\nu(z e^{\pm i\pi}) = (-)^\nu K_\nu(z) \mp i\pi I_\nu(z) \:.
  \label{Bat1}
\end{align}
As the $K$-Bessel function has a branch cut along the negative real
axis, it is convenient to adopt the convention $\arg(z) \in
(-\pi,\pi]$ for the complex variable $z$.  According to this
convention, reversing the sign of $z = r e^{i\theta }$ yields 
\begin{equation}
  -z =
  \begin{cases}
    r e^{i(\theta - \pi)}  & \text{for }\theta \in (0,\pi] \:, \\
    r e^{i(\theta + \pi)}  & \text{for }\theta \in (-\pi,0] \:,
  \end{cases}
  \label{minusz}
\end{equation}
such that $-z$ also has its argument in $(-\pi,\pi]$.
Combining Eqs.\ \eqref{Bat1} and \eqref{minusz} gives the relation
\begin{align}
  I_\nu(z) &= \frac{i\eta(z)}{\pi} \big( (-)^\nu K_\nu(z) - K_\nu(-z) \big)
  \label{IKK}
\end{align}
with
\begin{align}
  \eta(z) =
  \begin{cases}
    +1 & \text{for }\arg(z) \in (0,\pi]\:, \\
    -1 & \text{for }\arg(z) \in (-\pi,0]\:.
  \end{cases}
  \label{etaz}
\end{align}
Alternative definitions for $\eta(z)$ are $i\eta(z) =
\sqrt{z}/\sqrt{-z}$ or $\eta(z) = \sign(\im z)$, the latter only for
$z \notin \mathbb{R}$.  Substituting Eq.~\eqref{Kas} in
Eq.~\eqref{IKK} gives the asymptotic formula\footnote{Note that the
  asymptotic formula (9.7.1) in Ref.~\cite{Abram:1964} only contains
  the second term of Eq.~\eqref{Ias} and cannot be used here as it is
  only valid for $|\arg z|<\pi/2$.}
\begin{align}
  I_\nu(z) &\sim \frac{i\eta(z)}{\sqrt{2\pi}} \left( (-)^\nu
    \frac{e^{-z}}{\sqrt{z}} - \frac{e^z}{\sqrt{-z}}\right) =
  \frac{1}{\sqrt{2\pi}} \left(  (-)^\nu
    \frac{e^{-z}}{\sqrt{-z}} + \frac{e^z}{\sqrt{z}} \right) \:,
\label{Ias}
\end{align}
where we used $i\eta(z) = \sqrt{z}/\sqrt{-z}$ to derive the last
expression.  With Eqs.~\eqref{Kas} and \eqref{Ias}, the asymptotic
behavior of the integrand in Eq.~\eqref{calI} is proportional to
\begin{align}
  \frac{|z|^{2\nu+1}}{z^2-\hm^2}
  e^{-\textstyle\frac{x^2}{2\ha}}
  (z^*)^{-\nu+k} 
  \Bigg(
  (-)^{\nu+k} \frac{e^{-z^*}}{\sqrt{-z^*}}
  + \frac{e^{z^*}}{\sqrt{z^*}} 
  \Bigg) \:,
  \label{micZ1by1asym}
\end{align}
where $z=x+iy$. Along the $x$-direction the integrand of the
two-dimensional integral decreases like a Gaussian with width
$\sqrt{\ha}$. However, in the $y$-direction the integrand oscillates
very rapidly inside an envelope that goes like $y^{\nu+k-3/2}$.
Eqs.~\eqref{qu0} and \eqref{phfacNfflavors_arbmass} contain terms with
$\nu+k\ge 1$, for which the integral \eqref{calI} will diverge unless
some particular cancellations occur due to the oscillatory behavior of
the integrand.  As the integral represents an observable quantity in
random matrix theory we do expect such cancellations to obtain a
finite result.

An instructive numerical exercise is the direct computation of the
two-dimensional integral over the generalized Laguerre polynomials for
finite $N$, as given in Eq.~\eqref{Z1by1} for the quenched
case. Although Mathematica can only handle the numerical integration
for $N\lesssim 30$ because of the strong oscillations, the results for
$N=1,\ldots, 30$ show a clear convergence towards a finite microscopic
limit.  (These numerical results also agree with the simulations
presented in Sec.~\ref{Sec:Results}.) This clearly indicates that
Eq.~\eqref{micZ1by1} is perfectly sane, even though the evaluation of
the integral is nontrivial.

\subsection{Change of integration path and transformation of
  variables} 

The main problem is to find a way to integrate over the oscillatory
behavior in the $y$-direction.  For $\nu=0$ a method was devised in
Ref.~\cite{Splittorff:2007ck} in which the integration along the real
$y$-axis in the original $(x,y)$-plane was deformed to an integration
path in the complex $y$-plane. This results in a well-behaved
one-dimensional integral.  Here, we show how this derivation can be
generalized to $\nu \neq 0$, where proper care has to be taken of an
additional singularity occurring in the integration domain.

The two terms $e^{-z^*}$ and $e^{z^*}$ in Eq.~\eqref{micZ1by1asym}
behave differently for $|y|\to\infty$.  The first term decreases
exponentially in the upper half of the complex $y$-plane and diverges
exponentially in the lower half, whereas the second term behaves the
other way around.  Rather than treating these two terms separately, we
can use Eq.~\eqref{IKK} to simplify Eq.~\eqref{calI}.  Because of
the $z\to-z$ symmetry of the integrand in \eqref{calI}, the two
terms in \eqref{IKK} give the same contribution to the integral, and
we obtain
\begin{equation}
  \label{561}
  \I{k}
  = - \frac{i e^{-2\ha}}{2\pi^2 \ha\hm^\nu}
  \int_\mathbb{C} \frac{d^2 z}{z^2-\hat m^2} 
  |z|^{2(\nu+1)}
  e^{-\textstyle\frac{z^2+{z^*}^2}{8\ha}}
  K_\nu\left(\frac{|z|^2}{4\ha}\right) 
  \eta(z^*) 
  (-)^{\nu+k} 
  (z^*)^{-\nu+k} K_{\nu+k}(z^*) \:.
\end{equation}
We now deform the $y$-integration path from the real $y$-axis to the
path shown in Fig.~\ref{Fig:contoury}, where $y=y_r + i y_i$ and
$\epsilon\to0$ after the integration over $d^2z$.  This can be done
since the integrand vanishes exponentially for $|y|\to\infty$ along
the deformed path and since the integrand has no singularities between
the real $y$-axis and the deformed path.  
\FIGURE[t]{
  \includegraphics[width=10cm]{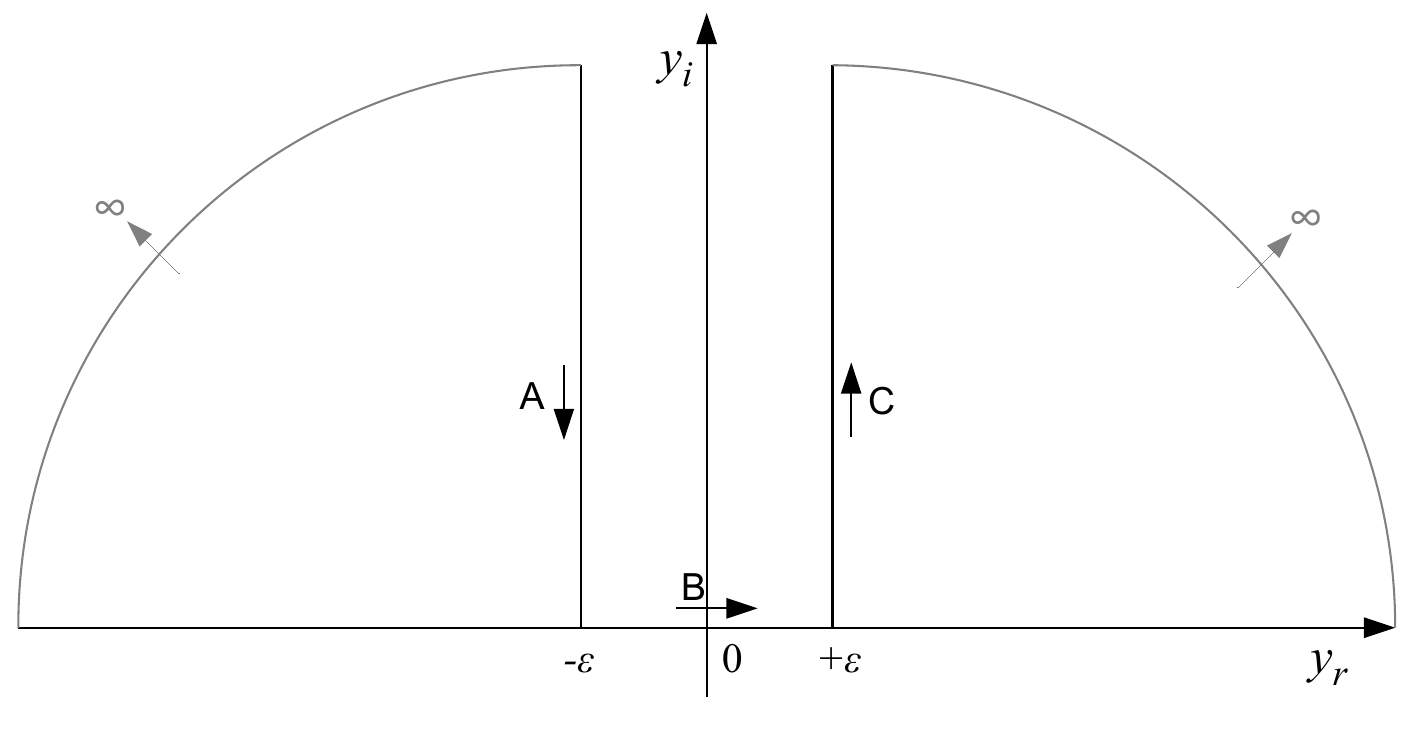}
  \caption{Deformation of the $y$-integration from the original
    integration along the $y_r$-axis to a path on which the
    integrand vanishes sufficiently rapidly for $|y|\to\infty$.}
  \label{Fig:contoury}
}
Writing $y=is\mp\epsilon$ on parts A and C of the path, respectively,
we find for Eq.~\eqref{561}
\begin{align}
  \I{k} 
  &= \lim_{\epsilon\to 0^+} \int_{-\infty}^\infty dx 
  \left\{ i \int_{\infty}^0 ds \, f(x,is-\epsilon)  +
    \int_{-\epsilon}^\epsilon dy_r \, f(x,y_r) + i \int_0^\infty ds \,
    f(x,is+\epsilon) \right\} 
  \notag\\
  &= \lim_{\epsilon\to 0^+} i\int_{-\infty}^\infty dx \int_0^\infty ds
  \, \left[ f_+(x,is) - f_-(x,is) \right]  \:,
  \label{Z1by1xs}
\end{align}
where we introduced the notation $f_\pm(x,is) = f(x,is\pm\epsilon)$ with
\begin{align}
  f(x,i s)
  = - \frac{i e^{-2\ha}}{2\pi^2 \ha\hm^\nu}
  \frac{(-)^{\nu+k}(x-s)^{\nu+1}}{(x-s)^2-\hat m^2} 
  & 
  e^{-\textstyle\frac{x^2+s^2}{4\ha}}
  K_\nu\left(\frac{x^2-s^2}{4\ha}\right)
  \eta(x+s) (x+s)^{k+1} K_{\nu+k}(x+s)
  \:.
  \label{563}
\end{align}
Note that when continuing $y$ to the complex plane we have rewritten
the integrand as an explicit function of $x$ and $y$, since $z$ and
$z^*$ are no longer complex conjugate.  The second integral in the
first line of \eqref{Z1by1xs} gives zero in the $\epsilon\to 0$ limit
since the integrand is regular at $y=0$, which is most easily seen by
looking back at Eq.~\eqref{calI}.  As $\epsilon\to 0$ in
Eq.~\eqref{Z1by1xs} we are thus left with the difference of two
integrals over semi-infinite sheets infinitesimally close and parallel
to the $(x,s)$-plane, which we denote by $S_+$ and $S_-$ for $y_r>0$
and $y_r<0$, respectively.

The integrand \eqref{563} can be further simplified by introducing the
variable transformation
\begin{align}
  &\left\{
    \begin{aligned}
      t = x-s \\
      u = x+s
    \end{aligned}
  \right. 
  \qquad
  \qquad\text{or}
  \qquad
  \left\{
    \begin{aligned}
      x = (u+t)/2 \\
      s = (u-t)/2 
    \end{aligned}
  \right.
  \label{xstu}
\end{align}
with Jacobian $1/2$, yielding
\begin{align}
  \I{k}
  &= \lim_{\epsilon\to 0^+} \int_{-\infty}^\infty dt \int_t^\infty du 
  \left[ f_+(t,u) + f_-(t,u) \right] \:,
\label{Ztu}
\end{align}
where the integration limits in the transformed variables can be read
off from Fig.~\ref{Fig:integplane}.
\FIGURE[t]{
  \includegraphics[width=10cm,trim=0 10 0 0]{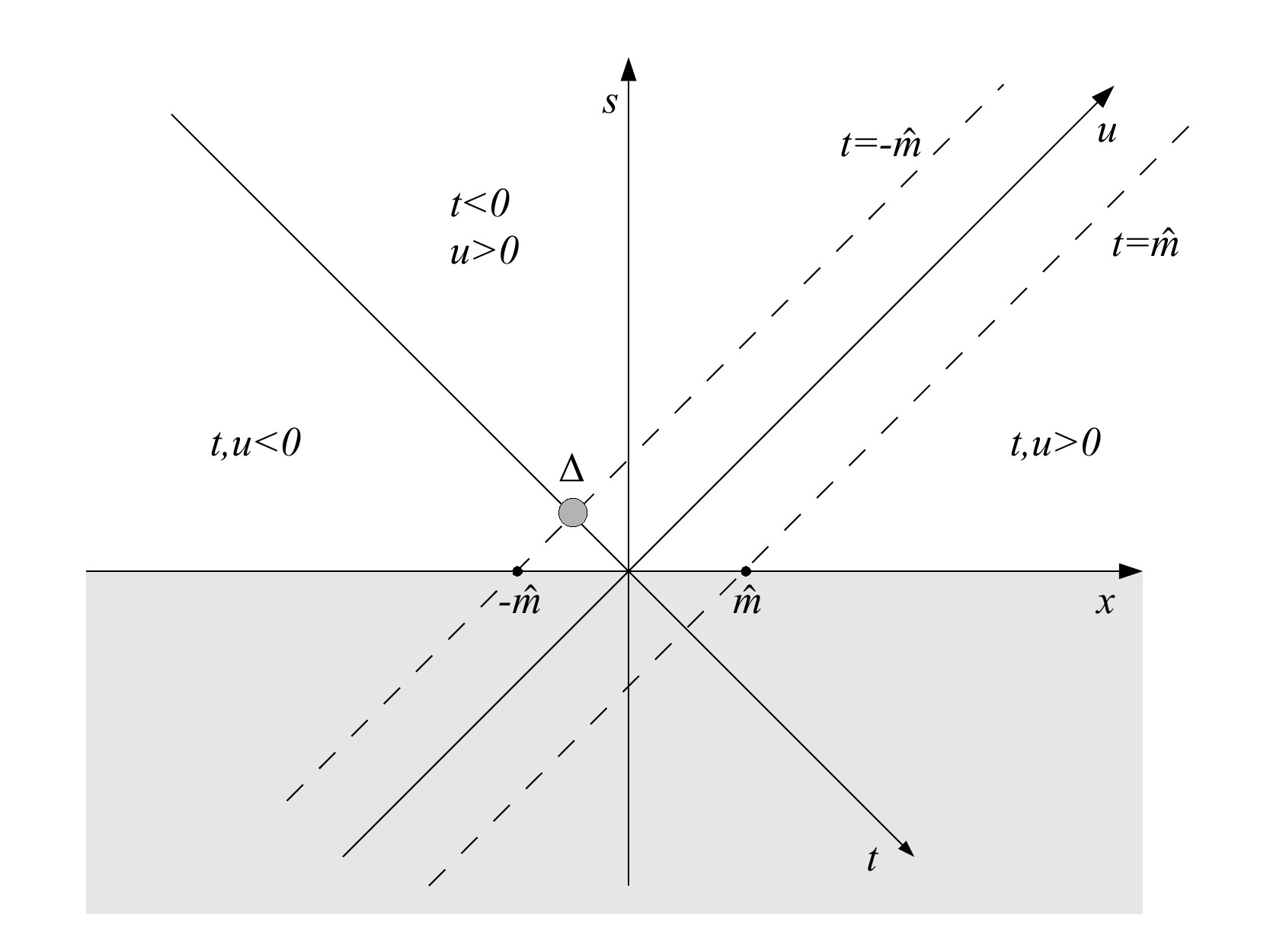}
  \caption{Structure of the integration plane. The integral is
    computed over the semi-infinite sheets $S_\pm$, with $s>0$, just
    above and below the $(x,s)$-plane. The shaded dot labeled by
    $\Delta$ at $u=0$, $t=-\hm$ indicates the singularity of the
    integrand responsible for a new contribution to the phase factor
    that arises when $\nu \neq 0$.}
  \label{Fig:integplane}
} 
We introduced the notation $f_{\pm}(t,u)=f(t_\pm,u_\pm)$ with
$t_\pm=x-s\pm i\epsilon$ and $u_\pm=x+s\mp i\epsilon$, corresponding
to the integrand on the two sheets $S_\pm$.  In the transformed
variables the integrand is given by
\begin{align}
f(t,u)
&= -\frac{e^{-2\ha}}{4\pi^2 \ha\hm^\nu}
\frac{(-)^{\nu+k}t^{\nu+1}u^{k+1}}{t^2-\hat m^2} 
e^{-\textstyle\frac{t^2+u^2}{8\ha}}
K_\nu\left(\frac{t u}{4\ha}\right)
K_{\nu+k}(u)
\:,
\label{ftu}
\end{align}
where we have absorbed the Jacobian and factors of $i$ in the
definition of $f$.  We have used $\eta(u_\pm)=\mp 1$ according to
Eq.~\eqref{etaz}, which changes the difference in Eq.~\eqref{Z1by1xs}
to a sum in Eq.~\eqref{Ztu}.

\subsection{Singularities of the integrand}

When taking the limit $\epsilon\to 0$ in Eq.~\eqref{Ztu} one has to
take into account the singularities of the integrand in the
$(t,u)$-plane.  There are two mass-pole lines parallel to the $u$-axis
at $t=\pm\hat m$ as well as singularities and branch cuts of the
$K$-Bessel functions for zero and negative real argument, respectively
(see Fig.~\ref{Fig:integplane}).

For future use we first analyze the branch cut of the Bessel functions
along the negative real axis.  When going from one Riemann sheet to
the next across the branch cut, $K_\nu(z)$ changes by $ (-)^{\nu+1}
2\pi i I_\nu(z)$ \cite[Eq.~(7.11.45)]{BatemanII}.  For $x \in
\mathbb{R}$ we have
\begin{align}
  \label{eq:cut}
  \lim_{\epsilon\to 0^+} K_\nu(x \pm i \epsilon) =
  \begin{cases}
    K_\nu(|x|) &\text{for } x>0 \:, \\
    (-)^\nu K_\nu(|x|) \mp i \pi I_\nu(|x|) &\text{for } x<0 \:.
  \end{cases}
\end{align}
Consider separately the three sectors of the integration region
indicated in Fig.~\ref{Fig:integplane}.  Using \eqref{eq:cut} together
with $t_\pm = t \pm i \epsilon$, $u_\pm = u \mp i \epsilon$, and
$(tu)_\pm=tu \pm i\epsilon$ (with $\epsilon>0$ because $s>0$), we
can write the product $K_\nu(t u) K_{\nu+k}(u)$ (with the factor
$1/4\ha$ omitted for simplicity) appearing in \eqref{ftu} as
\begin{align}
  \lim_{\epsilon\to 0^+} \bigl[ K_\nu(t u) K_{\nu+k}(u) \bigr]_\pm 
  =
  \begin{cases}
    K_\nu(|tu|) K_{\nu+k}(|u|) & \text{for } t,u>0 \:, \\
    (-)^\nu K_\nu(|tu|) K_{\nu+k}(|u|) \mp i \pi I_\nu(|tu|)
    K_{\nu+k}(|u|) & \text{for } t<0<u \:, \\ 
    (-)^{\nu+k} K_\nu(|tu|) K_{\nu+k}(|u|) \pm i \pi K_\nu(|tu|)
    I_{\nu+k}(|u|) & \text{for } t, u<0 \:. 
  \end{cases}
  \label{KetaK}
\end{align}
Note that the fourth quadrant, where $t>0$ and $u<0$, does not overlap
with the region of integration and thus does not need to be considered.

We now turn to the mass-pole factor in Eq.~\eqref{ftu}, which can be
written as
\begin{equation}
  \frac{1}{(t \pm i\epsilon)^2-\hm^2} 
  = \frac{1}{2\hm}\left( \frac{1}{t\pm i\epsilon-\hm} - \frac{1}{t\pm
      i\epsilon+\hm} \right) \:.
\label{masspole}
\end{equation}
This enables us to apply the Sokhatsky-Weierstrass theorem
\begin{align}
  \lim_{\epsilon\to 0^+} \int_a^b \frac{f(x)}{x\pm i\epsilon} dx = \mp
  i \pi f(0) + {\PV} \int_a^b \frac{f(x)}{x} dx\:,
\label{SW-theorem}
\end{align}
where ${\PV}$ denotes the Cauchy principal value integral, to perform
the $t$-integral in Eq.~\eqref{Ztu}, which thus becomes a sum of the
residues at the mass poles $t=\pm\hm$ and a principal value integral
over the complete $t$-axis.  In App.~\ref{Sec:PV} we show that the
principal value part of the $t$-integral in Eq.~\eqref{Ztu} vanishes
because of symmetry properties so that the $t$-integral is entirely
determined by the mass-pole contributions.  The residue contributions
from the mass poles yield two one-dimensional integrals over $u$ along
the $t=\pm\hm$ lines, resulting in
\begin{align}
  \I{k}
  &= -\frac{i\pi}{2\hm} \Bigg\{ \lim_{\epsilon\to 0^+}  
  \int_{\hm}^\infty \! du 
  \left[ g(\hm,u_+) - g(\hm,u_-) \right] 
  -\lim_{\epsilon\to 0^+}  \int_{-\hm}^\infty du 
  \left[ g(-\hm,u_+) - g(-\hm,u_-) \right] 
  \Bigg\} \:,
  \label{Zgtu}
\end{align}
where
\begin{align}
  g(t,u) &= (t^2-m^2)f(t,u) = -\frac{e^{-2\ha}}{4\pi^2 \ha\hm^\nu}
  (-)^{\nu+k}t^{\nu+1}u^{k+1}
  e^{-\textstyle\frac{t^2+u^2}{8\ha}}
  K_\nu\left(\frac{t u}{4\ha}\right) K_{\nu+k}(u)
  \:.
  \label{gtu}
\end{align}
While the first integral in Eq.~\eqref{Zgtu} is well-behaved, for
$\nu>0$ the integrand of the second integral (for which $t=-\hm$) is
singular at $u=0$.  This singularity is labeled by $\Delta$ in
Fig.~\ref{Fig:integplane}.  To avoid this singularity when taking the
$\epsilon\to0$ limit, we deform the integration path on $S_\pm$
near zero as shown in Fig.~\ref{Fig:integline} and then take
$\epsilon\to0$.  This yields \FIGURE[t]{
  \includegraphics[width=10cm]{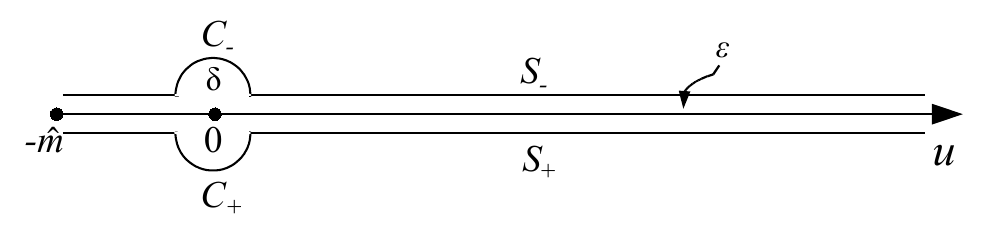}
  \caption{Deformed integration paths over $u$ for $t=-\hm$.}
  \label{Fig:integline}
}
\begin{align}
  \I{k}
  = -\frac{i\pi}{2\hm} \Bigg\{ \int_{\hm}^\infty du \, G(\hm,u) 
  - \lim_{\delta\to0^+} \Biggl[ \int_{-\hm}^{-\delta}
  +\int_\delta^\infty \Biggr]du \, G(-\hm,u) 
  \Bigg\}  + \Csing_{\nu,k}(\ha,\hm) 
  \label{Zu}
\end{align}
with $G(t,u) = \lim_{\epsilon\to 0^+} [g(t,u_+) - g(t,u_-)]$.  The
contribution of the singularity at $u=0$, which for simplicity will be
called the $\Delta$-term, is
\begin{align}
  \Csing_{\nu,k}(\ha,\hm) \equiv \frac{i\pi}{2\hm} \lim_{\delta\to
    0^+} \Bigg\{ \int_{C_+} du \, g(-\hm,u) -
  \int_{C_-} du \, g(-\hm,u) \Bigg\} \:,
  \label{Csingdef}
\end{align}
where $C_\pm$ denotes the two small semicircles shown in
Fig.~\ref{Fig:integline}.  We shall see in Sec.~\ref{Sec:singularity}
that this term makes a contribution to the phase factor for $\nu\ne0$.

\subsection{Contribution of the branch cut discontinuity}

The integrand in the curly braces of Eq.~\eqref{Zu} involves
differences that can be simplified using Eq.~\eqref{KetaK},
\begin{align}
  \lim_{\epsilon\to 0^+} \bigl[ K_\nu(t u) K_{\nu+k}(u) \bigr]_+ - \bigl[
  K_\nu(t u) K_{\nu+k}(u) \bigr]_-
  =
  \begin{cases}
    0 &\text{for }  t,u>0 \:,\\
    - 2 i \pi I_\nu(|tu|) K_{\nu+k}(|u|) &\text{for }  t<0<u \:,\\
    + 2 i \pi K_\nu(|tu|) I_{\nu+k}(|u|) &\text{for }  t, u<0 \:,
  \end{cases}
  \label{sumkern}
\end{align}
where we have again omitted the factor $1/4\ha$.
The only contributions in Eq.~\eqref{sumkern} come from the branch cut
discontinuity of the $K$-Bessel function for negative real arguments.
The first case in Eq.~\eqref{sumkern} vanishes as neither Bessel
function has a negative argument and no branch cut discontinuity is
encountered.  We now use Fig.~\ref{Fig:integplane} to distinguish the
various integration regions of Eq.~\eqref{Zu}. For the $t=\hm$ integral
we always have $t,u>0$, and according to Eq.~(\ref{sumkern}) this
integral vanishes.  Substituting Eq.~\eqref{sumkern} in Eq.~\eqref{Zu}
with integrand \eqref{gtu}, we are left with
\begin{align}
  \I{k}
  = \frac{e^{-2\ha-\frac{\hm^2}{8\ha}}}{4\ha}
  \Bigg\{  &
  \int_0^\infty du \, 
  (-)^{k} u^{k+1} 
  e^{-\textstyle\frac{u^2}{8\ha}}
  I_\nu\left(\frac{|\hm u|}{4\ha}\right)
  K_{\nu+k}(|u|)
  \notag\\
  &+\int_{-\hm}^{0} du \,
  (-)^{k+1} u^{k+1} 
  e^{-\textstyle\frac{u^2}{8\ha}}
  K_\nu\left(\frac{|\hm u|}{4\ha}\right)
  I_{\nu+k}(|u|) 
  \Bigg\}
  +  \Csing_{\nu,k}(\ha,\hm) 
  \:,
\end{align}
where the integration limits $\pm\delta$ in the curly braces of
Eq.~\eqref{Zu} were set to zero as the integrands are regular for $u
\to 0$.  This expression can be simplified by transforming $u \to -u$
in the second integral, and we obtain
\begin{align}
  \I{k}
  &= \frac{e^{-2\ha-\frac{\hm^2}{8\ha}}}{4\ha} \Bigg\{  
  \int_0^\infty du \, 
  (-)^k u^{k+1} 
  e^{-\textstyle\frac{u^2}{8\ha}}
  I_\nu\left(\frac{\hm u}{4\ha}\right)
  K_{\nu+k}(u)
  +\int_0^{\hm} du \,
  u^{k+1} 
  e^{-\textstyle\frac{u^2}{8\ha}}
  K_\nu\left(\frac{\hm u}{4\ha}\right)
  I_{\nu+k}(u) 
  \Bigg\} \notag \\
  &\quad + \Csing_{\nu,k}(\ha,\hm) 
  \:.
  \label{simplepole}
\end{align}
For $\nu=0$ this expression is identical to the result given
previously in Ref.~\cite{Splittorff:2007ck}, as the $\Delta$-term
vanishes in this case. For $\nu \neq 0$ the additional contribution is
important and will be computed below, see Eq.~\eqref{Csingnuk} for the
result.  We shall see that it even dominates for small $\hm$.

\subsection{Contribution of the Bessel function singularity}
\label{Sec:singularity}

The semicircles around the singularity in Eq.~\eqref{Csingdef} run in
opposite directions on the sheets $S_+$ and $S_-$.  Reversing the
direction of integration on one of the sheets changes the difference
in Eq.~\eqref{Csingdef} to a sum.

The $K$-Bessel functions in the integrand can be split into a
meromorphic part with a pole at $u=0$ and a part containing the branch
cut.  For $\delta \neq 0$ the integrals along $C_\pm$ would not only
receive contributions from the singularity at $u=0$ but also from the
branch cuts, lying on both sides of the origin, which were already
included in the line integrals given in Eq.~\eqref{simplepole}. When
$\delta\to 0$ the branch cut contributions will vanish from the
integral \eqref{Csingdef}, and only the meromorphic part of
the integrand will contribute to the $\Delta$-term. Therefore we can
rewrite the $\Delta$-term \eqref{Csingdef} as an integral over a
closed contour $\Gamma_0$ enclosing the singularity of the
meromorphic part $\tilde g(-\hm,u)$ of the integrand.  We thus obtain
\begin{align}
  \Csing_{\nu,k}(\ha,\hm) = \frac{i\pi}{2\hm} \oint_{\Gamma_0} du \,
  \tilde g(-\hm,u) \:,
  \label{csing}
\end{align}
where $\Gamma_0$ consists of $C_+$ and $C_-$ and is traversed in the
counterclockwise direction.

For $z\to0$, $K_\nu(z)$ diverges logarithmically for $\nu=0$ and like
$z^{-\nu}$ for $\nu>0$. Upon multiplication by $u^{k+1}$, for $\nu=0$
the integrand \eqref{gtu} is not singular at $u=0$, and thus the
$\Delta$-term vanishes in this case.  For $\nu>0$ the integrand has a
pole of order $(2\nu-1)$ at $u=0$.  Using the residue theorem, we
obtain
\begin{align}
  \Csing_{\nu,k}(\ha,\hm) = \frac{i\pi}{2\hm}\: 2\pi i\: \Res_{u=0}
  \tilde g = -\frac{\pi^2}{\hm}a_1\:,
  \label{residue}
\end{align}
where $a_1$ is the coefficient of the $u^{-1}$ term in the Laurent
expansion of $\tilde g(-\hm,u)$ around zero.  To find this
coefficient, we neglect the $u$-independent terms in Eq.~\eqref{gtu}
and write the $u$-dependent terms as
\begin{align}
  u f(u) g(u) h (u) = u \underbrace{\exp\Big(-\frac{u^2}{8\ha}\Big)}_{=f} 
  \underbrace{K_\nu\Big(\frac{-\hm u}{4\ha}\Big)}_{=g}
  \underbrace{\Big[ u^k K_{\nu+k}(u) \Big]}_{=h} \:.
  \label{ufgh}
\end{align}
As the product $u g(u) h(u)$ has a singularity of order $2\nu-1$, we
need to perform the Taylor expansion of the exponential to order
$2\nu-2$ to find all contributions to the simple pole.  Hence
\begin{align}
  f(u) \equiv \exp\left(-\frac{u^2}{8\ha}\right) 
  = \sum_{\ell=0}^{\nu-1} \frac{1}{\ell!}
  \left(-\frac{u^2}{8\ha}\right)^\ell + {\cal O}(u^{2\nu}) \:. 
  \label{expseries}
\end{align}
For $\nu>0$ the series expansion of $K_\nu(z)$ for small $z$ is
\cite[Eq.~(9.6.11)]{Abram:1964}
\begin{align}
  K_\nu(z) = \frac{2^{\nu-1}}{z^\nu} \sum_{k=0}^{\nu-1}
  \frac{(\nu-1-k)!}{k!} \left( -\frac{z^2}{4}\right)^k + {\cal
    O}(z^\nu)\:. 
  \label{Kseries}
\end{align}
The functions $g$ and $h$ containing the Bessel functions each need to
be expanded to order $\nu-2$ using Eq.~\eqref{Kseries}, as the
remaining part of the integrand is as singular as $u^{1-\nu}$. So
\begin{align}
  g(u) \equiv K_\nu(bu) 
  = \frac{2^{\nu-1}}{(bu)^\nu} \sum_{i=0}^{\nu-1}
  \frac{(\nu-1-i)!}{i!} \left( -\frac{(bu)^2}{4}\right)^i + {\cal
    O}(u^\nu) 
  \label{guseries}
\end{align}
with $b=-\hm/4\ha$ and 
\begin{align}
  h(u) \equiv u^{k} K_{\nu+k}(u) 
  =  \frac{2^{\nu+k-1}}{u^\nu} \sum_{j=0}^{\nu-1}
  \frac{(\nu+k-1-j)!}{j!} \left( -\frac{u^2}{4}\right)^j + {\cal
    O}(u^\nu)\:. 
  \label{huseries}
\end{align}
Putting things together yields a triple sum, given by the product of
Eqs.~\eqref{expseries}, \eqref{guseries}, and \eqref{huseries}.  Since
we are only looking for the coefficient of $u^{-1}$, we obtain the
condition $1+2\ell+(2i-\nu)+(2j-\nu)=-1$, which eliminates one of the
sums by setting $\ell = \nu-1-i-j$.  Since $\ell\ge0$ we also have
$i+j\le\nu-1$ and thus
\begin{align}
  \Res_{u=0} \big[ u f(u) g(u) h(u) \big]
  &= - \sum_{i,j=0}^{i+j\le\nu-1}
  \frac{(\nu-1-i)!(\nu+k-1-j)!}{(\nu-1-i-j)! i! j!}
  2^{\nu+k+1-3i+j}\hm^{-\nu+2i}\ha^{-i+j+1}
  \:.
\end{align}
Combining this expression with the $u$-independent terms in
Eq.~\eqref{gtu} and substituting it in Eq.~\eqref{residue} yields
\begin{align}
  \Csing_{\nu,k}(\ha,\hm) &= e^{-2\ha-\frac{\hm^2}{8\ha}}  
  \frac{ (-)^k2^{\nu+k-1}}{\hm^\nu} 
  \sum_{i,j=0}^{i+j\le\nu-1}
  \frac{(\nu-1-i)!(\nu+k-1-j)!}{(\nu-1-i-j)! i! j!}
  \left(\frac{\hm^{2}}{8\ha}\right)^i \left(2\ha\right)^j
  \:,
  \label{Csingnuk}
\end{align}
which is a bivariate polynomial of degree $\nu-1$ in $\hm^2/\ha$ and
$\ha$ which contains $\nu(\nu+1)/2$ terms.  Our final result for
$\I{k}$ is thus
\begin{align}
  \I{k}
  &= \frac{e^{-2\ha-\frac{\hm^2}{8\ha}}}{4\ha} \Bigg\{  
  \int_0^\infty du \, 
  (-)^k u^{k+1} 
  e^{-\textstyle\frac{u^2}{8\ha}}
  I_\nu\left(\frac{\hm u}{4\ha}\right)
  K_{\nu+k}(u)  
  +\int_0^{\hm} du \,
  u^{k+1}
  e^{-\textstyle\frac{u^2}{8\ha}}
  K_\nu\left(\frac{\hm u}{4\ha}\right)
  I_{\nu+k}(u)
  \Bigg\} \notag\\
  &\quad+ e^{-2\ha-\frac{\hm^2}{8\ha}} 
  \frac{(-)^k 2^{\nu+k-1}}{\hm^\nu} 
  \sum_{i,j=0}^{i+j\le\nu-1}
  \frac{(\nu-1-i)!(\nu+k-1-j)!}{(\nu-1-i-j)! i! j!}
  \left(\frac{\hm^{2}}{8\ha}\right)^i \left(2\ha\right)^j
  \:.
  \label{Z1total}
\end{align}
As mentioned earlier, for $\nu <0$ one just needs to replace $\nu$ by
$|\nu|$.  Our general expression can also be used for $\nu=0$ as the
double sum vanishes and the correct $\nu=0$ result is reproduced.

Note that for $k=0$, Eq.~\eqref{Z1total} has been computed in
Ref.~\cite{Osborn:2008jp} using a different method.  The result given
in Eq.~(65) of that reference looks rather different from our result
but agrees numerically with Eq.~\eqref{Z1total} for $k=0$ after
adjusting some prefactors.  However, for the calculation of the phase
factor using Eqs.~\eqref{qu0} and \eqref{phfacNfflavors_arbmass} we
need the more general result of Eq.~\eqref{Z1total} for arbitrary $k$.
The method of Ref.~\cite{Osborn:2008jp} does not straightforwardly
extend to $k\ne0$ since it uses orthogonality relations that only hold
for $k=0$.

\section{Explicit results}
\label{sec:results}

\subsection{Quenched case}
\label{Sec:Quenched}

We now write down the explicit form of the phase factor for the
quenched case using the solution for the Cauchy transform integral
derived in the previous section.  Recalling Eq.~\eqref{qu0}, we obtain
\begin{align}
  \phs_{N_f=0}
  &= 
  \frac{e^{-2\ha-\frac{\hm^2}{8\ha}}}{4\ha} 
  \Bigg\{ 
  \int_0^\infty du \,  u
  e^{-\textstyle\frac{u^2}{8\ha}}
  I_\nu\left(\frac{\hm u}{4\ha}\right) 
  \big[ 
  \hm I_{\nu+1}\left(\hm\right) K_{\nu}(u) 
  +
  I_\nu\left(\hm\right) u
  K_{\nu+1}(u)  
  \big] 
  \notag\\
  &\quad+\int_0^{\hm} du \, u
  e^{-\textstyle\frac{u^2}{8\ha}}
  K_\nu\left(\frac{\hm u}{4\ha}\right)
  \big[
  \hm I_{\nu+1}\left(\hm\right) I_{\nu}(u)
  -
  I_\nu\left(\hm\right) u I_{\nu+1}(u)
  \big]
  \Bigg\}
  +\Csing_{\nu}^{\Nf=0}(\ha,\hm) 
  \:,
  \label{phfacquenched}
\end{align}
where
\begin{align}
  \Csing_{\nu}^{\Nf=0}(\ha,\hm) 
  & = e^{-2\ha-\frac{\hm^2}{8\ha}} 
  \frac{ 2^{\nu-1}}{\hm^\nu} 
  \sum_{i,j=0}^{i+j\le\nu-1}
  \frac{(\nu-1-i)!(\nu-1-j)!}{(\nu-1-i-j)! i! j!}
  \left(\frac{\hm^{2}}{8\ha}\right)^i \left(2\ha\right)^j 
  \begin{vmatrix}
    1 & -2(\nu-j) \\
    I_\nu\left(\hm\right) & \hm I_{\nu+1}\left(\hm\right)
  \end{vmatrix}
  \label{Csingnu}
  \:. 
\end{align}
The contribution of both terms to the total phase factor will be
illustrated in the numerical results of Sec.~\ref{Sec:Results}.

\subsection{One and two dynamical flavors with equal masses}
\label{sec:equalmass}

We now give explicit expressions for the one- and two-flavor case.
For one dynamical fermion with mass equal to the valence quark mass,
the phase factor from Eq.~\eqref{phfacNfflavors} is given by
\begin{align}
\phs_{N_f=1}
= \frac{1}{ 2\hm  I_\nu(\hm)} 
 \begin{vmatrix}
\I{0} & \I{1} & \I{2}\\[1mm]
I_{\nu,0}(\hm) & I_{\nu,1}(\hm) & I_{\nu,2}(\hm)\\[1mm]
I'_{\nu,0}(\hm) & I'_{\nu,1}(\hm) & I'_{\nu,2}(\hm)
\label{1flavor2}
\end{vmatrix} \:.
\end{align}
Substituting the solution \eqref{Z1total} derived in
Sec.~\ref{Sec:Analytic} for the complex integral we find
\begin{align}
  \label{phfac1flav} 
  \phs_{N_f=1}
  &= \frac{e^{-2\ha-\frac{\hm^2}{8\ha}}}{8\ha\hm I_\nu(\hm)}
  \left\{  
  \int_0^\infty du \, 
  u 
  e^{-\textstyle\frac{u^2}{8\ha}}
  I_\nu\left(\frac{\hm u}{4\ha}\right)
  \begin{vmatrix}
    K_\nu(u) & - u K_{\nu+1}(u) & u^2 K_{\nu+2}(u)\\[1mm]
    I_{\nu,0}(\hm) & I_{\nu,1}(\hm) & I_{\nu,2}(\hm)\\[1mm]
    I'_{\nu,0}(\hm) & I'_{\nu,1}(\hm) & I'_{\nu,2}(\hm)
  \end{vmatrix}\right.  \\
  & \hspace{25mm}\left. +\int_0^{\hm} du \, u 
  e^{-\textstyle\frac{u^2}{8\ha}}
  K_\nu\left(\frac{\hm u}{4\ha}\right) 
  \begin{vmatrix}
    I_\nu(u) & u I_{\nu+1}(u) & u^2 I_{\nu+2}(u)\\[1mm]
    I_{\nu,0}(\hm) & I_{\nu,1}(\hm) & I_{\nu,2}(\hm)\\[1mm]
    I'_{\nu,0}(\hm) & I'_{\nu,1}(\hm) & I'_{\nu,2}(\hm)
  \end{vmatrix}\: \right\}
  \notag\\
  &\quad+ \frac{2^{\nu-2} e^{-2\ha-\frac{\hm^2}{8\ha}}}{\hm^{\nu+1} I_\nu(\hm)}
  \sum_{i,j=0}^{i+j\le\nu-1}
  \frac{(\nu-1-i)!(\nu-1-j)!}{(\nu-1-i-j)! i! j!}
  \left(\frac{\hm^{2}}{8\ha}\right)^i \left(2\ha\right)^j 
  \begin{vmatrix}
    1 & -2(\nu-j)\quad & 4(\nu-j)_2\\[1mm]
    I_{\nu,0}(\hm) & I_{\nu,1}(\hm) & I_{\nu,2}(\hm)\\[1mm]
    I'_{\nu,0}(\hm) & I'_{\nu,1}(\hm) & I'_{\nu,2}(\hm)
  \end{vmatrix} \notag
\end{align}
with the Pochhammer symbol $(a)_n$ defined in Eq.~\eqref{eq:poch}.
Using $ I'_\nu(z) = I_{\nu-1}(z) - \nu I_\nu(z)/z $
\cite[Eq.~(9.6.26)]{Abram:1964} the derivatives in
Eq.~\eqref{phfac1flav} can be explicitly computed,
\begin{align}
I'_{\nu,k}(\hm) = [\hm^k I_{\nu+k}(\hm)]' 
= \hm^{k-1} \Big[ \hm I_{\nu+k-1}(\hm) - \nu I_{\nu+k}(\hm) \Big] \:.
\end{align}
For two dynamical fermions with masses equal to that of the valence
quark, Eq.~\eqref{phfacNfflavors} yields
\begin{align}
  \phs_{N_f=2}
  &= \frac{1}{ 8\hm^2} 
  \frac{ \begin{vmatrix}
      \I{0} & \I{1} & \I{2} & \I{3}\\[1mm]
      I_{\nu,0}(\hm) & I_{\nu,1}(\hm) & I_{\nu,2}(\hm) & I_{\nu,3}(\hm) \\[1mm]
      I'_{\nu,0}(\hm) & I'_{\nu,1}(\hm) & I'_{\nu,2}(\hm) & I'_{\nu,3}(\hm) \\[1mm]
      I''_{\nu,0}(\hm) & I''_{\nu,1}(\hm) & I''_{\nu,2}(\hm) & I''_{\nu,3}(\hm)
    \end{vmatrix}
  }{
    \begin{vmatrix}
      I_{\nu,0}(\hm) & I_{\nu,1}(\hm) \\[1mm]
      I'_{\nu,0}(\hm) & I'_{\nu,1}(\hm) 
    \end{vmatrix}
  }\:.
\end{align}
One can show that
\begin{align}
  \begin{vmatrix}
    I_{\nu,0}(\hm) & I_{\nu,1}(\hm) \\[1mm]
    I'_{\nu,0}(\hm) & I'_{\nu,1}(\hm) 
  \end{vmatrix}
  = \hm \left[I_{\nu }(\hm)^2-I_{\nu -1}(\hm) I_{\nu +1}(\hm)\right] \:,
\end{align}
and using the solution \eqref{Z1total} for the complex integral we find
\begin{align}
  \phs_{N_f=2}
  &= \frac{e^{-2\ha-\frac{\hm^2}{8\ha}}}{32\ha\hm^3 
    \left[I_{\nu }(\hm)^2-I_{\nu -1}(\hm) I_{\nu +1}(\hm)\right]}
  \notag\\
  &\quad\times \left\{  
    \int_0^\infty du \, u 
    e^{-\textstyle\frac{u^2}{8\ha}}
    I_\nu\left(\frac{\hm u}{4\ha}\right)
    \begin{vmatrix}
      K_\nu(u) & - u K_{\nu+1}(u) & u^2 K_{\nu+2}(u) & -u^3 K_{\nu+3}(u) \\[1mm]
      I_{\nu,0}(\hm) & I_{\nu,1}(\hm) & I_{\nu,2}(\hm) & I_{\nu,3}(\hm) \\[1mm]
      I'_{\nu,0}(\hm) & I'_{\nu,1}(\hm) & I'_{\nu,2}(\hm) & I'_{\nu,3}(\hm) \\[1mm]
      I''_{\nu,0}(\hm) & I''_{\nu,1}(\hm) & I''_{\nu,2}(\hm) & I''_{\nu,3}(\hm)
    \end{vmatrix}\right.
  \notag\\
  &\qquad\qquad+\left.\int_0^{\hm} du \, u 
    e^{-\textstyle\frac{u^2}{8\ha}}
    K_\nu\left(\frac{\hm u}{4\ha}\right)
    \begin{vmatrix}
      I_\nu(u) & u I_{\nu+1}(u) & u^2 I_{\nu+2}(u) & u^3 I_{\nu+3}(u) \\[1mm]
      I_{\nu,0}(\hm) & I_{\nu,1}(\hm) & I_{\nu,2}(\hm) & I_{\nu,3}(\hm) \\[1mm]
      I'_{\nu,0}(\hm) & I'_{\nu,1}(\hm) & I'_{\nu,2}(\hm) & I'_{\nu,3}(\hm) \\[1mm]
      I''_{\nu,0}(\hm) & I''_{\nu,1}(\hm) & I''_{\nu,2}(\hm) & I''_{\nu,3}(\hm)
    \end{vmatrix} \:\right\}
  \notag\\
  &\quad+ \frac{2^{\nu-4}e^{-2\ha-\frac{\hm^2}{8\ha}}}{\hm^{\nu+3} \left[I_{\nu }(\hm)^2-I_{\nu -1}(\hm) I_{\nu +1}(\hm)\right]}
  \sum_{i,j=0}^{i+j\le\nu-1}
  \frac{(\nu-1-i)!(\nu-1-j)!}{(\nu-1-i-j)! i! j!}
  \left(\frac{\hm^{2}}{8\ha}\right)^i \left(2\ha\right)^j \notag\\
  &\hspace{30mm}\times \begin{vmatrix}
    1 & -2(\nu-j)\quad & 4(\nu-j)_2\quad & -8(\nu-j)_3 \\[1mm]
    I_{\nu,0}(\hm) & I_{\nu,1}(\hm) & I_{\nu,2}(\hm) & I_{\nu,3}(\hm) \\[1mm]
    I'_{\nu,0}(\hm) & I'_{\nu,1}(\hm) & I'_{\nu,2}(\hm) & I'_{\nu,3}(\hm) \\[1mm]
    I''_{\nu,0}(\hm) & I''_{\nu,1}(\hm) & I''_{\nu,2}(\hm) & I''_{\nu,3}(\hm)
  \end{vmatrix} 
  \label{phfac2f}
\end{align}
with the Pochhammer symbol $(a)_n$ from Eq.~\eqref{eq:poch}.  

In Fig.~\ref{Fig:unq-m-dep} we compare the mass dependence of the
average phase factor in the quenched case, given by
Eq.~\eqref{phfacquenched}, with the predictions for one and two
dynamical flavors from Eqs.~\eqref{phfac1flav} and \eqref{phfac2f}.
Although the sign problem becomes less severe as the fermion mass
increases, the dynamical quarks have a negative effect as they clearly
reduce the value of the phase factor.  Note that, surprisingly, for
$\nu=0$ and small mass ($\hm\lesssim1$) the effect of the dynamical
quarks seems to be reversed (this is true for any value of $\ha$).
This is related to the fact that for $\nu=1$ the chiral limit of the
average phase factor is equal to $e^{-2\ha}$, independently of the
number of flavors, see Eq.~\eqref{eq:ch} below.  This is also verified
in the middle plot of Fig.~\ref{Fig:unq-m-dep}.
\FIGURE{
  \includegraphics[type=pdf,ext=.pdf,read=.pdf,width=0.32\textwidth]{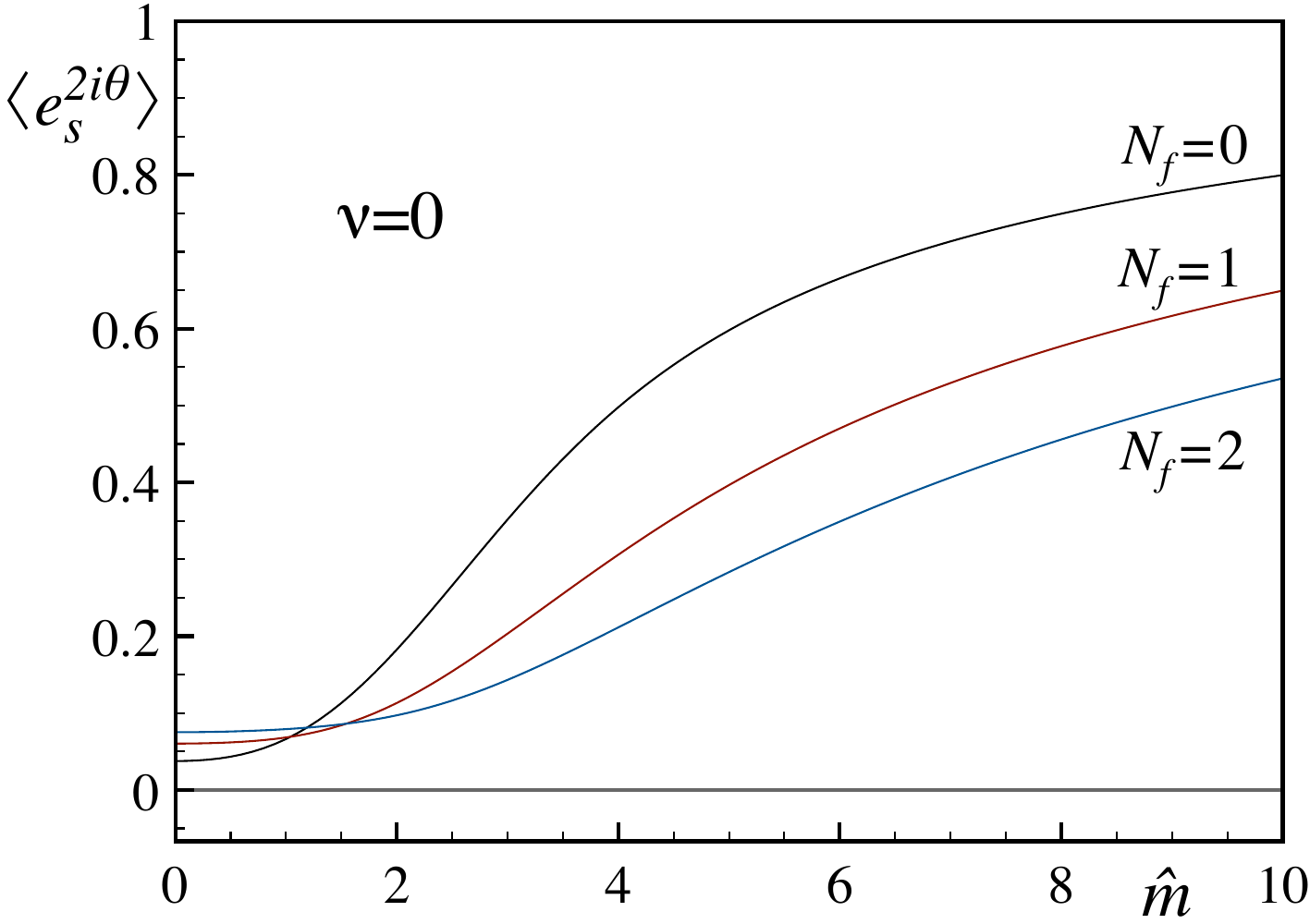}
  \includegraphics[type=pdf,ext=.pdf,read=.pdf,width=0.32\textwidth]{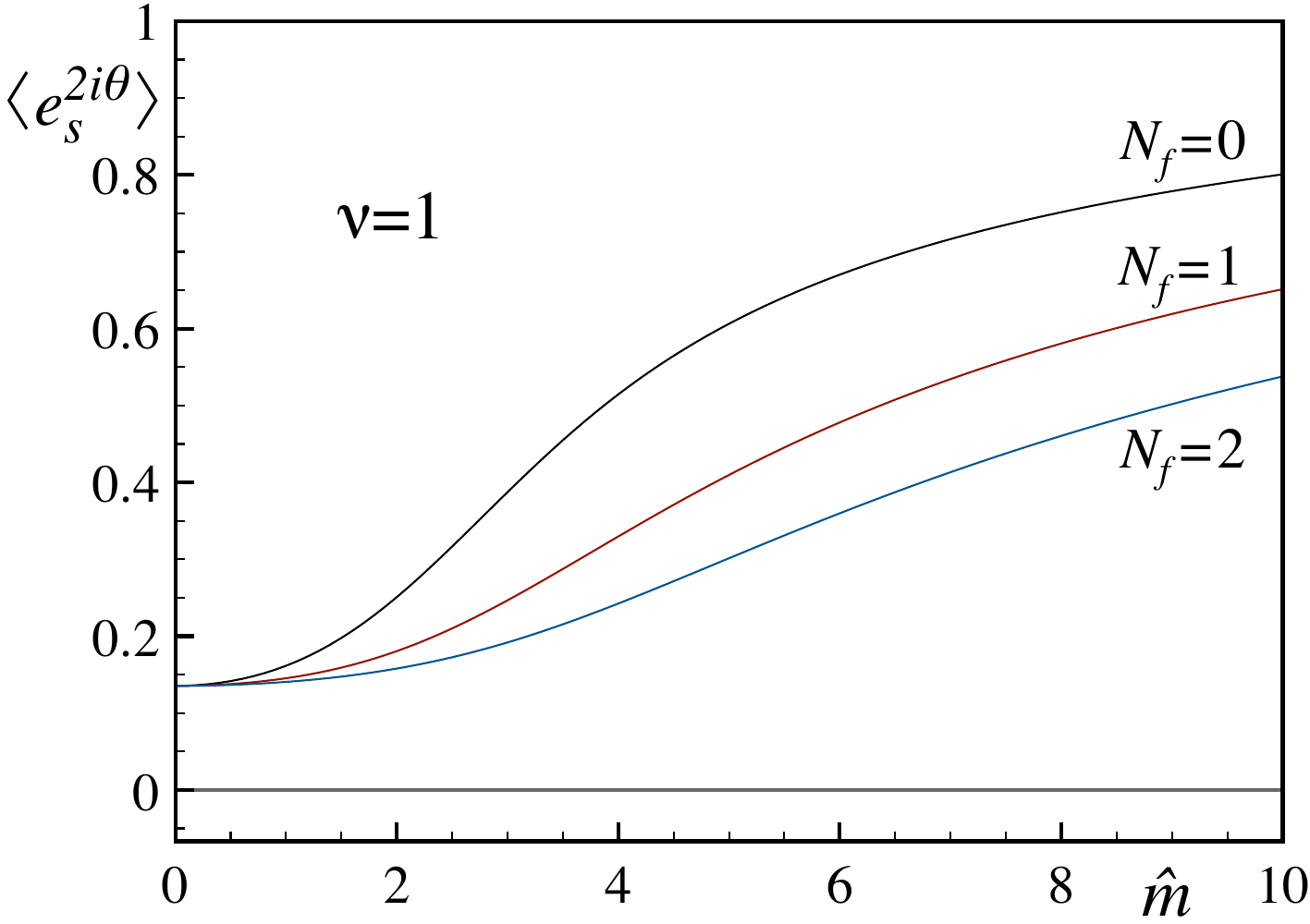}
  \includegraphics[type=pdf,ext=.pdf,read=.pdf,width=0.32\textwidth]{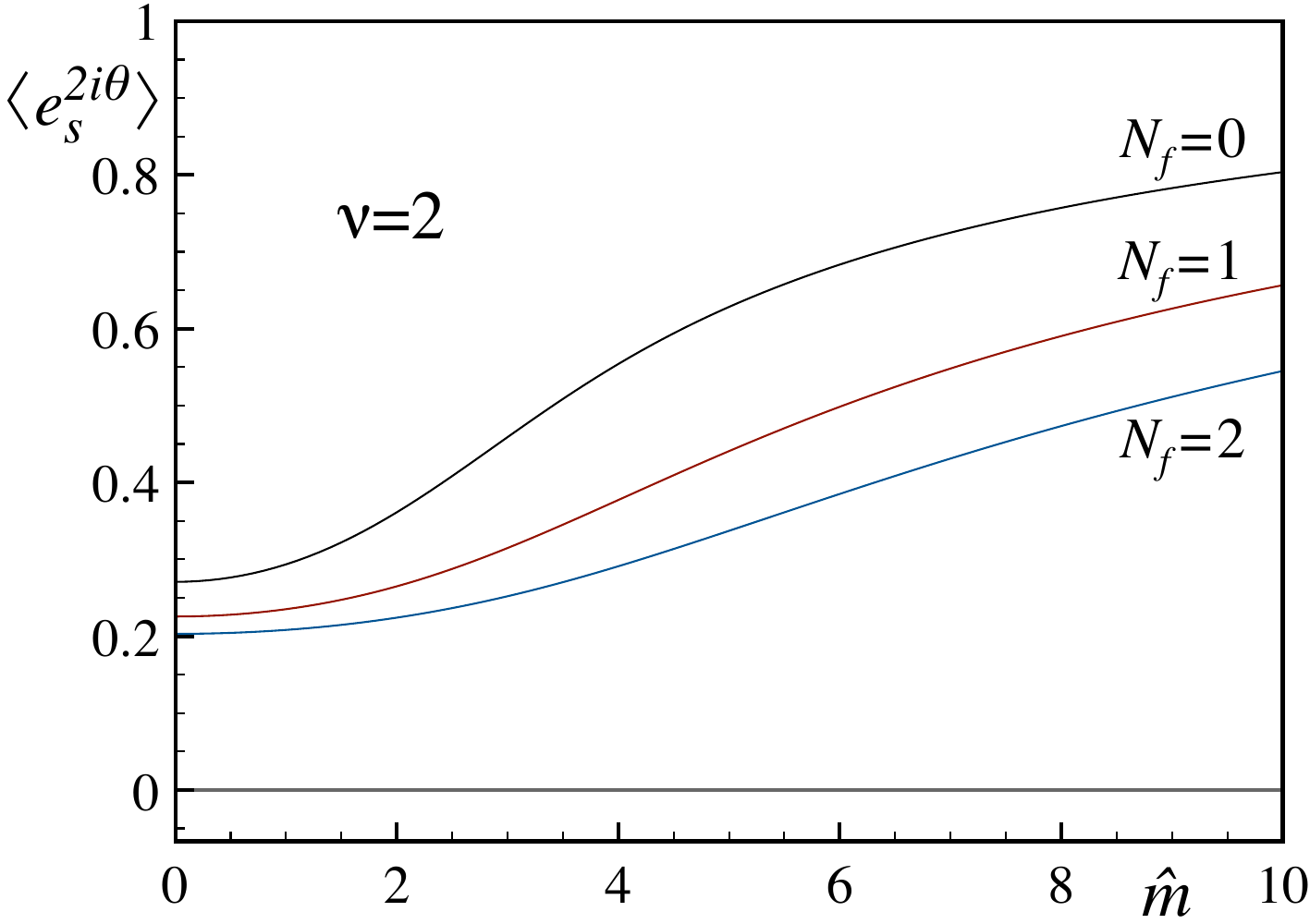}
  \caption{Average phase factor of the fermion determinant for
    $\Nf=0,1,2$ as a function of the fermion mass $\hm$ for $\ha=1.0$
    and $\nu=0,1,2$.  We have verified that the $\hm\to0$ limit of the
    curves agrees with the chiral limit computed analytically in
    Sec.~\ref{sec:Chiral}.}
  \label{Fig:unq-m-dep}
}

\subsection{Chiral limit}
\label{sec:Chiral}

In the limit $\hm \to 0$ the phase factor \eqref{phapp} can be
simplified and written in terms of special functions.  The limit has
to be derived differently for $\nu=0$ and for $\nu \neq 0$, and the
general derivation for arbitrary $\Nf$ can be found in
App.~\ref{app:chiral}.  For trivial topology the result is given by
Eq.~\eqref{phfacchiralnu=0},
\begin{equation}
  \phs_{\nu=0,\hm=0}
  = (\Nf+1) (2\ha)^{N_f+1} \Gamma(-N_f-1,2\ha)  \:,
  \label{lim_m=0_nu=0}
\end{equation}
where $\Gamma(a,z)$ is the incomplete gamma function.  For nontrivial
topology the result is given by Eq.~\eqref{H12},
\begin{align}
  \label{eq:ch}
  \phs_{\nu>0,\hm=0}
  &= e^{-2\ha}
  \sum_{j=0}^{\nu-1}
  \frac{(\nu-j)_{\Nf+1}}{(\nu)_{\Nf+1}}
  \frac{\left(2\ha\right)^j}{j!} \:,
\end{align}
where the Pochhammer symbol $(a)_n$ is defined in Eq.~\eqref{eq:poch}.
Note that for $\nu=1$ the phase factor is simply given by $e^{-2\ha}$,
independently of $N_f$.  The result \eqref{eq:ch} can also be
expressed in terms of incomplete gamma functions as shown in
Eq.~\eqref{H16},
\begin{align}
  \phs_{\hm=0}
  &= \frac{1}{(\nu+\Nf)!} \sum_{k=0}^{\Nf+1} (-)^k
  \binom{\Nf + 1}{k} 
  (\nu - k)_{\Nf+1} (2\ha)^k  \Gamma(\nu-k,2\ha)\:.
  \label{lim_m=0_nu}
\end{align}
The last expression is also valid for $\nu=0$, see
Eq.~\eqref{eq:chnu0}.  The quenched results are obtained by setting
$N_f=0$ in the above equations.

As discussed in App.~\ref{app:chiral}, for $\hm=0$ and $\nu \neq 0$
the phase of the determinant is exclusively given by the new
$\Delta$-term.  It is interesting to note that the contributions to
the phase of the determinant in the chiral limit originate from
different terms for $\nu=0$ and $\nu \neq 0$.

\subsection{Thermodynamic limit}
 
\FIGURE[b]{
  \label{fig:thermo}
  \hfill
  \includegraphics[type=pdf,ext=.pdf,read=.pdf,width=0.37\textwidth]{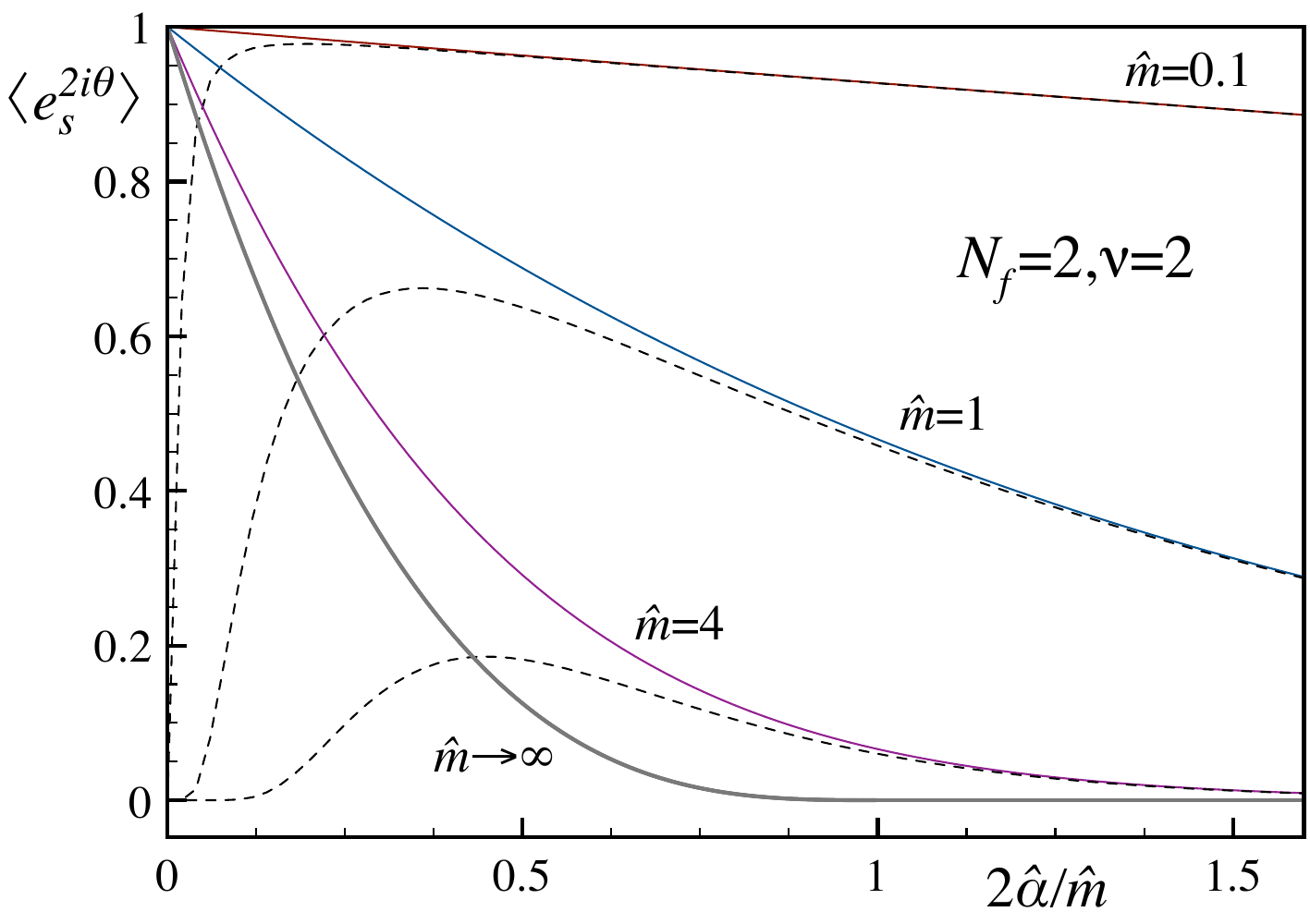}
  \hfill
  \includegraphics[type=pdf,ext=.pdf,read=.pdf,width=0.37\textwidth]{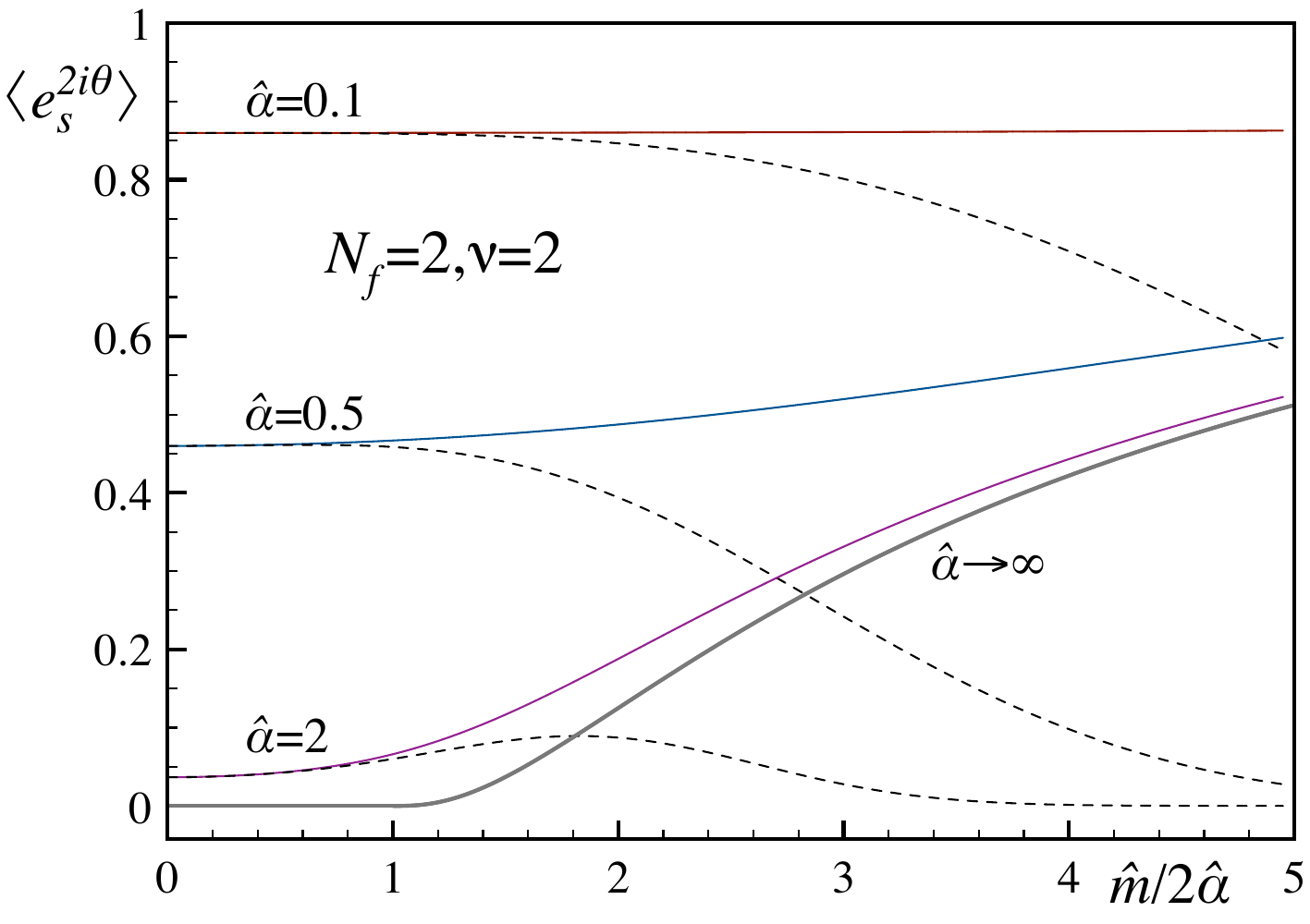}
  \hfill
  \caption{The left plot shows how the $\ha$-dependence of the phase
    factor changes as a function of $\hm$.  In the right plot the
    roles of $\ha$ and $\hm$ are reversed.  The curves $\hm\to\infty$
    and $\ha\to\infty$ correspond to the thermodynamic limit
    \eqref{G1}.  The dashed lines indicate the contribution of the
    $\Delta$-term to the phase factor (this contribution vanishes in
    the thermodynamic limit).}
}

The thermodynamic limit of the phase factor \eqref{phapp} is
the limit of that equation for $\ha=\mu^2F^2V\to\infty$
and $\hm=mV\Sigma\to\infty$.  As discussed in
Ref.~\cite{Splittorff:2007ck}, this limit depends on whether
$2\ha/\hm$ (or, equivalently, $2\mu/m_\pi$) is smaller or larger than
1. We find that for $2\ha<\hm$ the thermodynamic limit is given by 
\begin{align}
  \phs^\text{th}
  = \left(1-2\ha/\hm\right)^{\Nf+1}\:.
  \label{G1}
\end{align}
As expected, the thermodynamic limit does not depend on $\nu$, which
is a consistency check of our result.  Equation~\eqref{G1} agrees with
the special cases $\nu=0$ and $N_f=0,1,2$ considered in
Ref.~\cite{Splittorff:2007ck}.  The proof of our general result is
given in App.~\ref{app:thermo}.  Note that the contribution of the
$\Delta$-term vanishes in the thermodynamic limit.  For $2\ha>\hm$ the
phase factor is exponentially suppressed in the volume so that its
thermodynamic limit is zero.  An asymptotic large-volume expansion of
the phase factor could be computed for this case from
Eq.~\eqref{phapp}, analogously to Ref.~\cite{Splittorff:2007ck}.

In Fig.~\ref{fig:thermo} we show how the thermodynamic limit is
approached for the case of $N_f=2$ and $\nu=2$.

\subsection{Numerical simulations}
\label{Sec:Results}

To check our analytical results and especially the contribution of the
$\Delta$-term, which is new in the $\nu \neq 0$ case, we performed
numerical random matrix simulations to compute the average phase
factor of the fermion determinant in quenched \chrmt. The simulation
details can be found in App.~\ref{rmtsim}.

In order to keep the statistical error sufficiently small the
simulations were performed with samples of 100,000 random matrices.
Figure~\ref{Fig:phase_vs_alpha} shows the $\ha$-dependence of the
quenched average phase factor for $\nu=0,1,2,3,4,5$ with $\hm=0$ and
matrix size $N=20$. The simulation results are compared with the
chiral limit of the analytical results given in
Eqs.~\eqref{lim_m=0_nu=0} and \eqref{eq:ch} with $N_f=0$. %
The agreement is extremely good except for larger values of $\nu$
($\ge 3$) where the simulation results lie slightly, but
systematically, below the predictions.  This is merely a finite-$N$
effect, which is also clear from the comparison with the exact
finite-$N$ result shown for $\nu=5$.

\FIGURE{
  \includegraphics[width=0.32\textwidth]{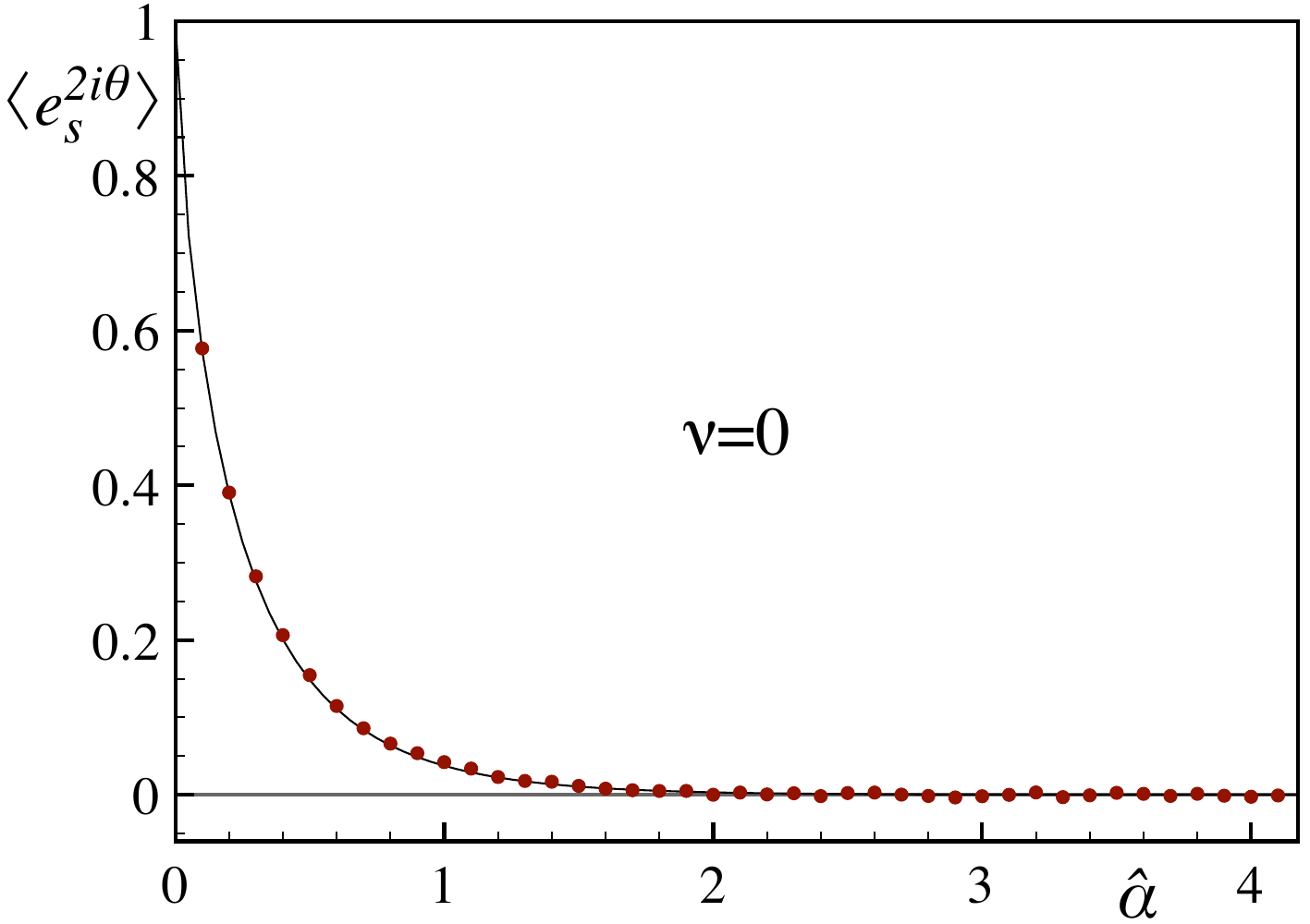}
  \includegraphics[width=0.32\textwidth]{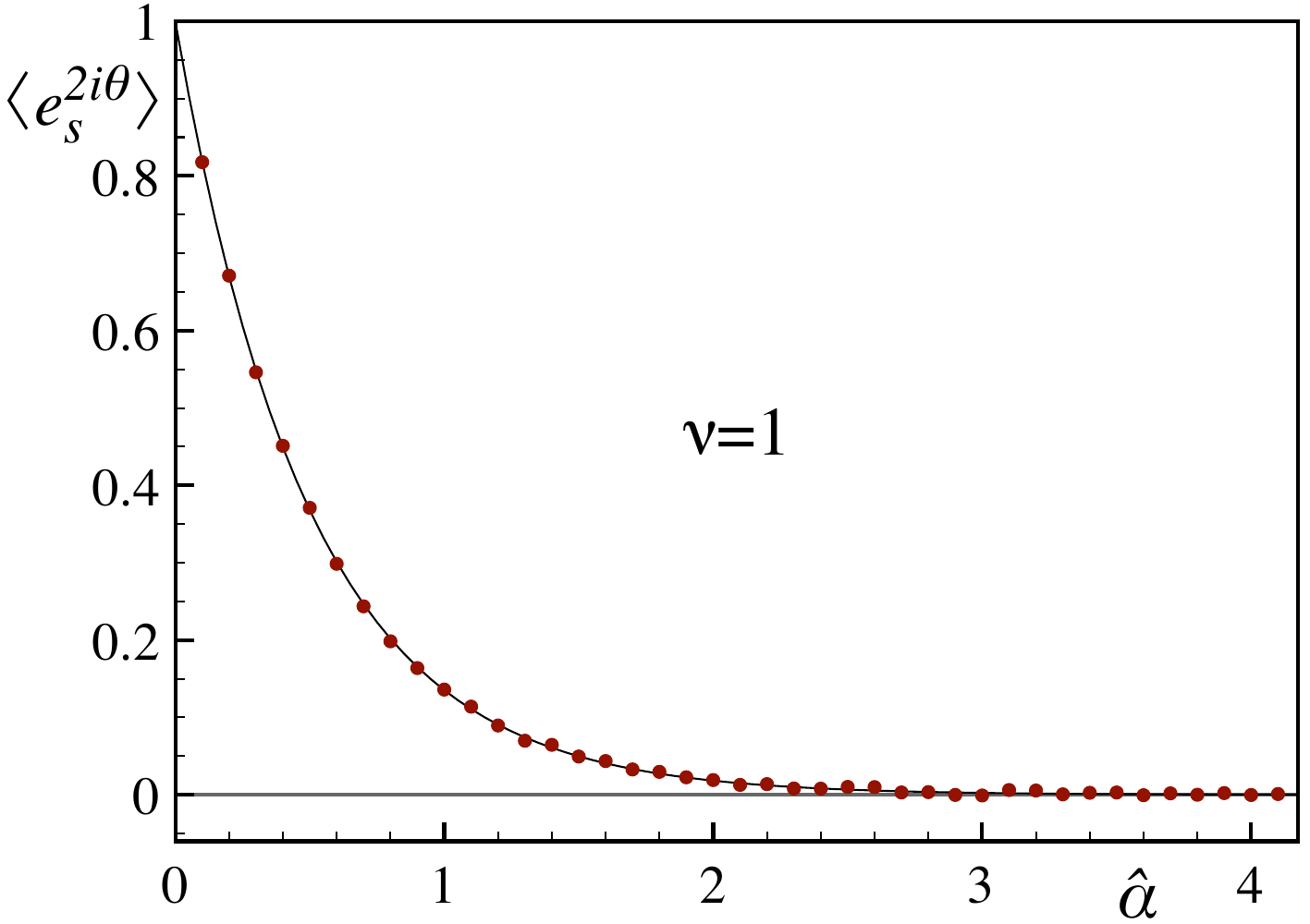}
  \includegraphics[width=0.32\textwidth]{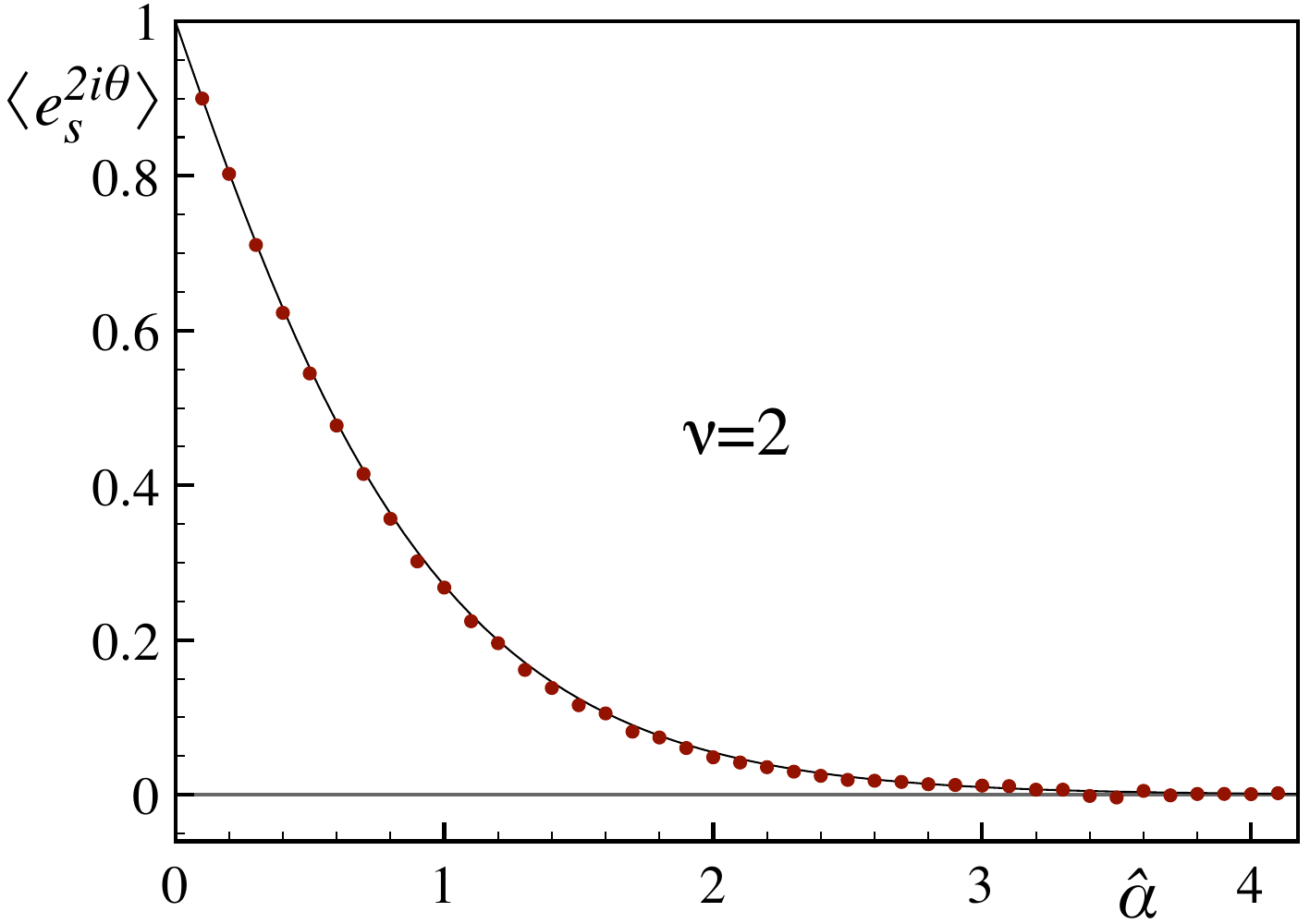}\\\hspace*{1mm}
  \includegraphics[width=0.32\textwidth]{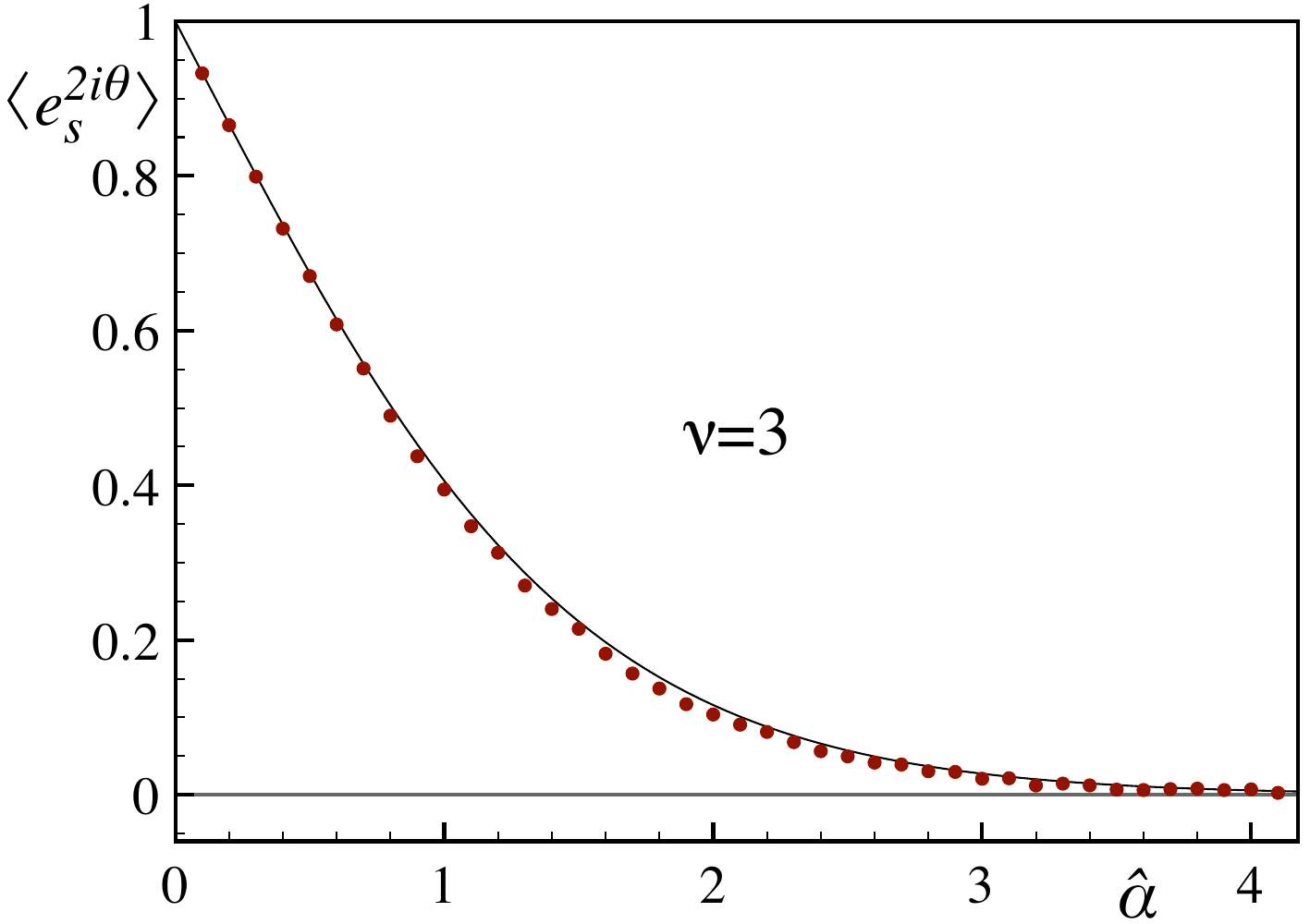}
  \includegraphics[width=0.32\textwidth]{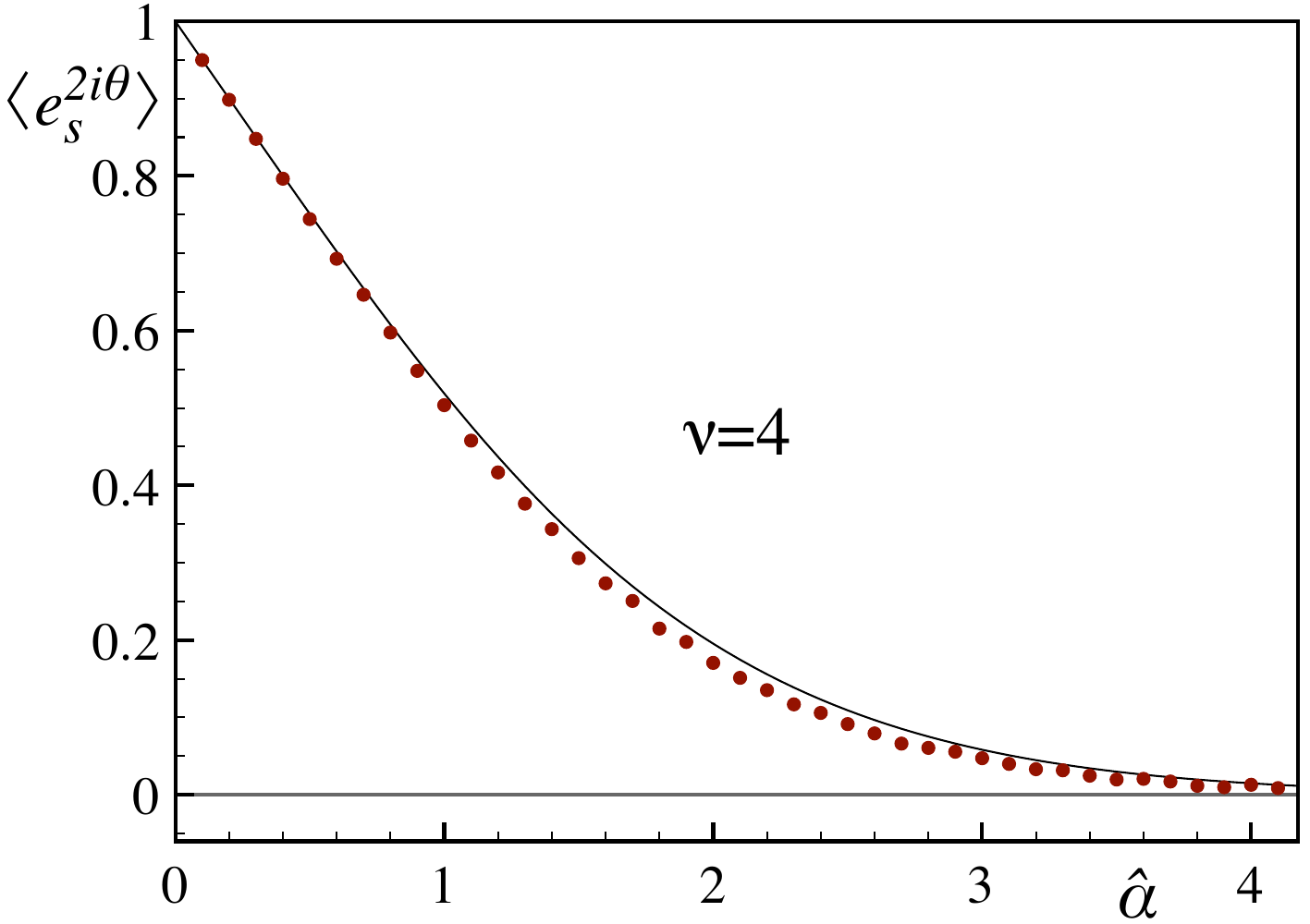}
  \includegraphics[width=0.32\textwidth]{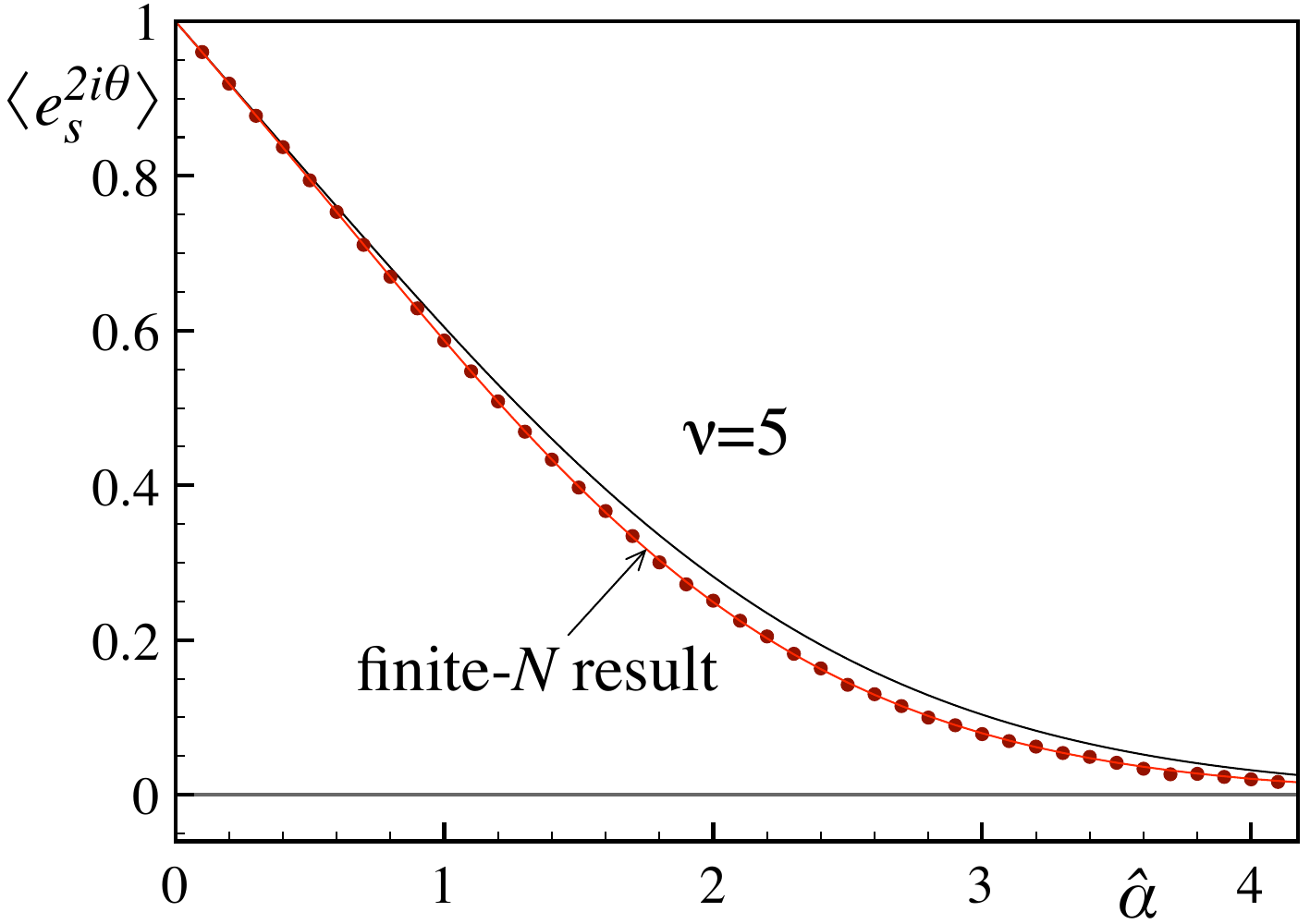}
  \caption{Average phase factor of the fermion determinant with
    $\hm=0$ in the quenched case for varying chemical potential
    parameter $\ha$ and $\nu=0,1,2,3,4,5$.  The full lines are the
    predictions of Eqs.~\protect\eqref{lim_m=0_nu=0} and
    \protect\eqref{eq:ch} for $N_f=0$.  For $\nu=5$ we also show the
    exact finite-$N$ result from Eq.~\eqref{Z1by1}.  The data points
    were computed from RMT simulations with matrix size $N=20$ and
    100,000 samples.  No error bars are shown since they are smaller
    than the data points.}
\label{Fig:phase_vs_alpha}
}

\FIGURE{
  \includegraphics[width=0.4\textwidth, trim=0 0 0 30]{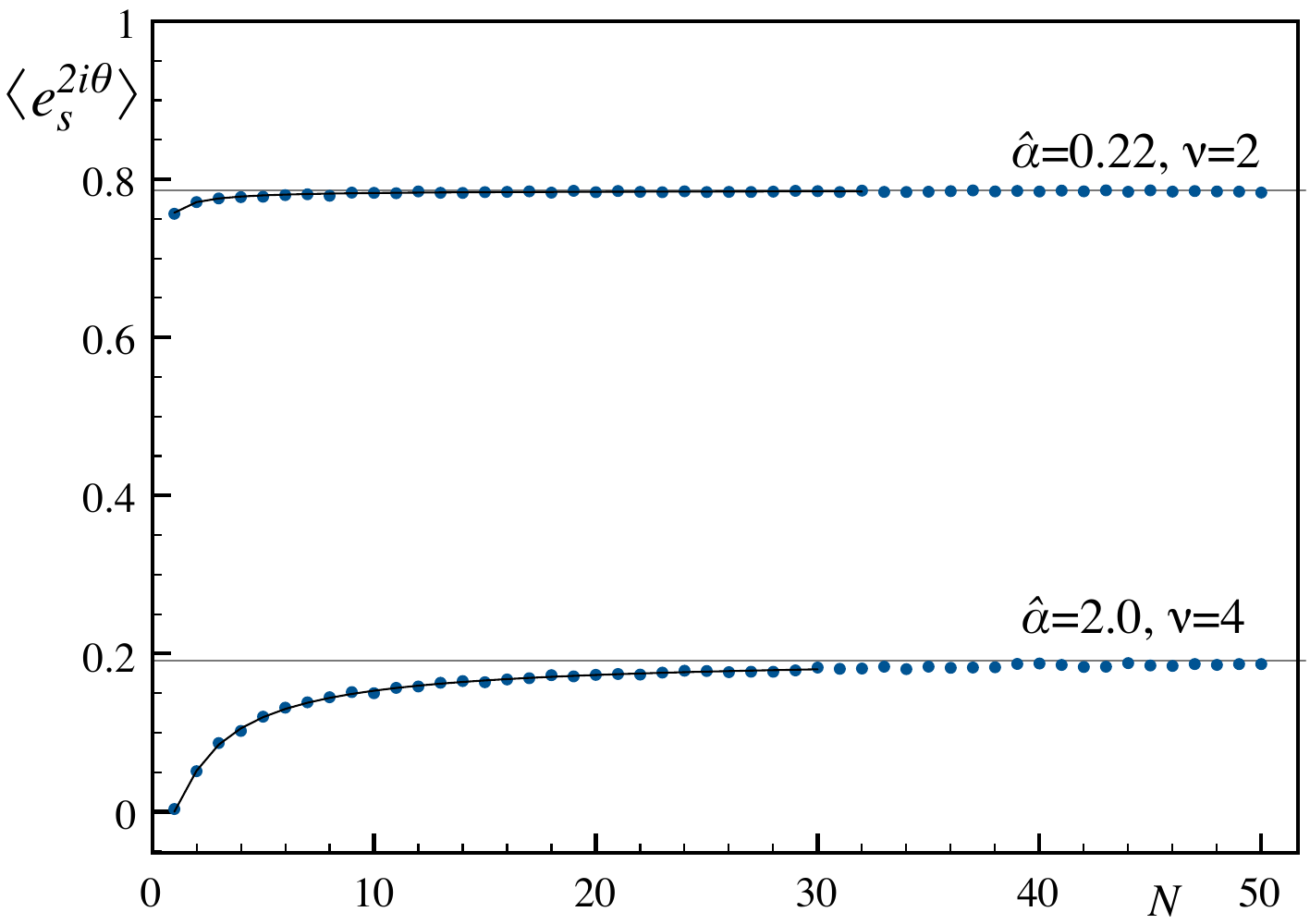}
  \caption{Average phase factor vs matrix size $N$. The horizontal
    lines show the analytical predictions in the microscopic limit
    ($N\to\infty$), while the solid lines going through the data
    points are the finite-$N$ results from Eq.~\eqref{Z1by1} for
    $N\le30$. The upper curve was computed with $\ha=0.22$, $\nu=2$,
    the lower with $\ha=2$, $\nu=4$, both for $N_f=0$ and
    $\hm=0$.\vspace{-8mm}} 
  \label{Fig:phase_vs_N}
}
\begin{sloppypar}
The convergence towards the microscopic limit is illustrated in
Fig.~\ref{Fig:phase_vs_N}, where we show the $N$-dependence of the
average phase factor for two typical cases: fast convergence for
$\ha=0.22$, $\nu=2$ versus slow convergence for $\ha=2$, $\nu=4$.  The
figure can help us determine how large the random matrices need to be
in order to reproduce the analytical results in the $N\to\infty$
limit.  In the figure we also show the $N$-dependence of the phase
factor from the theoretical framework, by numerically solving the
two-dimensional integral for the finite-$N$ expression \eqref{Z1by1},
which is expressed in terms of generalized Laguerre
polynomials.  We find perfect agreement with the data of
Fig.~\ref{Fig:phase_vs_N}, within statistical errors, for $N$ from $1$
to about $30$, at which point the integrals oscillate too strongly,
prohibiting Mathematica from performing the numerical integration
with sufficient accuracy.
\end{sloppypar}

From the study of the $N$-dependence we conclude that for $\nu=0,1,2$
it suffices to take $N=20$. However, for $\nu=3,4,5$ we increased the
matrix size to $N=80$ to be close enough to the microscopic limit. As
can be seen in Fig.~\ref{Fig:phase_vs_alpha_N=80}, this increase in
$N$ results in excellent agreement between simulations and analytical
predictions for larger $\nu$ as well.  Note that the $\hm=0$ case is a
key test for the new $\nu \neq 0$ contribution \eqref{Csingnu}, as
only this term contributes in the chiral limit.
\FIGURE[t]{
  \includegraphics[width=0.32\textwidth]{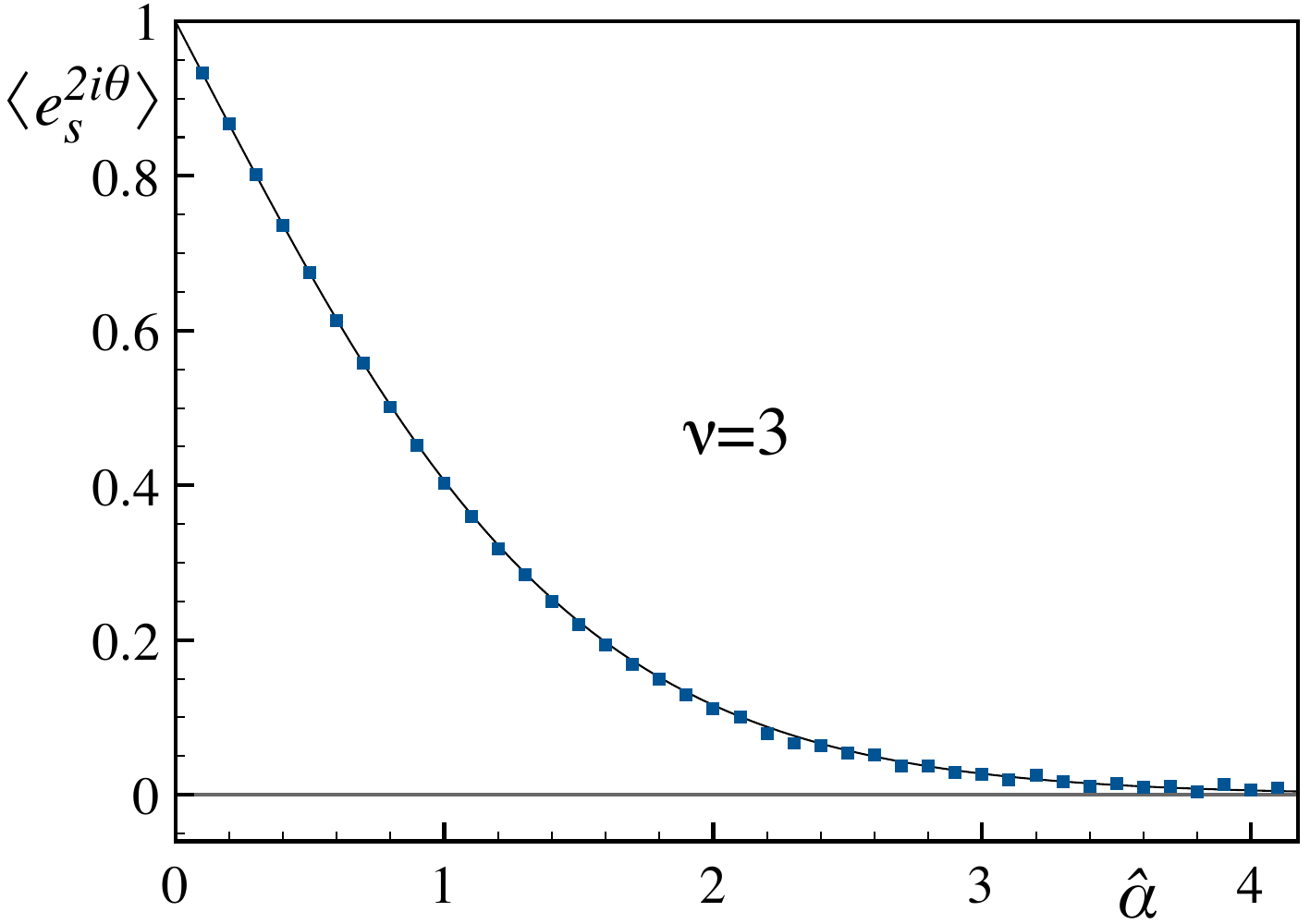}
  \includegraphics[width=0.32\textwidth]{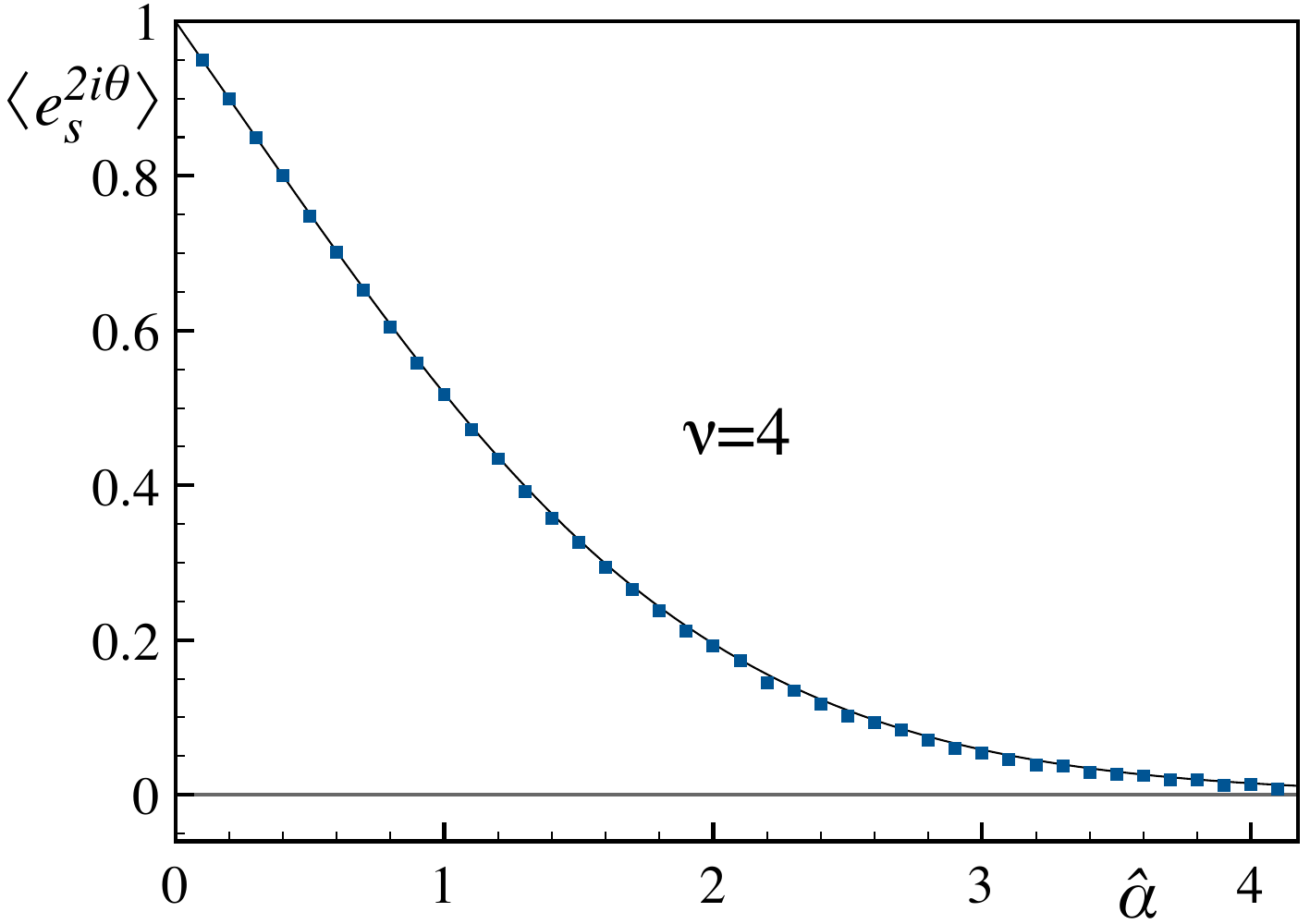}
  \includegraphics[width=0.32\textwidth]{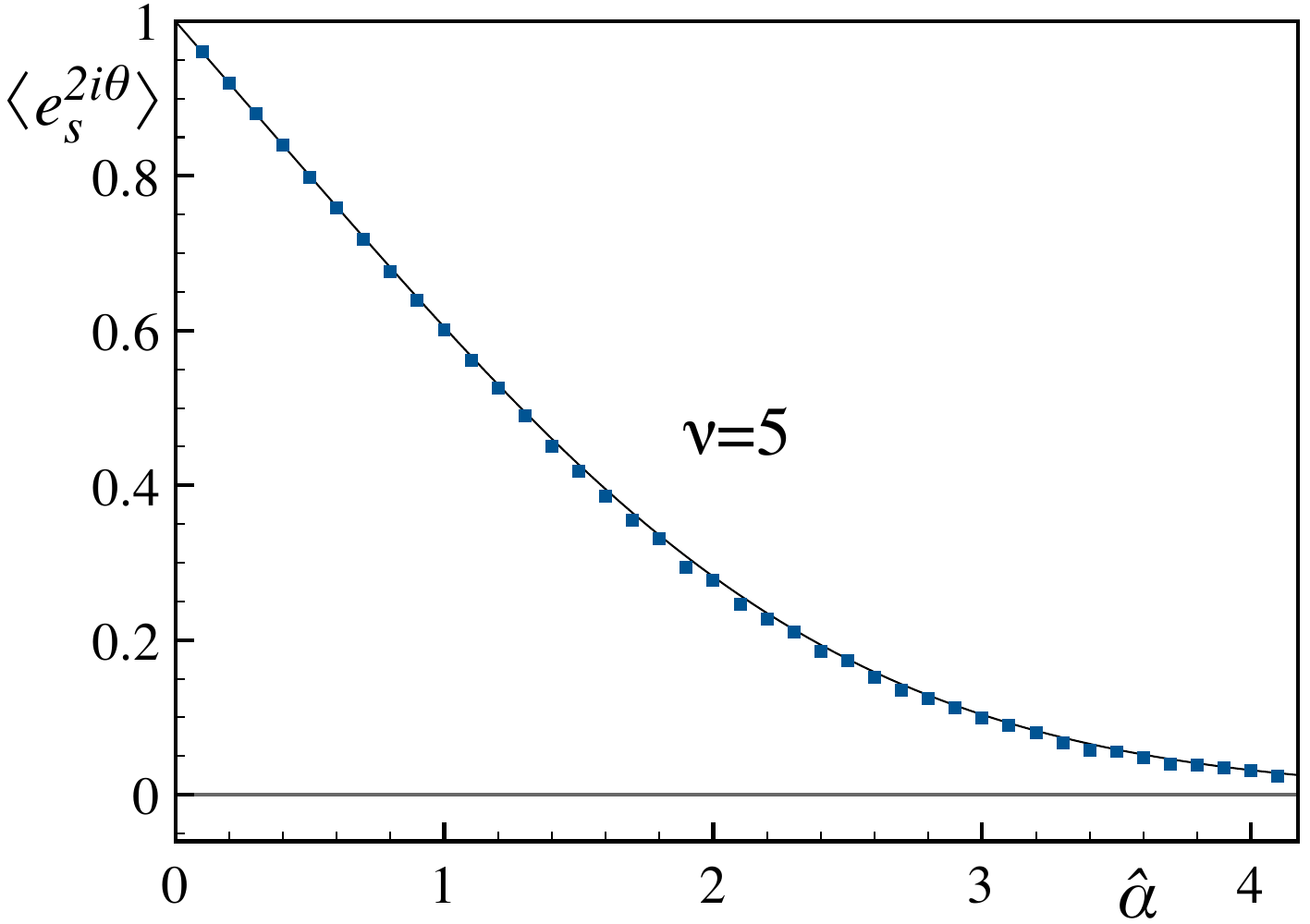}
  \caption{Average phase factor of the fermion determinant as a
    function of the chemical potential parameter $\ha$ for $\nu=3,4,5$
    with $N_f=0$ and $\hm=0$ as in
    Fig.~\protect\ref{Fig:phase_vs_alpha}, but with increased $N=80$.}
\label{Fig:phase_vs_alpha_N=80}
}

From Figs.~\ref{Fig:phase_vs_alpha} and \ref{Fig:phase_vs_alpha_N=80}
we conclude that, as expected, the average phase factor becomes unity
when the chemical potential vanishes, as the Dirac operator then
becomes anti-Hermitian and the determinant real. For large $\ha$ the
average phase factor goes to zero, pointing to the increasing sign
problem in dynamical simulations at large chemical potential. Observe
that for increasing topology the sign problem seems to be delayed, as
it sets in at a larger value of the chemical potential.  

\FIGURE[b]{
  \includegraphics[type=pdf,ext=.pdf,read=.pdf,width=0.32\textwidth]{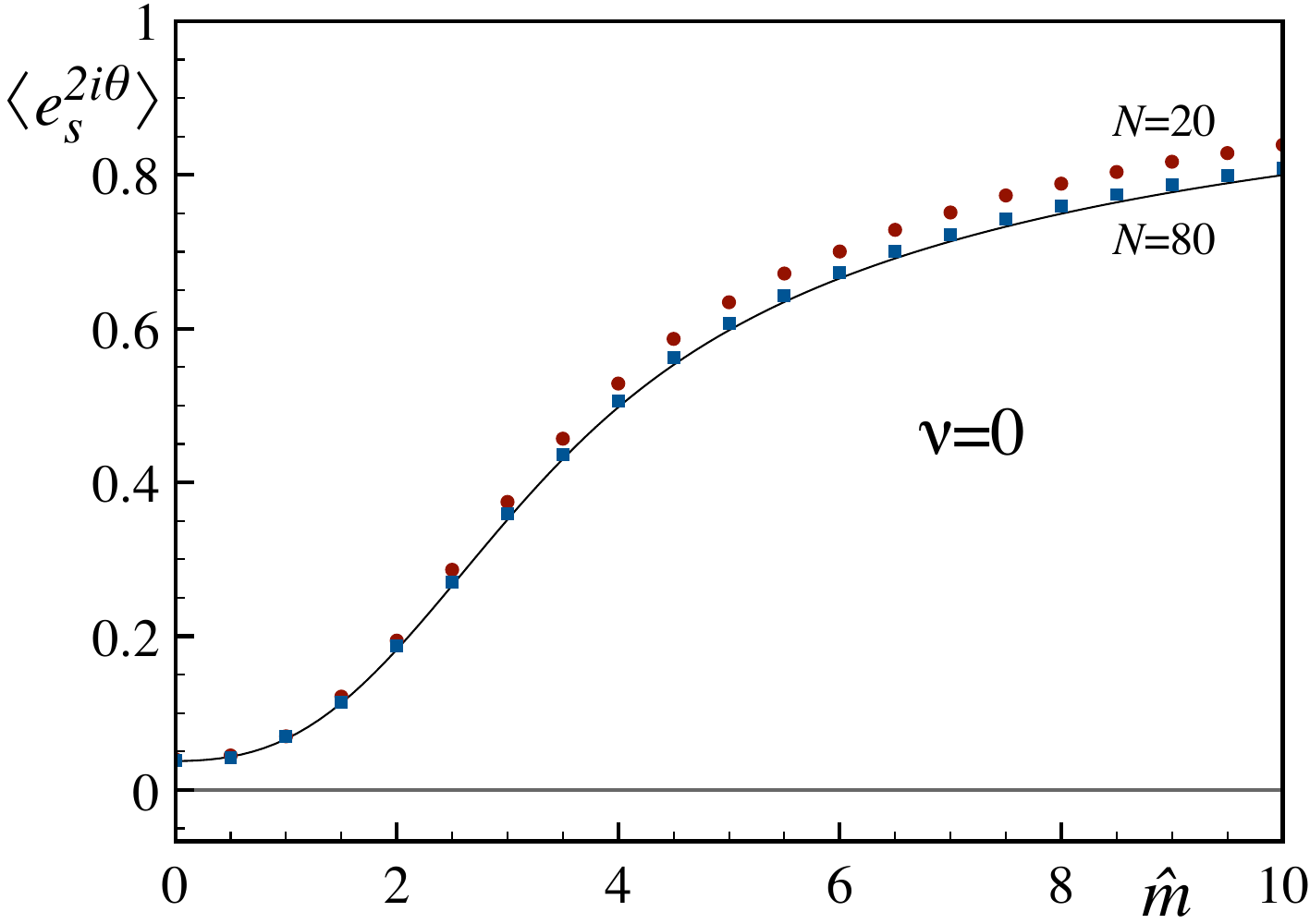}
  \includegraphics[type=pdf,ext=.pdf,read=.pdf,width=0.32\textwidth]{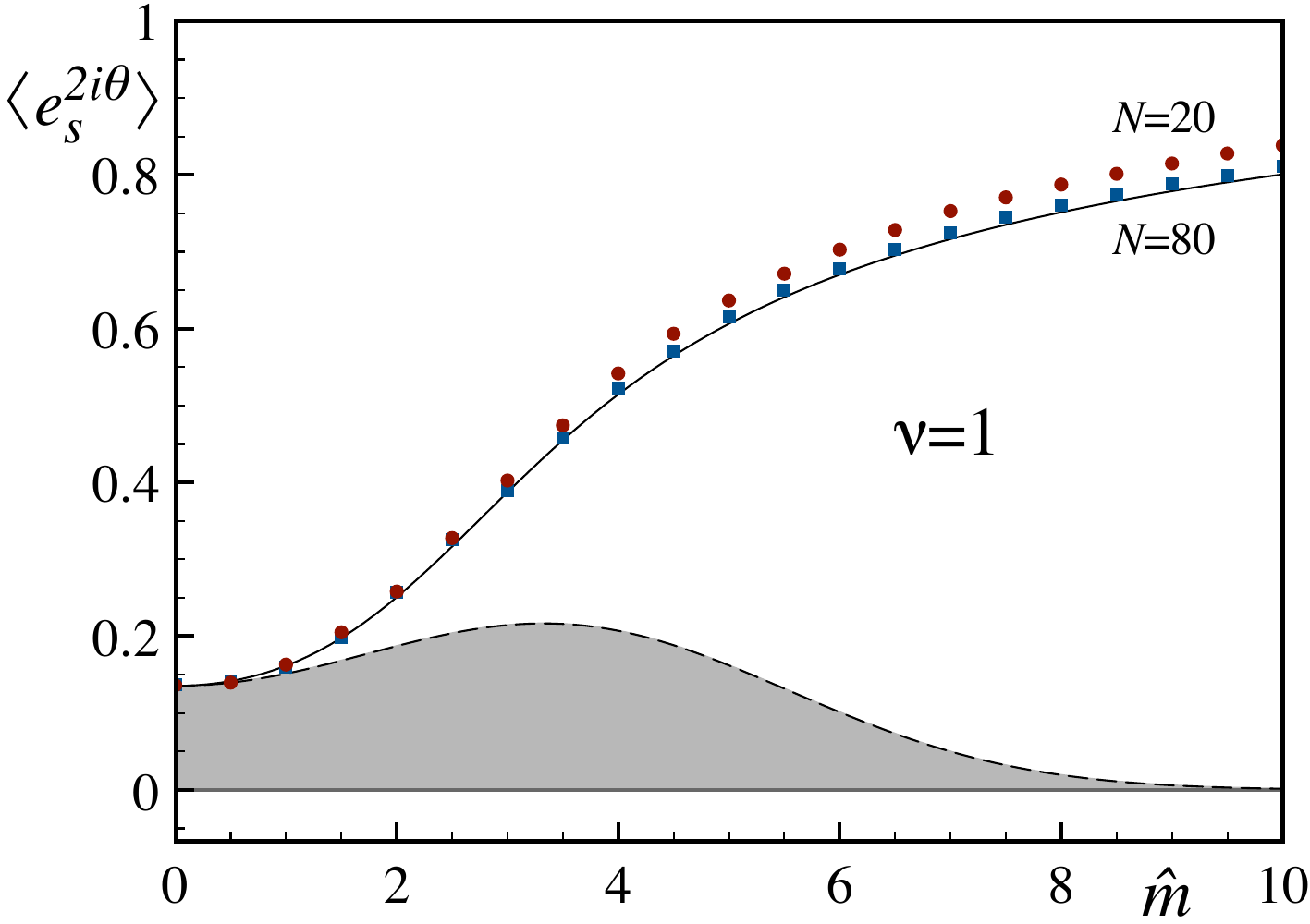}
  \includegraphics[type=pdf,ext=.pdf,read=.pdf,width=0.32\textwidth]{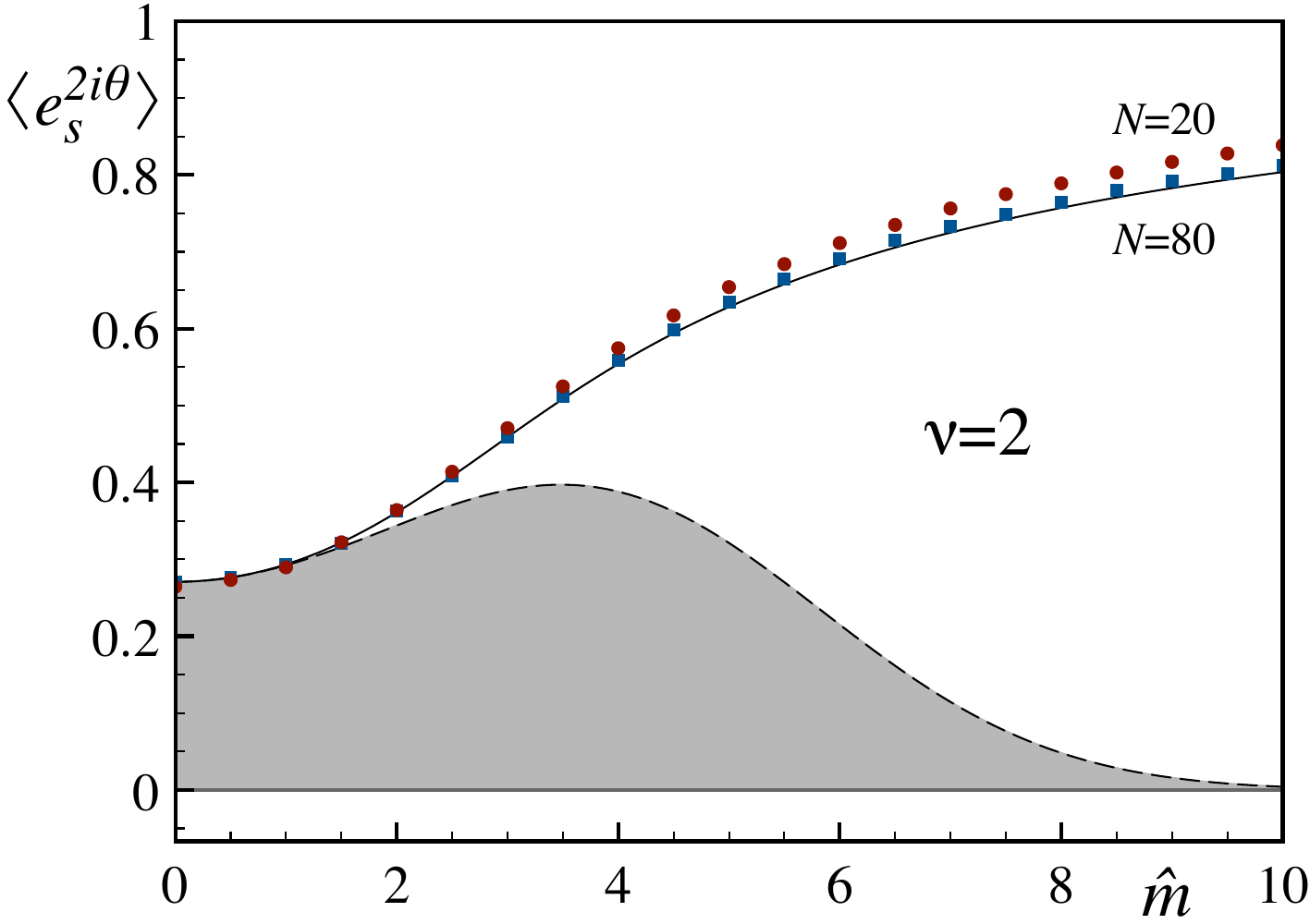}\\\hspace*{1mm}
  \includegraphics[type=pdf,ext=.pdf,read=.pdf,width=0.32\textwidth]{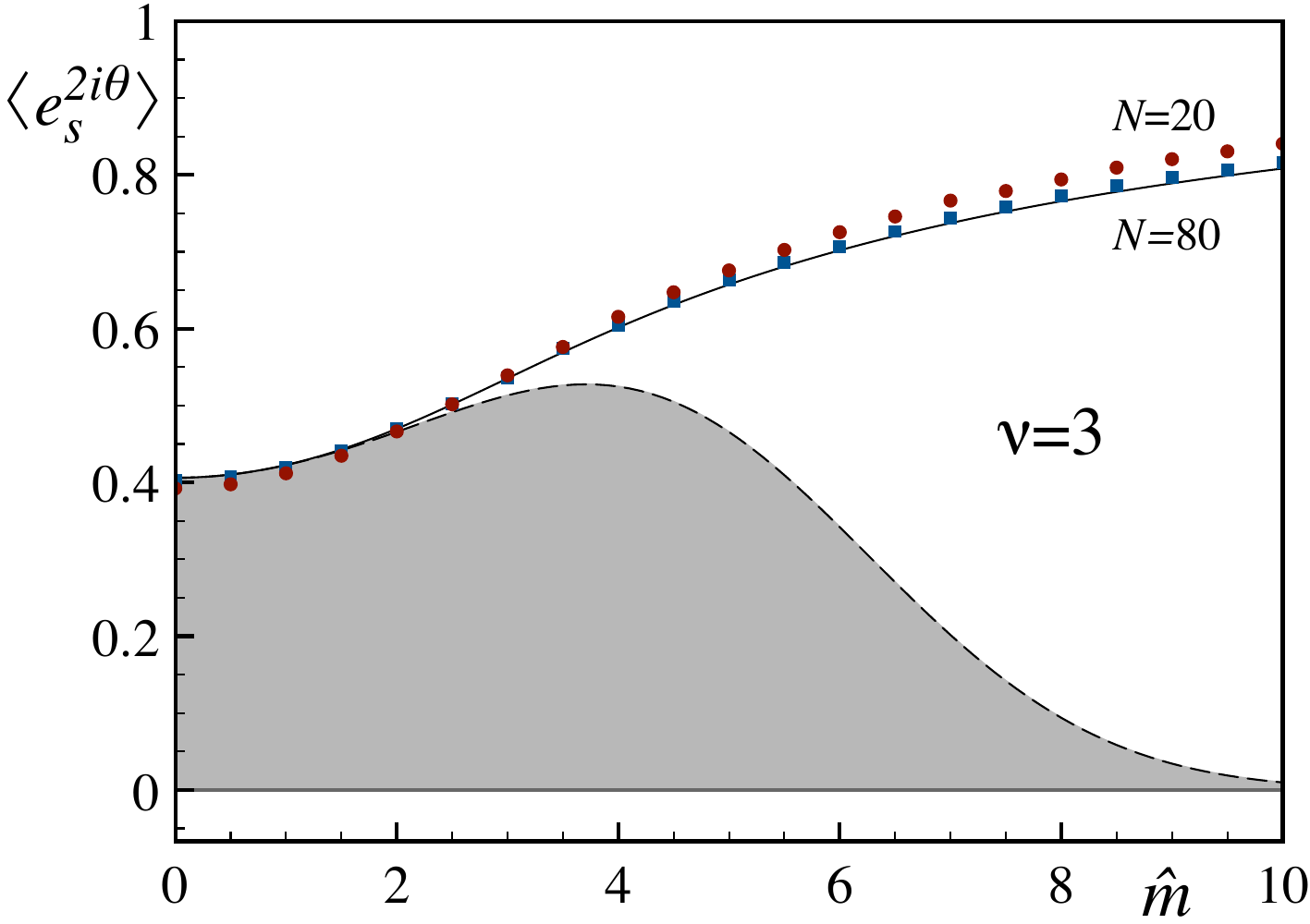}
  \includegraphics[type=pdf,ext=.pdf,read=.pdf,width=0.32\textwidth]{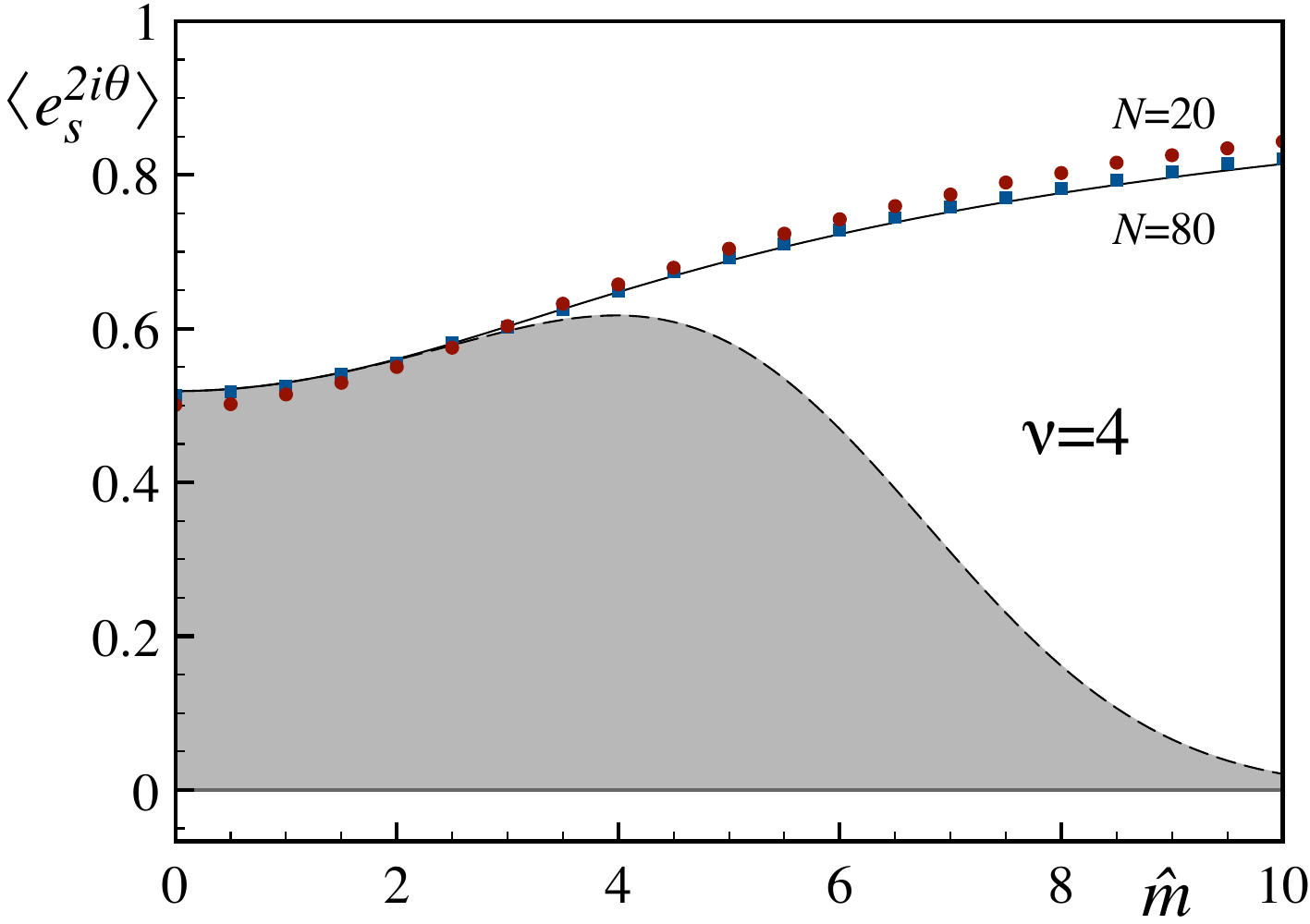}
  \includegraphics[type=pdf,ext=.pdf,read=.pdf,width=0.32\textwidth]{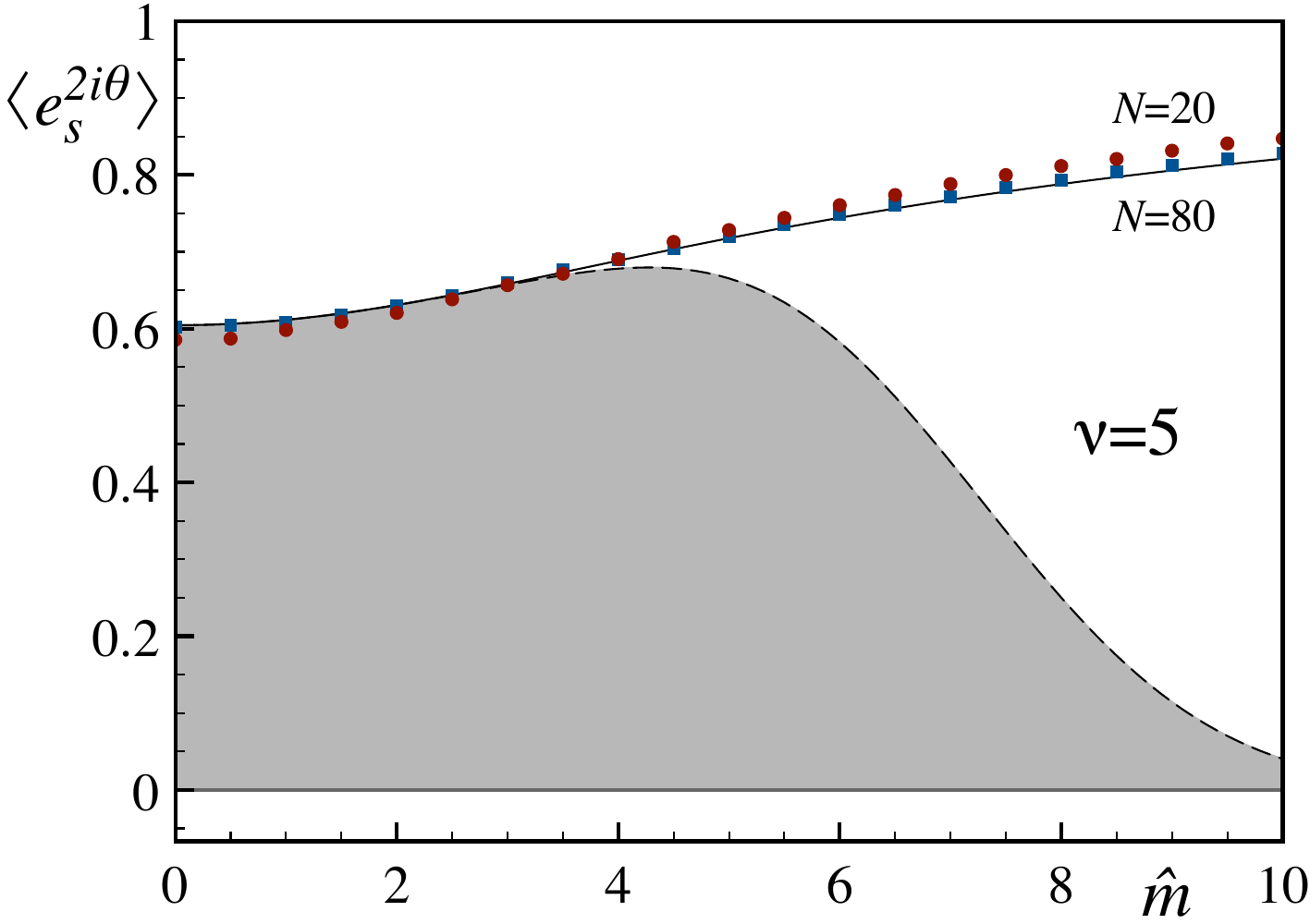}
  \caption{Average phase factor of the fermion determinant as a
    function of the fermion mass $\hm$ for $N_f=0$, $\ha=1.0$, and
    $\nu=0,1,2,3,4,5$. The simulations were performed with $N=20$
    (red points) and $N=80$ (blue squares). The latter data are
    already very close to the RMT-predictions for $N\to\infty$ (full
    lines).  The filled area shows the contribution
    of the $\Delta$-term specific to $\nu \neq 0$. (Note that we
    generated different random samples for each $\hm$ to avoid
    misleading correlations between measurements.)}
  \label{Fig:phase_vs_m}
}
In Fig.~\ref{Fig:phase_vs_m} we verify the mass-dependence of the
analytical formula \eqref{phfacquenched} and compare its predictions
with the results from random matrix simulations as a function of $\hm$
for fixed $\ha$.  Note that the convergence to the microscopic limit
slows down as the mass increases. This is noticeable in
Fig.~\ref{Fig:phase_vs_m}, where the $N=20$ data (red points) show a
systematic deviation from the RMT-predictions. When increasing the
size to $N=80$ (blue squares) the agreement improves
substantially.  The importance of the new 
$\Delta$-term for $\nu \neq 0$ is highlighted in
Fig.~\ref{Fig:phase_vs_m} by the gray area, which corresponds to the
contribution of the $\Delta$-term in Eq.~\eqref{phfacquenched}. This
clearly shows how this term dominates for small masses.  We also
observe in Fig.~\ref{Fig:phase_vs_m} that for fixed $\hm$ and $\ha$
the sign problem becomes less severe as the topological charge is
increased.  

Figures~\ref{Fig:phase_vs_alpha}--\ref{Fig:phase_vs_m} demonstrate
that the numerical simulations confirm the analytical prediction
\eqref{phfacquenched} for general topology in the quenched case.  We
are currently investigating the implementation of un\-quen\-ched
random matrix simulations at nonzero chemical potential using the
analytical information we have obtained about the sign problem.

\section{Conclusions}
\label{Sec:concl}

Dynamical lattice simulations of QCD at nonzero baryon density are
plagued by the sign problem caused by the oscillating fermion
determinant. To investigate this problem it is helpful to employ the
equivalence between chiral perturbation theory in the
$\epsilon$-regime of QCD and chiral random matrix theory, which also
holds at nonzero chemical potential.  As the average phase factor of
the fermion determinant is an important clue in the study of the sign
problem, it is a valuable quantity to compute in the framework of
chiral random matrix theory.

In this paper we derived an analytical formula for the average phase
factor of the fermion determinant in quenched and unquenched chiral
random matrix theory for general topology. The formula is a nontrivial
extension of the result previously published by Splittorff and
Verbaarschot for zero topology \cite{Splittorff:2007ck}. For nonzero
topology a new contribution shows up, which dominates the phase factor
for small valence quark mass. The new formula suggests that the
severity of the sign problem is reduced as the topological charge
increases.  We also computed the chiral and thermodynamic limits from
our general formula.

The quenched formula was verified by random matrix simulations in
different regimes of mass and chemical potential and for different
values of the topological charge.  Excellent agreement was found
between theory and simulations.  We are currently in the process of
comparing the RMT predictions derived in this paper to lattice QCD
data computed with the overlap operator at nonzero chemical potential.

\acknowledgments{ This work was supported in part by DFG grant
  FOR465-WE2332/4-2. JB would like to thank G.~Akemann for his
  hospitality and helpful suggestions on a visit to Brunel University.
  TW would like to thank the Theory Group of the INPS at KEK Tsukuba,
  where part of this work was carried out, for their hospitality and
  support.  We would also like to thank T.~Kaltenbrunner and C.~Lehner
  for motivating discussions.  }

\appendix

\section{Microscopic limit of the orthogonal polynomials}
\label{App:micpnu}

In this section we define and compute the microscopic limits of the
orthogonal polynomials \eqref{pknu}, the normalization factor
\eqref{rknu}, and the weight function \eqref{wnu}.

The microscopic limit of the orthogonal polynomials \eqref{pknu} is
defined as
\begin{align}
  \hp^\nu(\hz;\ha) &\equiv \lim_{N\to\infty}
  \frac{e^N}{(2N)^{\nu+1/2}} p_{N-1}^\nu(\hz/2N;\ha/2N)
  = \lim_{N\to\infty} \frac{e^N}{(2N)^{\nu+1/2}}
  \left(1-\frac{\ha}{2N} \right)^{N-1}
  \frac{(N-1)!}{N^{N-1}} \, L_{N-1}^\nu\left(-\frac{\hz^2}{4N}\right)\:.
  \label{App:limpN}
\end{align}
Using the definition of the exponential function,
\begin{equation}
  \lim_{N\to\infty} \left(1-\frac{\ha}{2N} \right)^{N-1} =
  e^{-\ha/2}\:,
  \label{limexp2}
\end{equation}
and Stirling's formula, from which we obtain
\begin{equation}
  \frac{(N-1)!}{N^{N-1}} = \frac{N!}{N^N}
  \stackrel{N\to\infty}{\longrightarrow} \sqrt{2\pi N} e^{-N}\:,
\label{limNNN}
\end{equation}
we find
\begin{equation}
  \hp^\nu(\hz;\ha) = 
  \lim_{N\to\infty} \sqrt{\pi} e^{-\ha/2} (2N)^{-\nu}
  L_{N-1}^\nu\left(-\frac{\hz^2}{4N}\right) \:. 
  \label{limp}
\end{equation}
The $N \to \infty$ limit of the Laguerre polynomial in
Eq.~\eqref{limp} is given by \cite[Eq.~(8.982.2)]{Grad:1980}
\begin{align}
  \lim_{N\to\infty} N^{-\nu} L_{N}^\nu\left(-\frac{z^2}{4N}\right) 
  = 2^\nu z^{-\nu} I_\nu(z) \:,
\end{align}
and hence Eq.~\eqref{limp} becomes
\begin{align}
  \hp^\nu(\hz;\ha) 
  &= \sqrt{\pi} e^{-\ha/2}
  \hz^{-\nu} I_\nu\left(\hz\right) \:. 
  \label{App:limp2}
\end{align}
Next, the microscopic limit of the normalization factor \eqref{rknu}
is defined as
\begin{align}
  \hr^\nu(\ha) &\equiv \lim_{N\to\infty} (2N)^2e^{2N}r_{N-1}^\nu(\ha/2N) =
  \lim_{N\to\infty} \frac{(2N)^2e^{2N}}{N^{2N+\nu}} \pi \frac{\ha}{2N}
  \left(1+\frac{\ha}{2N}\right)^{2N-2+\nu } (N-1)! (N-1+\nu )! \notag\\
  &= 4\pi^2 \ha e^{\ha} \:,
\label{App:micrknu}
\end{align}
where we again used Eqs.~\eqref{limexp2} and \eqref{limNNN}.  Finally,
the microscopic limit of the weight function \eqref{wnu} is defined as
\begin{align}
  \hw^\nu(\hz, \hz^*;\ha) 
  &=\lim_{N\to\infty} (2N)^{2\nu+2} w^\nu(\hz/2N, \hz^*/2N;\ha/2N)
  \notag\\ 
  &= \lim_{N\to\infty} |\hz|^{2(\nu+1)} 
  \exp\left(-\frac{N(1-\ha/2N)}{4\ha/2N}\frac{\hat
      z^2+\hz^{*2}}{4N^2}\right)  
  K_\nu\left(\frac{N(1+\ha/2N)}{2\ha/2N} \frac{|\hat
      z|^2}{4N^2}\right) \notag\\
  &= |\hz|^{2(\nu+1)} 
  \exp\left(-\frac{\hz^2+\hz^{*2}}{8\ha}\right) 
  K_\nu\left(\frac{|\hz|^2}{4\ha}\right) \:.
\label{App:micw}
\end{align}

\section{Principal value integral}
\label{Sec:PV}

In this section we show that the principal value integral originating
from the application of the Sokhatsky-Weierstrass theorem
\eqref{SW-theorem} to Eq.~\eqref{Ztu} vanishes because of symmetry
considerations.

For $\nu>0$, the integrand \eqref{ftu} of the principal value integral
over $t$ is singular along the line $u=0$ (for which $t<0$ in the
integration region). Therefore we split the $u$-integral into a
principal value part and a line integral circumventing the singularity
of the Bessel functions at $u=0$.  (To simplify the notation we do
this also for $\nu=0$ even though there is no singularity in this
case.) For any $t<0$, the $u$-integration in Eq.~\eqref{Ztu} is thus
rewritten as
\begin{align}
  \int_t^\infty du f_\pm(t,u)= \lim_{\delta\to 0^+}\left\{
    \int_t^{-\delta} + \int_{C_\pm} +
    \int_\delta^\infty \right\} du \, f_\pm(t,u)\:,  
\end{align}
where $C_\pm$ denotes the small semicircles shown in
Fig.~\ref{Fig:integline}.  The principal value integral over $t$ in
Eq.~\eqref{Ztu} therefore becomes
\begin{align}
  \label{PVint}
  \PV_t \int_{-\infty}^\infty dt \lim_{\epsilon\to 0^+} \int_t^\infty
  du \, [f_+(t,u)+f_-(t,u)]  
  &= \PV_{tu} \int_{-\infty}^\infty dt \int_t^\infty du \, F(t,u) \\
  \notag
  &\quad + \PV_t \int_{-\infty}^{0} dt  
  \lim_{\delta\to 0^+}  \Bigg\{ \int_{C_+} du \, f_+(t,u)
  + \int_{C_-} du \, f_-(t,u)
  \Bigg\} \:,
\end{align}
where $F(t,u) = \lim_{\epsilon\to 0^+} [f_+(t,u) + f_-(t,u)]$.  The
total principal value integral $\PV_{tu}$ can be split, for any $t$,
as
\begin{align}
  \PV_{tu} \int_{-\infty}^\infty dt \int_t^{\infty} du \, F(t,u)
  = \underbrace{\PV_{tu} \int_{-\infty}^\infty dt \int_0^{\infty} du
    \, F(t,u)}_{=A} 
  + \underbrace{\PV_{tu} \int_{-\infty}^\infty dt \int_t^0 du \,
    F(t,u)}_{=B} \:. 
  \label{provezero}
\end{align}
From Eq.~\eqref{ftu} we see that the dependence of the integrand on
$t$ and $u$ is given by
\begin{align}
  f(t,u) \propto \frac{t^{\nu+1}u^{k+1}}{t^2-\hm^2}
  e^{-\textstyle\frac{t^2+u^2}{8\ha}}
  K_\nu\left(\frac{tu}{4\ha}\right)
  K_{\nu+k}(u) \:,
\label{kernel}
\end{align}
and according to Eq.~\eqref{KetaK} the sum of the contributions on the
upper and lower sheets, $S_+$ and $S_-$, is proportional to (again
with the factor $1/4\ha$ omitted)
\begin{align}
  \lim_{\epsilon\to 0^+} \big[K_\nu(t u) K_{\nu+k}(u)\big]_+ +
  \big[K_\nu(t u) K_{\nu+k}(u)\big]_- 
  =
  \begin{cases}
    2 K_\nu(|tu|) K_{\nu+k}(|u|) &\text{for }  t,u>0 \:, \\
    2 (-)^\nu K_\nu(|tu|) K_{\nu+k}(|u|)  &\text{for }  t<0<u \:, \\
    2 (-)^{\nu+k} K_\nu(|tu|) K_{\nu+k}(|u|)  &\text{for } t, u<0 \:.
  \end{cases}
  \label{diffkern}
\end{align}
From Eqs.~\eqref{kernel} and \eqref{diffkern} we see that the
integrand of $A$ in Eq.~\eqref{provezero} is odd in $t$, and hence the
principal value of that integral vanishes. Integral $B$ of
Eq.~\eqref{provezero} can be rewritten as
\begin{align}
  \PV_{tu} \int_{-\infty}^\infty dt \int_t^0 du \, F(t,u) 
  &= \PV_{tu} \Big\{
  \int_{-\infty}^0 dt \int_t^0 du \, F(t,u) - \int_{0}^\infty dt
  \int_0^t du \, F(t,u) \Big\} \notag\\ 
  &= \PV_{tu} \Big\{\int_0^{\infty} dt \int_0^t du \, F(-t,-u) -
  \int_0^{\infty} dt \int_0^t du \, F(t,u) \Big\} \:.
  \label{eq:intB}
\end{align}
From Eqs.~\eqref{kernel} and \eqref{diffkern} we find for $u,t>0$
\begin{align}
F(-t,-u) = (-)^{\nu+1}(-)^{k+1}(-)^{\nu+k} F(t,u) = F(t,u)\:,
\label{Fsym}
\end{align}
and hence \eqref{eq:intB} and the total principal value integral
\eqref{provezero} vanish because of symmetry considerations.

Next we treat the last term in Eq.~\eqref{PVint}. For $\nu=0$ the
integrand \eqref{kernel} is regular at $u=0$, and thus the integral
trivially vanishes as $\delta \to 0$. For $\nu>0$ one can expand the
integrand in pole terms behaving like
\begin{align}
  u^{-2(\nu-j)+1}\:, \qquad j=0,\ldots, \nu-1
\end{align}
for $u\to 0$ by using the small-argument expansion \eqref{Kseries} of
the $K$-Bessel functions.  The $u$-integration around the pole flows
in opposite directions for the integration on the lower and upper
sheet.  The simple pole gives contributions with opposite signs on the
two sheets so that their sum is zero. The higher-order poles of the
integrand are also odd in $u$, and one can show that the integrals
along $C_+$ and $C_-$ vanish individually, so that
\begin{align}
  \PV_t \int_{-\infty}^{0} dt  
  \lim_{\delta\to 0^+}  \Bigg\{ \int_{C_+}du \, f_+(t,u)
  + \int_{C_-} du \, f_-(t,u) \Bigg\} = 0\:. 
\label{PVsing}
\end{align}
This completes the proof that the integral \eqref{PVint} vanishes.

\section{Chiral limit}
\label{app:chiral}

In the limit $\hm \to 0$ the phase factor can be simplified and
written in terms of special functions.  Our starting point is
Eq.~\eqref{phapp}, and we need to compute the chiral limit of the
Wronskian \eqref{G.4}, which contains the function $I_{\nu,k}$ defined
in Eq.~\eqref{eq:Inuk}.  For small argument $\hm$ the leading order
term of the $I$-Bessel function is \cite[Eq.~(9.6.10)]{Abram:1964}
\begin{align}
  I_\nu(\hm) \sim \frac{\hm^\nu}{2^\nu\nu!} \:,
  \label{Iorigin2}
\end{align}
from which we obtain
\begin{align}
  I_{\nu,k}(\hm) \sim \frac{\hm^{\nu+2k}}{2^{\nu+k}(\nu+k)!}
  \label{InukLO}
\end{align}
and its $p$-th derivative
\begin{align}
  I_{\nu,k}^{(p)}(\hm) \sim
  \frac{(\nu+2k)!}{2^{\nu+k}(\nu+k)!(\nu+2k-p)!}\hm^{\nu+2k-p} \:. 
  \label{InukpLO}
\end{align}
Substituting these expressions in the Wronskian \eqref{G.4} and using
properties of the determinant gives the leading-order result
\begin{align}
W_n(k_1,\ldots,k_n) \sim \frac{\hm^{n\nu+2\sum_i k_i-n(n-1)/2}} 
{2^{n\nu+\sum_i k_i-n(n-1)/2}\prod_{i} (\nu+k_i) !} \Delta_n(k_1,
\ldots, k_n) \:, 
\label{Wchiral}
\end{align}
where $\Delta_n(k_1, \ldots, k_n)$ is a Vandermonde determinant.  From
this we compute the chiral limit of the denominator in
Eq.~\eqref{phapp},
\begin{align}
W_{N_f}(0, 1,\ldots,\Nf-1) \sim \frac{\hm^{\Nf\nu+\Nf(\Nf-1)/2}} 
{2^{\Nf\nu}\prod_{i=0}^{\Nf-1} (\nu+i) !} \prod_{\ell=1}^{\Nf-1}
\ell ! \:, 
\label{denomchiral}
\end{align}
where we used the identities
\begin{align}
  \label{eq:vand}
  \Delta_\Nf(0,1,\ldots,\Nf-1) = \prod_{\ell=1}^{\Nf-1} \ell!
  \quad\text{and}\quad
  \sum_{i=0}^{\Nf-1} i = \frac12\Nf(\Nf-1) \:.
\end{align}
From Eq.~\eqref{Wchiral} it is easy to see that in the limit $\hm \to
0$ only the Wronskian corresponding to the $k=\Nf+1$ term in
Eq.~\eqref{numerator} will contribute to leading order, while all
other terms will be of higher order. The phase factor \eqref{phapp}
can therefore be written as
\begin{align}
  \phs_{\hm=0}
  = \lim_{\hm\to 0} \frac{ (-)^{\Nf+1} }{(2\hm)^\Nf \Nf!} 
  \frac{ W_{\Nf+1}(0, 1, \ldots, \Nf)}{W_\Nf(0, 1, \ldots, \Nf-1)}
   \I{\Nf+1} \:.
\end{align}
After substituting \eqref{denomchiral} we find
\begin{align}
  \phs_{\hm=0}
  =  \frac{ (-)^{\Nf+1} }{2^{\nu+\Nf}(\nu+\Nf)!} 
  \lim_{\hm\to 0} \hm^{\nu}  \I{\Nf+1} \:.
\label{phfacchiral}
\end{align}
The chiral limit of $\I{k}$ has to be derived differently for $\nu=0$
and for $\nu \neq 0$.  For $\nu=0$ and $\hm \to 0$ only the first
integral in Eq.~\eqref{Z1total} contributes as the $\Delta$-term is
absent and the second integral vanishes because the integration range
is empty. Hence,
\begin{align}
{\cal H}_{0,k}(\ha,0) = \frac{e^{-2\ha}}{4\ha}
\int_0^\infty du \, (-)^k u^{k+1} e^{-\frac{u^2}{8\ha}} K_k(u) \:.
\label{H0kchiral}
\end{align}
The integral in Eq.~\eqref{H0kchiral} can be evaluated analytically in terms of an incomplete gamma function.
Assuming that $n$ is a nonnegative integer and $\re z>0$, we find from
Ref.~\cite[Eq.~(6.631.3)]{Grad:1980}
\begin{align}
  {\cal I}_n(z) &\equiv \int_0^\infty du \, u^{n+1} 
  e^{-\textstyle\frac{u^2}{4z}} K_n(u)
  = 2^{n} n! \, z^{\frac{n+1}{2}} e^{z/2}
  W_{-\frac{n+1}{2},\frac{n}{2}}(z) \:, 
  \label{II2}
\end{align}
where $W_{\lambda,\mu}$ is a Whittaker function with integral
representation \cite[Erratum of Eq.~(9.222.1)]{Grad:1980}
\begin{align}
  W_{\lambda,\mu}(z) =
  \frac{z^{\mu+\frac{1}{2}}e^{-z/2}}{\Gamma(\mu-\lambda+\frac{1}{2})} 
  \int_0^\infty dt \, e^{-zt}
  t^{\mu-\lambda-\frac{1}{2}}(1+t)^{\mu+\lambda-\frac{1}{2}} \:. 
  \label{Whittaker}
\end{align}
Setting $\lambda=-(n+1)/2$ and $\mu=n/2$ in this equation and
substituting in Eq.~\eqref{II2} gives
\begin{align}
  {\cal I}_n(z) = 2^{n} z^{n+1} \int_0^\infty dt \, e^{-zt}
  \frac{t^n}{1+t} 
  = 2^{n} z^{n+1} n! e^z \Gamma(-n,z) \:, 
  \label{I2}
\end{align}
where the last equality follows from \cite[Eq.~(3.383.10)]{Grad:1980}
and the incomplete gamma function is given by
\begin{align}
  \label{eq:inc}
  \Gamma(a,z)=\int_z^\infty dt \: t^{a-1}e^{-t}\:.
\end{align}
After substituting the integral \eqref{I2} in Eq.~\eqref{H0kchiral} we find
\begin{align}
{\cal H}_{0,k}(\ha,0) = 
(-)^k 2^{k-1} k! (2\ha)^k \Gamma(-k,2\ha) \:.
\end{align}
Substituting the latter in Eq.~\eqref{phfacchiral} with $\nu=0$ yields
\begin{align}
  \phs_{\nu=0,\hm=0}
  = (\Nf+1) (2\ha)^{N_f+1} \Gamma(-N_f-1,2\ha)  \:.
\label{phfacchiralnu=0}
\end{align}
For $\nu \neq 0$ we first analyze the two integrals in
Eq.~\eqref{Z1total} for $\hm \to 0$.  For small argument $\hm$ and
$\nu>0$ the $I$-Bessel function goes to zero as given by
Eq.~\eqref{Iorigin2}, and hence the first integral in
Eq.~\eqref{Z1total} vanishes because the integrand goes to zero like
$\hm^{\nu}$.  The second integral in Eq.~\eqref{Z1total} trivially
vanishes because the integration range is empty.  Therefore, the limit
of the average phase factor for $\hm\to 0$ is completely determined by
the new $\Delta$-term \eqref{Csingnuk}.  It is interesting to note
that the contributions to the phase of the determinant in the chiral
limit originate from different terms for $\nu=0$ and $\nu \neq 0$.
Equation~\eqref{phfacchiral} becomes
\begin{align}
  \phs_{\nu>0,\hm=0}
  &=  \frac{ e^{-2\ha} }{(\nu+\Nf)!} \:\lim_{\hm\to0}
  \sum_{i,j=0}^{i+j\le\nu-1}
  \frac{(\nu-1-i)!(\nu+\Nf-j)!}{(\nu-1-i-j)! i! j!}
  \left(\frac{\hm^{2}}{8\ha}\right)^i \left(2\ha\right)^j \:.
\end{align}
In this double sum only the terms with $i=0$ will contribute when
$\hm\to 0$ so that
\begin{align}
  \phs_{\nu>0,\hm=0}
  &= e^{-2\ha}
  \sum_{j=0}^{\nu-1}
  \frac{(\nu-j)_{\Nf+1}}{(\nu)_{\Nf+1}}
  \frac{\left(2\ha\right)^j}{j!} \:,
  \label{H12}
\end{align}
where we introduced the Pochhammer symbol 
\begin{align}
  \label{eq:poch}
  (a)_n\equiv a(a+1)\cdots(a+n-1)\quad\text{with}\quad (a)_0=1
\end{align}
to simplify the notation.  For $\nu=1$ we notice the intriguing fact
that the chiral limit of the phase factor is independent of the number
of flavors, i.e., $\bigl\langle
e^{2i\theta}_s\bigr\rangle_{\nu=1,\hm=0} = e^{-2\ha}$.

Eq.~\eqref{H12} can be expressed in terms of incomplete gamma
functions. 
To do so we note that using the Vandermonde convolution \cite{Comtet74}
\begin{align}
(a+b)_n = \sum_{k=0}^n \binom{n}{k} (a)_{n-k} (b)_k
\label{conj3}
\end{align}
and the identity
\begin{align}
(-a)_n = (-1)^n (a-n+1)_n \:,
\end{align}
we can rewrite the coefficients of Eq.~\eqref{H12} as
\begin{align}
\label{eq:pochsum}
\frac{(\nu-j)_{\Nf+1}}{(\nu)_{\Nf}} = 
\sum_{k=0}^{\Nf+1} (-)^k \binom{\Nf+1}{k} 
\frac{(j-k+1)_k}{(\nu+\Nf+1-k)_{k}}  \:.
\end{align}
After substituting this expression in Eq.~\eqref{H12} we find
\begin{align}
\phs_{\nu>0,\hm=0}
  &= e^{-2\ha}
\sum_{k=0}^{\min(\Nf+1,\nu-1)} (-)^k \binom{\Nf+1}{k} 
\frac{1}{(\nu+\Nf+1-k)_{k}}   
  \sum_{j=k}^{\nu-1}
(j-k+1)_k  \frac{\left(2\ha\right)^j}{j!} \:.
  \label{eq:poch2}
\end{align}
Because $(j-k+1)_k=0$ for $k>j$ only terms with $k\le j$ contribute to
the sum in Eq.~\eqref{eq:pochsum}. Therefore the second sum in
Eq.~\eqref{eq:poch2} starts with $j=k$. Moreover, because $j\le \nu-1$
only terms with $k \le \nu-1$ contribute, which explains the upper
limit of the first sum.  Using the series expansion of the incomplete
gamma function with positive integer first argument
\cite[Eq.~(8.352.2)]{Grad:1980},
\begin{align}
  \Gamma(n,z) = (n-1)! e^{-z} \sum_{j=0}^{n-1} \frac{z^j}{j!} \:,
  \label{incGamma2}
\end{align}
Eq.~\eqref{eq:poch2} can be simplified to
\begin{align}
  \phs_{\nu>0,\hm=0}
  &= \frac{1}{(\nu+\Nf)!} \sum_{k=0}^{\min(\Nf+1,\nu-1)} (-)^k
 \binom{\Nf + 1}{k} 
  (\nu - k)_{\Nf+1} (2\ha)^k  \Gamma(\nu-k,2\ha)\:.
  \label{H16.0}
\end{align}
Because $(\nu-k)_{\Nf+1}=0$ for $k \ge \nu$ the sum in
Eq.~\eqref{H16.0} can be extended to $\Nf+1$ for any $\nu\ge 1$.  In
this case, for $n\le0$ Eq.~\eqref{incGamma2} should be replaced by the
usual integral definition \eqref{eq:inc} of $\Gamma(n,z)$.
Thus
\begin{align}
  \phs_{\nu>0,\hm=0}
  &= \frac{1}{(\nu+\Nf)!} \sum_{k=0}^{\Nf+1} (-)^k
 \binom{\Nf + 1}{k} 
  (\nu - k)_{\Nf+1} (2\ha)^k  \Gamma(\nu-k,2\ha)
  \label{H16}
\end{align}
for any $\Nf$ and $\nu \ge 1$.
Even though Eq.~\eqref{H16} was derived for $\nu>0$, it can formally
be continued to $\nu=0$.  In this case we have $(-k)_{N_f+1}=0$ for
$k=0,\ldots,N_f$ so that only the term with $k=N_f+1$ contributes to
the sum.  We obtain
\begin{align}
  \label{eq:chnu0}
  \lim_{\nu\to 0} \phs_{\nu>0,\hm=0}
  &=   (\Nf+1) (2\ha)^{\Nf+1}  \Gamma(-\Nf-1,2\ha)\:.
\end{align}
This reproduces, somewhat surprisingly, the correct $\nu=0$ result
\eqref{phfacchiralnu=0}, even though it originates from a different
term in Eq.~\eqref{Z1total}.  

\section{Thermodynamic limit}
\label{app:thermo}

The thermodynamic limit is defined by $\ha=\mu^2F^2V\to\infty$,
$\hm=mV\Sigma\to\infty$, and $\hm_f=m_fV\Sigma\to\infty$ for
$f=1,\ldots,N_f$.  As in Sec.~\ref{sec:equalmass}, we assume for
simplicity that $\hm=\hm_1=\ldots=\hm_{N_f}$.  In the following we
show that for $2\ha/\hm < 1$, the phase factor in the thermodynamic
limit is given by
\begin{align}
\phs^\text{th} = \left(1-\frac{2\ha}{\hm}\right)^{\Nf+1} \:.
\label{conjecture}
\end{align}
The starting point is Eq.~\eqref{phapp} for the phase factor in the
unquenched case with $\Nf$ dynamical quarks of equal masses.  We now
compute the thermodynamic limit of the Wronskian
$W_{n}(k_1,\ldots,k_n)$.  Starting from the definition $I_{\nu,k}(\hm)
= \hm^k I_{\nu+k}(\hm)$ it is easy to show that the $p$-th derivative
is given by
\begin{align}
I_{\nu,k}^{(p)}(\hm) = \sum_{q=0}^p \binom{p}{q} \frac{k!}{(k-p+q)!}
m^{k-p+q} I_{\nu+k}^{(q)}(\hm) \:, 
\label{deriv}
\end{align}
where the latter expression only contains derivatives of the
$I$-Bessel function.  After substituting this expansion in the
Wronskian determinant and using the asymptotic expansion
\begin{align}
I_\nu(\hm) 
= \frac{e^{\hm}}{\sqrt{2\pi\hm}} \left( 1 + \sum_{j=1}^\infty 
\frac{a_j(\nu)}{\hm^j}\right)+e^{-\hm}(\cdots)
\label{fullexp}
\end{align}
of the $I$-Bessel function one can show, using basic properties of
determinants, that in the thermodynamic limit only the $q=0$ term of
Eq.~\eqref{deriv} contributes to leading order so that
\begin{align}
W_{n}(k_1,\ldots,k_n) 
& \sim
\left(\frac{e^{\hm}}{\sqrt{2\pi}}\right)^n \hm^{\sum_i k_i -n^2/2} 
\Delta_n(k_1,\dots,k_n) 
\end{align}
for arbitrary $\{k_i\}$, where $\Delta_n(k_1,\ldots,k_n)$ is a
Vandermonde determinant.  Using this expression and
Eq.~\eqref{eq:vand}, the Wronskian in the denominator of
Eq.~\eqref{phapp} becomes
\begin{align}
W_\Nf(0, 1, \ldots, \Nf-1)
\sim  \left(  \frac{e^{\hm}}{\sqrt{2\pi\hm}} \right)^\Nf
\prod_{\ell=1}^{\Nf-1} \ell! \:.
\label{denominator}
\end{align}
In the thermodynamic limit $\Ith{k}$ can be computed using Eq.~(C.2)
of Ref.~\cite{Splittorff:2007ck}, resulting in
\begin{align}
\Ith{k} \sim \sqrt{\frac{\pi}{2\hm}} (-)^k \hm^k \left(1- \frac{4\ha}{\hm}\right)^{k} e^{-\hm}
\:,
\label{Ith}
\end{align}
which is independent of $\nu$.  To compute the thermodynamic limit of
the Wronskians in Eq.~\eqref{numerator} we use 
\begin{align}
\Delta_{\Nf+1}(0, \ldots, k-1, k+1, \ldots, \Nf+1)
= \frac{\prod_{\ell=1}^{\Nf+1} \ell!}{(\Nf+1-k)! k!} 
\quad\text{and}\quad
\sum_{{\genfrac{}{}{0pt}{1}{i=0}{i\neq k}}}^{\Nf+1} i =
\frac12(\Nf+2)(\Nf+1) - k
\end{align}
to show that
\begin{align}
W_{\Nf+1}(0, \ldots, k-1, k+1, \ldots, \Nf+1)
&\sim \left(\frac{e^{\hm}}{\sqrt{2\pi}}\right)^{\Nf+1} \hm^{\frac{\Nf+1}{2} - k} 
 \frac{ \prod_{\ell=1}^{\Nf+1} \ell! }{(\Nf+1-k)! k!}\:.
\label{Wth2}
\end{align}
After substituting Eqs.~\eqref{Ith} and \eqref{Wth2}, the thermodynamic
limit of Eq.~\eqref{numerator} becomes
\begin{align}
{\cal W}_{\Nf}(\ha,\hm) \sim
\frac{1}{2} 
\left(\frac{e^{\hm}}{\sqrt{2\pi}}\right)^{\Nf} \hm^{\frac{\Nf}{2}}
\left(\prod_{\ell=1}^{\Nf+1} \ell! \right) 
 \sum_{k=0}^{\Nf+1} 
 \frac{1}{(\Nf+1-k)! k!}  \left(1- \frac{4\ha}{\hm}\right)^{k} 
  \:.
\label{numerator2}
\end{align}
From the binomial theorem we know that
\begin{align}
\sum_{k=0}^n \frac{x^k}{(n-k)!k!} = \frac{(1+x)^n}{n!} \:,
\end{align} 
and hence
\begin{align}
{\cal W}_{\Nf}(\ha,\hm) \sim
2^\Nf 
\left(\frac{e^{\hm}}{\sqrt{2\pi}}\right)^{\Nf} \hm^{\frac{\Nf}{2}} 
\left(\prod_{\ell=1}^{\Nf} \ell! \right) 
 \left(1 - \frac{2\ha}{\hm}\right)^{\Nf+1} \:.
\label{numerator3}
\end{align}
After substituting Eqs.~\eqref{numerator3} and \eqref{denominator} in Eq.~\eqref{phapp} we find the thermodynamic limit for the phase factor,
\begin{align}
\phs^\text{th} = 
\frac{1}{(2\hm)^\Nf \Nf!}
\frac{2^\Nf \left(\frac{e^{\hm}}{\sqrt{2\pi}}\right)^{\Nf} \hm^{\frac{\Nf}{2}} 
\left(\prod_{\ell=1}^{\Nf} \ell! \right)  
 \left(1 - \frac{2\ha}{\hm}\right)^{\Nf+1} }
 { \left(\prod_{\ell=1}^{\Nf-1} \ell! \right)  \left(  \frac{e^{\hm}}{\sqrt{2\pi\hm}} \right)^\Nf }
 \:,
\end{align}
which simplifies to $(1-2\ha/\hm)^{N_f+1}$ as given in
Eq.~\eqref{conjecture}.

\section{Numerical random matrix simulations}
\label{rmtsim}

In this appendix we describe the numerical simulations of random
matrices used to verify the analytical results derived in the main
body of the paper, for both trivial and nontrivial topology.  This
procedure also illustrates the potential usefulness of numerical
simulations in cases where analytical results would not be immediately
accessible.

We performed numerical simulations of random matrices in the chiral
GUE with chemical potential for the quenched case.  As mentioned in
Eq.~\eqref{Dran}, these random matrices can be constructed as
\begin{equation}
  D(\mu) = 
  \begin{pmatrix}
    0 & i\Phi + \mu \Psi \\
    i \Phi^\dagger + \mu \Psi^\dagger & 0
  \end{pmatrix} \:,
  \label{Dranbis}
\end{equation}
where $\Phi$ and $\Psi$ are complex $(N+\nu) \times N$ matrices
generated according to the Gaussian weight function
\begin{equation}
  w(X) \propto \exp(-N \Trace X^\dagger X) 
  = \exp\Big(-N \sum_{k\ell} |X_{k\ell}|^2\Big) 
  = \prod_{k\ell} \exp\Big(-N (\re X_{k\ell})^2\Big) \exp\Big(-N(\im
  X_{k\ell})^2\Big) \:.
\end{equation}
The last expression shows that the real and imaginary parts of each
matrix element are i.i.d.\ random numbers drawn from the Gaussian
distribution
\begin{equation}
  w(x) \propto \exp\left(-N x^2\right)
\end{equation}
with standard deviation $1/\sqrt{2N}$. 

As we want to investigate the microscopic limit of the theory, we will
keep $\ha = 2N\alpha$ and $\hm = 2N m$ constant, while taking $N$
large enough to approach the microscopic limit, in which $N \to
\infty$.  Hence, when generating and diagonalizing the matrices from
Eq.~\eqref{Dranbis}, the chemical potential will be scaled as $\mu =
\sqrt{\ha/2N}$. Each random matrix $D(\mu)$ is then diagonalized and
the real part of the phase factor of its determinant computed with
\begin{align}
  \cos 2\theta = \cos \left[ 2 \sum_{i=1}^{2N+\nu} \arg(\lambda_i+m) \right]\:,
\label{cos2theta}
\end{align}
where the $\lambda_i$ are the eigenvalues and $m = \hm/2N$.  For a
sample with $N_\text{s}$ random matrices the real part of the average
phase factor will be given by
\begin{align}
  \overline{\cos2\theta}(\nu,\ha,\hm) = \frac{1}{N_\text{s}}
  \sum_{j=1}^{N_\text{s}} \cos 2\theta_j \:, 
  \label{sampleaverage}
\end{align}
where $\theta_j$ is the phase of the determinant of the $j$-th random
matrix in the sample, given by Eq.~\eqref{cos2theta}.  For simplicity
we have omitted the subscript $s$ (for the microscopic limit) on
$\cos2\theta$.  The average of the imaginary part will be zero within
the statistical error because of the symmetry properties of the
ensemble and is therefore disregarded in our analysis.

The chiral symmetry of the matrix can be used to improve the
efficiency of the computer implementation by transforming the
$(2N+\nu)\times(2N+\nu)$ diagonalization problem to one of size
$N\times N$. Let us first rewrite Eq.~\eqref{Dranbis} as
\begin{equation}
  D(\mu) = 
  \begin{pmatrix}
    0 & A \\
    B & 0
  \end{pmatrix} \:.
\end{equation}
The eigenvalue equation
\begin{equation}
  D(\mu) v = \lambda v
\end{equation}
can then be written as
\begin{align}
  \begin{pmatrix}
    0 & A \\
    B & 0
  \end{pmatrix}
  \begin{pmatrix}
    v_1 \\
    v_2
  \end{pmatrix}
  = \lambda
  \begin{pmatrix}
    v_1 \\
    v_2
  \end{pmatrix} 
  \:, \quad \text{or} \quad
  \left\{
    \begin{aligned}
      A v_2 &= \lambda v_1 \:,\\
      B v_1 &= \lambda v_2 
    \end{aligned}
  \right. 
  \label{eigeq}
\end{align}
with complex eigenvalues $\lambda$ and eigenvector decomposition
$v=(v_1, v_2)$, where $v_1$ and $v_2$ are complex vectors with $(N+\nu)$
and $N$ elements, respectively.  Without fine-tuning $D(\mu)$ has
exactly $\nu$ zero modes that obey the two homogeneous linear systems
\begin{equation}
  \left\{
    \begin{aligned}
      A v_2 &= 0\:,\\
      B v_1 &= 0 \:.
    \end{aligned}
  \right. 
  \label{zeromode}
\end{equation}
The first system contains $N+\nu$ linear equations with $N$ variables,
and the second one $N$ equations with $N+\nu$ variables.  If both $A$
and $B$ are of rank $N$ (no fine-tuning), the first homogeneous system
only has solutions $v_2=0$, while the second system is
underconstrained and has $\nu$ linearly independent solutions for
$v_1$. Hence the zero modes will be represented by $\nu$ eigenvectors
$(v_{1k},0)$ with $k=1,\ldots,\nu$.

Moreover, it is clear from Eq.~\eqref{eigeq} that each nonzero
eigenvalue $\lambda$ with eigenvector $(v_1,v_2)$ will be paired with
an eigenvalue $-\lambda$ with eigenvector $(v_1,-v_2)$. This is a
consequence of the chiral symmetry of the problem.

These properties of the spectrum of $D(\mu)$ will now be used to
transform the diagonalization problem from order $2N+\nu$ to order
$N$.  For this, let us first multiply Eq.~\eqref{eigeq} from the left
with $D(\mu)$,
\begin{equation}
  D^2(\mu) v = 
  \begin{pmatrix}
    AB & 0 \\
    0  & BA
  \end{pmatrix}
  \begin{pmatrix}
    v_1 \\
    v_2
  \end{pmatrix}
  = \lambda^2
  \begin{pmatrix}
    v_1 \\
    v_2
  \end{pmatrix} \:,
\end{equation}
where $AB$ is an $(N+\nu)\times(N+\nu)$ matrix and $BA$ has dimension
$N\times N$. Clearly, each nonzero eigenvalue $\lambda^2$ of $AB$
(with eigenvector $v_1$) is also an eigenvalue of $BA$ (with
eigenvector $v_2$). However, $AB$ has $\nu$ additional eigenvalues,
which necessarily correspond to the zero modes $(v_{1k},0)$ satisfying
Eq.~\eqref{zeromode}.

This can be used to expedite the numerical simulations. It suffices to
diagonalize the $N\times N$ matrix $BA$ to find the $N$ nonzero
eigenvalues $\lambda_i^2$. We then know that the eigenvalues of
$D(\mu)$ are the $N$ pairs $(\lambda_i,-\lambda_i)$ supplemented by
$\nu$ eigenvalues equal to zero. The determinant for a fermion of mass
$m$ will then be given by
\begin{align}
  \det \left(D(\mu)+m\right) = m^\nu \prod_{i=1}^N (m^2 - \lambda_i^2)\:,
\end{align}
and the real part of its phase factor is
\begin{align}
  \cos 2\theta = \cos \left[ 2 \sum_{i=1}^{N} \arg(m^2-\lambda_i^2) \right] , 
\end{align}
which replaces Eq.~\eqref{cos2theta} in our practical simulations.
Note that the cost of the additional multiplication of an
$N\times(N+\nu)$ by an $(N+\nu)\times N$ matrix to construct the
product $BA$ is negligible compared to the cost of the
diagonalization.

\sloppypar
\bibliography{biblio} 

\providecommand{\href}[2]{#2}\begingroup\raggedright\begin{thebibliography}{10}

\bibitem{Stephanov:2007fk}
M.~A. Stephanov, {\it {QCD phase diagram: An overview}},  {\em PoS} {\bf
  LAT2006} (2006) 024, [\href{http://xxx.lanl.gov/abs/hep-lat/0701002}{{\tt
  hep-lat/0701002}}].

\bibitem{Ambjorn:2002pz}
J.~Ambjorn, K.~N. Anagnostopoulos, J.~Nishimura, and J.~J.~M. Verbaarschot,
  {\it {The factorization method for systems with a complex action -a test in
  Random Matrix Theory for finite density QCD-}},  {\em JHEP} {\bf 10} (2002)
  062, [\href{http://xxx.lanl.gov/abs/hep-lat/0208025}{{\tt hep-lat/0208025}}].

\bibitem{Ambjorn:2004jk}
J.~Ambjorn, K.~N. Anagnostopoulos, J.~Nishimura, and J.~J.~M. Verbaarschot,
  {\it {Non-commutativity of the zero chemical potential limit and the
  thermodynamic limit in finite density systems}},  {\em Phys. Rev.} {\bf D70}
  (2004) 035010, [\href{http://xxx.lanl.gov/abs/hep-lat/0402031}{{\tt
  hep-lat/0402031}}].

\bibitem{Troyer:2004}
M.~Troyer and U.-J. Wiese, {\it {Computational complexity and fundamental
  limitations to fermionic quantum Monte Carlo simulations}},  {\em Phys. Rev.
  Lett.} {\bf 94} (2005) 170201,
  [\href{http://xxx.lanl.gov/abs/cond-mat/0408370}{{\tt cond-mat/0408370}}].

\bibitem{Osborn:2005ss}
J.~C. Osborn, K.~Splittorff, and J.~J.~M. Verbaarschot, {\it {Chiral symmetry
  breaking and the Dirac spectrum at nonzero chemical potential}},  {\em Phys.
  Rev. Lett.} {\bf 94} (2005) 202001,
  [\href{http://xxx.lanl.gov/abs/hep-th/0501210}{{\tt hep-th/0501210}}].

\bibitem{Imachi:2006mw}
M.~Imachi, Y.~Shinno, and H.~Yoneyama, {\it {Sign problem and MEM in lattice
  field theory with the theta term}},  {\em Prog. Theor. Phys.} {\bf 115}
  (2006) 931--949, [\href{http://xxx.lanl.gov/abs/hep-lat/0602009}{{\tt
  hep-lat/0602009}}].

\bibitem{Fukushima:2006uv}
K.~Fukushima and Y.~Hidaka, {\it {A model study of the sign problem in the
  mean-field approximation}},  {\em Phys. Rev.} {\bf D75} (2007) 036002,
  [\href{http://xxx.lanl.gov/abs/hep-ph/0610323}{{\tt hep-ph/0610323}}].

\bibitem{Ejiri:2007ga}
S.~Ejiri, {\it {On the existence of the critical point in finite density
  lattice QCD}},  {\em Phys. Rev.} {\bf D77} (2008) 014508,
  [\href{http://xxx.lanl.gov/abs/0706.3549}{{\tt arXiv:0706.3549}}].

\bibitem{Aarts:2008wh}
G.~Aarts, {\it {Can stochastic quantization evade the sign problem? -- the
  relativistic Bose gas at finite chemical potential}},
  \href{http://xxx.lanl.gov/abs/0810.2089}{{\tt arXiv:0810.2089}}.

\bibitem{Shuryak:1992pi}
E.~V. Shuryak and J.~J.~M. Verbaarschot, {\it {Random matrix theory and
  spectral sum rules for the Dirac operator in QCD}},  {\em Nucl. Phys.} {\bf
  A560} (1993) 306--320, [\href{http://xxx.lanl.gov/abs/hep-th/9212088}{{\tt
  hep-th/9212088}}].

\bibitem{Verbaarschot:2000dy}
J.~J.~M. Verbaarschot and T.~Wettig, {\it {Random matrix theory and chiral
  symmetry in QCD}},  {\em Ann. Rev. Nucl. Part. Sci.} {\bf 50} (2000)
  343--410, [\href{http://xxx.lanl.gov/abs/hep-ph/0003017}{{\tt
  hep-ph/0003017}}].

\bibitem{Verbaarschot:2005rj}
J.~J.~M. Verbaarschot, {\it {QCD, chiral random matrix theory and
  integrability}},  \href{http://xxx.lanl.gov/abs/hep-th/0502029}{{\tt
  hep-th/0502029}}.

\bibitem{Basile:2007ki}
F.~Basile and G.~Akemann, {\it {Equivalence of QCD in the epsilon-regime and
  chiral random matrix theory with or without chemical potential}},  {\em JHEP}
  {\bf 12} (2007) 043, [\href{http://xxx.lanl.gov/abs/0710.0376}{{\tt
  arXiv:0710.0376}}].

\bibitem{Stephanov:1996ki}
M.~A. Stephanov, {\it {Random matrix model of QCD at finite density and the
  nature of the quenched limit}},  {\em Phys. Rev. Lett.} {\bf 76} (1996)
  4472--4475, [\href{http://xxx.lanl.gov/abs/hep-lat/9604003}{{\tt
  hep-lat/9604003}}].

\bibitem{Akemann:2002ym}
G.~Akemann, {\it {Microscopic correlation functions for the QCD Dirac operator
  with chemical potential}},  {\em Phys. Rev. Lett.} {\bf 89} (2002) 072002,
  [\href{http://xxx.lanl.gov/abs/hep-th/0204068}{{\tt hep-th/0204068}}].

\bibitem{Akemann:2002js}
G.~Akemann, {\it {The solution of a chiral random matrix model with complex
  eigenvalues}},  {\em J. Phys.} {\bf A36} (2003) 3363,
  [\href{http://xxx.lanl.gov/abs/hep-th/0204246}{{\tt hep-th/0204246}}].

\bibitem{Osborn:2004rf}
J.~C. Osborn, {\it {Universal results from an alternate random matrix model for
  QCD with a baryon chemical potential}},  {\em Phys. Rev. Lett.} {\bf 93}
  (2004) 222001, [\href{http://xxx.lanl.gov/abs/hep-th/0403131}{{\tt
  hep-th/0403131}}].

\bibitem{Akemann:2007rf}
G.~Akemann, {\it {Matrix models and QCD with chemical potential}},  {\em Int.
  J. Mod. Phys.} {\bf A22} (2007) 1077--1122,
  [\href{http://xxx.lanl.gov/abs/hep-th/0701175}{{\tt hep-th/0701175}}].

\bibitem{Akemann:2003wg}
G.~Akemann and T.~Wettig, {\it {QCD Dirac operator at nonzero chemical
  potential: Lattice data and matrix model}},  {\em Phys. Rev. Lett.} {\bf 92}
  (2004) 102002, [\href{http://xxx.lanl.gov/abs/hep-lat/0308003}{{\tt
  hep-lat/0308003}}].

\bibitem{Bloch:2006cd}
J.~Bloch and T.~Wettig, {\it {Overlap Dirac operator at nonzero chemical
  potential and random matrix theory}},  {\em Phys. Rev. Lett.} {\bf 97} (2006)
  012003, [\href{http://xxx.lanl.gov/abs/hep-lat/0604020}{{\tt
  hep-lat/0604020}}].

\bibitem{Akemann:2007yj}
G.~Akemann, J.~Bloch, L.~Shifrin, and T.~Wettig, {\it {Individual complex Dirac
  eigenvalue distributions from random matrix theory and lattice QCD at nonzero
  chemical potential}},  {\em Phys. Rev. Lett.} {\bf 100} (2008) 032002,
  [\href{http://xxx.lanl.gov/abs/0710.2865}{{\tt arXiv:0710.2865}}].

\bibitem{Ginsparg:1981bj}
P.~H. Ginsparg and K.~G. Wilson, {\it {A remnant of chiral symmetry on the
  lattice}},  {\em Phys. Rev.} {\bf D25} (1982) 2649.

\bibitem{Narayanan:1993sk}
R.~Narayanan and H.~Neuberger, {\it {Chiral determinant as an overlap of two
  vacua}},  {\em Nucl. Phys.} {\bf B412} (1994) 574--606,
  [\href{http://xxx.lanl.gov/abs/hep-lat/9307006}{{\tt hep-lat/9307006}}].

\bibitem{Narayanan:1994gw}
R.~Narayanan and H.~Neuberger, {\it {A Construction of lattice chiral gauge
  theories}},  {\em Nucl. Phys.} {\bf B443} (1995) 305--385,
  [\href{http://xxx.lanl.gov/abs/hep-th/9411108}{{\tt hep-th/9411108}}].

\bibitem{Neuberger:1997fp}
H.~Neuberger, {\it {Exactly massless quarks on the lattice}},  {\em Phys.
  Lett.} {\bf B417} (1998) 141--144,
  [\href{http://xxx.lanl.gov/abs/hep-lat/9707022}{{\tt hep-lat/9707022}}].

\bibitem{Luscher:1998pqa}
M.~L{\"u}scher, {\it {Exact chiral symmetry on the lattice and the
  Ginsparg-Wilson relation}},  {\em Phys. Lett.} {\bf B428} (1998) 342--345,
  [\href{http://xxx.lanl.gov/abs/hep-lat/9802011}{{\tt hep-lat/9802011}}].

\bibitem{Bloch:2007xi}
J.~Bloch and T.~Wettig, {\it {Domain-wall and overlap fermions at nonzero quark
  chemical potential}},  {\em Phys. Rev.} {\bf D76} (2007) 114511,
  [\href{http://xxx.lanl.gov/abs/0709.4630}{{\tt arXiv:0709.4630}}].

\bibitem{Splittorff:2007ck}
K.~Splittorff and J.~J.~M. Verbaarschot, {\it {The QCD sign problem for small
  chemical potential}},  {\em Phys. Rev.} {\bf D75} (2007) 116003,
  [\href{http://xxx.lanl.gov/abs/hep-lat/0702011}{{\tt hep-lat/0702011}}].

\bibitem{Fyodorov:2002jw}
Y.~V. Fyodorov and E.~Strahov, {\it {An exact formula for general spectral
  correlation function of random Hermitian matrices}},  {\em J. Phys.} {\bf
  A36} (2003) 3203--3214, [\href{http://xxx.lanl.gov/abs/math-ph/0204051}{{\tt
  math-ph/0204051}}].

\bibitem{Strahov:2002zu}
E.~Strahov and Y.~V. Fyodorov, {\it {Universal results for correlations of
  characteristic polynomials: Riemann-Hilbert approach}},  {\em Commun. Math.
  Phys.} {\bf 241} (2003) 343--382,
  [\href{http://xxx.lanl.gov/abs/math-ph/0210010}{{\tt math-ph/0210010}}].

\bibitem{Akemann:2004zu}
G.~Akemann and A.~Pottier, {\it {Ratios of characteristic polynomials in
  complex matrix models}},  {\em J. Phys.} {\bf A37} (2004) L453--L460,
  [\href{http://xxx.lanl.gov/abs/math-ph/0404068}{{\tt math-ph/0404068}}].

\bibitem{Akemann:2004dr}
G.~Akemann, J.~C. Osborn, K.~Splittorff, and J.~J.~M. Verbaarschot, {\it
  {Unquenched QCD Dirac operator spectra at nonzero baryon chemical
  potential}},  {\em Nucl. Phys.} {\bf B712} (2005) 287--324,
  [\href{http://xxx.lanl.gov/abs/hep-th/0411030}{{\tt hep-th/0411030}}].

\bibitem{Bergere:2004cp}
M.~C. Bergere, {\it Biorthogonal polynomials for potentials of two variables
  and external sources at the denominator},
  \href{http://xxx.lanl.gov/abs/hep-th/0404126}{{\tt hep-th/0404126}}.

\bibitem{Abram:1964}
M.~Abramowitz and I.~A. Stegun, {\em Handbook of Mathematical Functions with
  Formulas, Graphs, and Mathematical Tables}.
\newblock Dover, New York, ninth {D}over printing, tenth {GPO} printing~ed.,
  1964.

\bibitem{BatemanII}
A.~Erd{\'e}lyi, ed., {\em Higher Transcendental Functions, vol II. Bateman
  Manuscript Project.}
\newblock McGraw-Hill Book Company, Inc., 1953.

\bibitem{Osborn:2008jp}
J.~C. Osborn, K.~Splittorff, and J.~J.~M. Verbaarschot, {\it {Chiral condensate
  at nonzero chemical potential in the microscopic limit of QCD}},  {\em Phys.
  Rev.} {\bf D78} (2008) 065029, [\href{http://xxx.lanl.gov/abs/0805.1303}{{\tt
  arXiv:0805.1303}}].

\bibitem{Grad:1980}
I.~Gradshteyn and I.~Ryzhik, {\em Table of Integrals, Series and Products}.
\newblock Academic Press, San Diego, U.S.A.; London, U.K., 5th~ed., 1994,
  errata:
  \href{http://www.mathtable.com/errata/gr6_errata.pdf}{http://www.mathtable.c%
om/errata/gr6{\_}errata.pdf}.

\bibitem{Comtet74}
L.~Comtet, {\em Advanced Combinatorics: The Art of Finite and Infinite
  Expansions}.
\newblock D. Reidel Publishing Company, 1974.

\end{thebibliography}\endgroup
\bibliographystyle{JHEP}

\end{document}